\newif\iffigs\figstrue

\documentclass[12pt]{article}
\iffigs
  \input epsf
\else
\fi

\batchmode
\newfont{\footscrfont}{rsfs10}
  \newfont{\footbbbfont}{msbm10}
  \newfont{\manfont}{manfnt}
\errorstopmode

\newif\ifscrf\scrftrue
\ifx\footscrfont\nullfont
  \scrffalse
\fi

\newif\ifamsf\amsftrue
\ifx\footbbbfont\nullfont
  \amsffalse
\fi

\def\ppnumber{\vbox{\baselineskip14pt\hbox{CU-TP 963}
\hbox{hep-th/yymmxxx}}}
\def\ppdate{January 2000}
\def\pplogo{\vbox{\kern-\headheight\kern -15pt
\halign{##&##\hfil\cr&{
\ppnumber}\cr\rule{0pt}{2.5ex}&\ppdate\cr}
}}

\makeatletter
\date{}
\def\dedicatory#1{\def\@date{\normalsize\it#1}}
\def\subjclass#1{\def\@thefnmark{}\@footnotetext{1991
    {\it Mathematics Subject Classification.} #1}}
\def\keywords#1{\def\@thefnmark{}\@footnotetext{
    {\it Key words and phrases.} #1}}

\def\ps@firstpage{\ps@empty \def\@oddhead{\hss\pplogo}%
  \let\@evenhead\@oddhead 
}
\def\maketitle{\par
 \begingroup
 \def\thefootnote{\fnsymbol{footnote}}
 \def\@makefnmark{\hbox
 to 0pt{$^{\@thefnmark}$\hss}}
 \if@twocolumn
 \twocolumn[\@maketitle]
 \else \newpage
 \global\@topnum\z@ \@maketitle \fi\thispagestyle{firstpage}\@thanks
 \endgroup
 \setcounter{footnote}{0}
 \let\maketitle\relax
 \let\@maketitle\relax
 \gdef\@thanks{}\gdef\@author{}\gdef\@title{}\let\thanks\relax}

\def\abstract{\if@twocolumn
\section*{Abstract}
\else \small
\begin{center}
{\bf ABSTRACT}
\end{center}
\quotation
\fi}

\def\thebibliography#1{\section*{References\@mkboth
 {REFERENCES}{REFERENCES}}\small\list
 {[\arabic{enumi}]}{\settowidth\labelwidth{[#1]}\leftmargin\labelwidth
 \advance\leftmargin\labelsep
 \usecounter{enumi}}
 \def\newblock{\hskip .11em plus .33em minus .07em}
 \sloppy\clubpenalty4000\widowpenalty4000
 \sfcode`\.=1000\relax}

\newif\iffn\fnfalse

\@ifundefined{reset@font}{\let\reset@font\empty}{} 
\long\def\@footnotetext#1{\insert\footins{\reset@font\footnotesize
    \interlinepenalty\interfootnotelinepenalty
    \splittopskip\footnotesep
    \splitmaxdepth \dp\strutbox \floatingpenalty \@MM
    \hsize\columnwidth \@parboxrestore
   \edef\@currentlabel{\csname p@footnote\endcsname\@thefnmark}\@makefntext
    {\rule{\z@}{\footnotesep}\ignorespaces
      \fntrue#1\fnfalse\strut}}}

\makeatother

\ifamsf
  \newfont{\bigbbbfont}{msbm10 scaled\magstep2}
  \newfont{\bbbfont}{msbm10 scaled\magstep1}  
  \newfont{\smallbbbfont}{msbm8}
  \newfont{\tinybbbfont}{msbm6}
  \newfont{\smallfootbbbfont}{msbm7}
  \newfont{\tinyfootbbbfont}{msbm5}
  \newfont{\biggthfont}{eufm10 scaled\magstep2}
  \newfont{\gthfont}{eufm10 scaled\magstep1}  
  \newfont{\smallgthfont}{eufm8}
  \newfont{\tinygthfont}{eufm6}
  \newfont{\footgthfont}{eufm10}
  \newfont{\smallfootgthfont}{eufm7}
  \newfont{\tinyfootgthfont}{eufm5}
\fi

\ifscrf
  \newfont{\scrfont}{rsfs10 scaled\magstep1}  
  \newfont{\smallscrfont}{rsfs7}
  \newfont{\tinyscrfont}{rsfs7}
  \newfont{\smallfootscrfont}{rsfs7}
  \newfont{\tinyfootscrfont}{rsfs7}
\fi

\ifamsf
  \newcommand{\Bbb}[1]{\iffn
      \mathchoice{\mbox{\footbbbfont #1}}{\mbox{\footbbbfont #1}}
      {\mbox{\smallfootbbbfont #1}}{\mbox{\tinyfootbbbfont #1}}\else
      \mathchoice{\mbox{\bbbfont #1}}{\mbox{\bbbfont #1}}
      {\mbox{\smallbbbfont #1}}{\mbox{\tinybbbfont #1}}\fi}
  
\else
  \def\bigbbbfont{\bf}
  \def\Bbb{\bf}
  
\fi

\ifscrf
  \newcommand{\Scr}[1]{\iffn
    \mathchoice{\mbox{\footscrfont #1}}{\mbox{\footscrfont #1}}
    {\mbox{\smallfootscrfont #1}}{\mbox{\tinyfootscrfont #1}}\else
    \mathchoice{\mbox{\scrfont #1}}{\mbox{\scrfont #1}}
    {\mbox{\smallscrfont #1}}{\mbox{\tinyscrfont #1}}\fi}
\else
  \def\Scr{\cal}
\fi

\def\C{{\Bbb C}}
\def\F{{\cal F}}

\def\P{{\Bbb P}}

\def\R{{\Bbb R}}
\def\Z{{\Bbb Z}}

\def\bearray{\begin{eqnarray}}
\def\eearray{\end{eqnarray}}
\def\bearraynn{\begin{eqnarray*}}
\def\eearraynn{\end{eqnarray*}}
\def\bfig{\begin{figure}}
\def\efig{\end{figure}}

\def\opeq#1{\advance\lineskip#1 \advance\baselineskip#1
        \advance\lineskiplimit#1}

\def\cM{{\Scr M}}

\def\cD{{\Scr D}}

\def\cMc{{\hfuzz=100cm\hbox to 0pt{$\;\overline{\phantom{X}}$}\cM}}
\def\barcD{{\hfuzz=100cm\hbox to 0pt{$\;\overline{\phantom{X}}$}\cD}}

\def\F#1#2{{}_{#1}F_{#2}}

\ifamsf

\else

\fi

\def\boldone{\relax{\rm 1\kern-.35em 1}}

\newtheorem{Proposition}{Proposition}[section]

\newtheorem{Theorem}{Theorem}[section]
\newtheorem{Lemma}{Lemma}[section]
\newtheorem{Corrolary}{Corrolary}[section]

\newcommand{\be}{\begin{equation}}
\newcommand{\ee}{\end{equation}}
\newcommand{\bea}{\begin{eqnarray}}
\newcommand{\eea}{\end{eqnarray}}

\newcommand{\bp}{\begin{Proposition}}
\newcommand{\ep}{\end{Proposition}}
\newcommand{\bt}{\begin{Theorem}}
\newcommand{\et}{\end{Theorem}}
\newcommand{\bl}{\begin{Lemma}}
\newcommand{\el}{\end{Lemma}}
\newcommand{\bc}{\begin{Corrolary}}
\newcommand{\ec}{\end{Corrolary}}
\newcommand{\nn}{\nonumber}

\def\HG_symbol[#1,#2]{
\left(
\begin{array}{c}
#1\\
#2
\end{array}
\right)
}

\def\foo[#1][#2,#3,#4,#5]{{\rm \bf #1}
\left(
\begin{array}{ccc}
#2&~~~~&#3\\
~~&~~~~&~~ \\
#4&~~~~&#5
\end{array}
\right)
}

\def\Foo[#1][#2,#3]{{\rm \bf #1}
\left(
\begin{array}{c}
#2\\
#3
\end{array}
\right)
}

\def\D[#1,#2][#3,#4]{{\rm D}^{#1}_{#2}
\left(
\begin{array}{c}
#3\\
#4
\end{array}
\right)
}

\def\I[#1,#2,#3,#4](#5){{\rm \bf I}
\left(
\begin{array}{ccc}
#1&~~~~&#2\\
~~&~~~~&~~ \\
#3&~~~~&#4
\end{array}
\right)
(#5)}

\def\M[#1,#2,#3,#4](#5){{\rm \bf G}
\left(
\begin{array}{ccc}
#1&~~~~&#2\\
~~&~~~~&~~ \\
#3&~~~~&#4
\end{array}
\right)
(#5)}

\def\G[#1,#2]{{\rm \Gamma}
\left(
\begin{array}{c}
#1\\
#2
\end{array}
\right)}

\def\B[#1,#2][#3,#4,#5,#6](#7){
\frac{1}{2\pi i}\int_{#1}{d#2~\G[{#3,#4},{#5,#6}]{#7}^{#2}}
}

\def\F[#1,#2][#3,#4](#5){_{#1}{\rm F}_{#2}
\left(
\begin{array}{c}
#3\\
#4
\end{array}
\right)
(#5)}

\usepackage{graphics}

\begin{document}

\title{Collapsing D-Branes in Calabi-Yau Moduli Space: I}

\author{B.~R.~Greene$^{a}$,\\C.~I.~Lazaroiu$^{b}$}


\maketitle

\vbox{
\centerline{Departments of Physics and Mathematics}
\centerline{Columbia University}
\centerline{New York, N.Y. 10027}
\medskip
\medskip
\bigskip
}

\abstract{We study the quantum volume of D-branes wrapped around
various cycles in  Calabi-Yau manifolds, as the manifold's moduli are varied.
In particular, we focus on the behaviour of these D-branes near 
phase transitions between distinct low energy physical descriptions
of the resulting string theory. Whereas previous studies have solely considered
quantum volumes in the context of two-cycles  in perturbative string theory or
D-branes in the specific example of the quintic hypersurface, 
we work more generally and find qualitatively new features. 
On the mathematical side, as we briefly note, our work has
some interesting implications for certain issues in arithmetics.
}

\vskip .6in

$^a$ greene@phys.columbia.edu

$^b$ lazaroiu@phys.columbia.edu

\pagebreak

\section*{Introduction}

Over the years, studies in quantum geometry have revealed
both quantitative and qualitative deviations from expectations
based on classical geometry. Such distinctions, in turn, have helped
us understand physical properties that are intrinsically ``stringy''. 
From the earliest studies, based solely on perturbative string theory,
we learned that a ``quantum'' Calabi-Yau is in some sense
a classical  Calabi-Yau manifold together with all of its rational
curves \cite{quintic}.  More
recently, with the advent of D-branes as non-perturbative 
probes, we realized that a ``quantum'' Calabi-Yau manifold 
is sensitive to all of its supersymmetric cycles since, regardless
of the cycle's dimension, there is some D-brane that can wrap around  
it (thinking collectively in type IIA and type IIB string theory).

In this paper, we focus on the notion of {\em volume} from a
string theoretic perspective \footnote{ Other approaches to quantum
gravity --- most notably, loop quantum gravity --- have found
interesting quantum mechanical features of volume \cite{ASR}
so perhaps this area may be a fruitful point of intersection.}.
There are two papers 
that provide the groundwork for this undertaking. The first is
\cite{small_distances1} 
which made use of mirror symmetry to calculate the instanton
action associated with strings whose worldsheet is wrapped around
a holomorphic two-cycle. 
The second is \cite{quantum_volumes}, in which another
approach to measuring quantum volumes --- based on 
the masses of wrapped D-branes --- was proposed.

Our goal in this paper is to begin a systematic union of the analyses
of those papers 
in the following sense: In \cite{small_distances1}
phase boundaries in the $N = 2$ moduli space 
were studied  with only the crudest
of tools --- perturbative string dynamics. This implies that 
\cite{small_distances1}
only had access to the quantum volume of two-cycles. A natural question
that immediately follows is: what can be learned from extending this 
analysis by probing phase boundaries with higher
dimensional D-branes?
In \cite{quantum_volumes} we learned, in one example (the quintic hypersurface 
of \cite{quintic}), of the kind of 
surprises that such a study may reveal: it was shown there that 
the six-cycle acquires zero quantum volume at the mirror of the conifold 
point, though the quantum volumes of two and four cycles remain nonzero
\footnote{This was also found independently
by \cite{PS}.}. This is a significant deviation from classical
expectations since on a manifold with only one K\"ahler modulus ---
a single ``breathing mode'' --- classical geometry requires that
vanishing of a six-cycle's volume entails the vanishing of all sub-cycle 
volumes.

We anticipate that some D-brane mass 
becomes zero at real codimension one along any phase boundary
(since such boundaries intersect the discriminant locus of
the mirror Calabi-Yau space), and the lesson of
\cite{quantum_volumes} is 
that classical geometric expectations regarding such
degenerations can be completely misleading. 
We will see that this lesson is quite general by finding 
disagreement between the classical and quantum pictures in a variety of 
one and two-parameter models. As an interesting 
aside we point out that the vanishing of certain D-brane masses at special
boundary points in $N = 2$ moduli space can be used as a tool for
generating arithmetic identities
which appear to be independent of known relations. While we have little
insight beyond that afforded by the physics of D-branes, this 
mathematical question seems to be ripe for further study.

The present work can be thought of as  complimentary to
the approach of \cite{Douglas_quintic}. The authors of that paper
use conformal field theory methods \cite{boundary_states, Ishibashi} to 
construct D-brane boundary states at the Landau-Ginsburg point
in the quintic moduli space, and then study the
behaviour of those states as one perturbs away from this limit. 
Their goal, which they largely accomplish, is to find the geometric
interpretation of such states in terms of wrapped D-branes
and extract quantum geometrical implications from a conformal field
theory analysis. In our work, we start at the ``other end'' of the moduli-space
(the large radius limit), in which the classical geometry
is manifest. We then move toward other phases (including but not limited
to the Landau-Ginsburg phase) and study what happens to wrapped
D-branes. In time, these two approaches should shed substantial light
on the non-perturbative structure of the string theory moduli space. 

In Section 1 we briefly review the results of \cite{small_distances1} and 
\cite{quantum_volumes} in the light of modern understanding of D-branes. 
We also give a short discussion of what has (and has not) 
been achieved in the literature. 
Section 2 introduces some old but largely overlooked
\footnote{An example where Meijer functions were used to perform the 
analytic continuation of periods can be found in \cite{Zaslow}. We thank 
E.~Zaslow for bringing this reference to our attention. Issues of analytic 
continuation were also recently discussed in \cite{Horja}, in connection with 
the Kontsevitch conjecture \cite{kontsevich}.}
mathematical methods that will greatly assist the calculations
we perform. In section 3 we discuss some
relevant issues regarding mirror symmetry and
integral structures. In Sections 4 through 7 we apply these methods to a 
number of compact and non-compact examples, 
investigating the quantum  volumes of two, four, and six-cycles
at various interesting points in their moduli space.
Finally, Section 8 discusses some open problems and presents 
our conclusions.
The two appendices give a brief primer on the notion of monodromy weight
filtration, as well as a list of useful identities.

\section{Quantum Volumes}

In this section we review some ideas relevant for the analysis of 
quantum volumes in string theory, and then
outline our strategy for implementing them.

\subsection{Perturbative Quantum Volumes}

String theory on Calabi-Yau manifolds provides a rich arena
for investigating quantum geometry. When a Calabi-Yau space
$X$ is ``large'', observables in the corresponding string model can
be expressed in terms of geometrical properties
of the compactification manifold. As this space decreases in size, 
the values of many observables continuously shift from their large radius
limits, thereby requiring shifted or deformed geometrical constructs
for their interpretation. Collectively, these constructs --- 
which reduce to classical geometric
counterparts in the limit of large scales --- yield what we mean by
``quantum'' geometry \footnote{We adhere to common terminology which
does not distinguish between ``stringy'' geometry and
``quantum'' geometry. More precisely, stringy geometry is concerned with the 
behaviour of compactified string theory for small string coupling constant,
i.e. with the moduli space of closed conformal field theory defined on the 
Riemann sphere. In this paper, we go beyond this framework by including 
D-branes, so strictly speaking we study the moduli space 
of {\em open} conformal field theory at genus zero, i.e. 
the moduli space of boundary conformal field theory on a 
sphere. }.

A well known example of this idea arises in perturbative string theory,
and provides the first approach to discussing quantum volumes.
If we consider the three-point correlation function of chiral primary
operators ${\cal O}_i$ associated with classes $A_i$ in $H^{1,1}(X,\C)$, 
we find:
\be
\label{threepoint}
<{\cal O}_i {\cal O}_j {\cal O}_k> = \int_X A_i \wedge A_j \wedge A_k +
\sum_{\Gamma}{\frac{{\bf q}^\Gamma}{1-{\bf q}^\Gamma}N_\Gamma(A_i,A_j,A_k)}~~,
\ee
where $\Gamma$ runs over the homology classes of rational curves in $X$, 
${\bf q}^\Gamma$ is a monomial in the variables $q_l=e^{2\pi i t_l}$, with 
$t_l$ the coordinates of the complexified K\"ahler class 
$B+iJ=\sum_{l}{t_lE_l}$ in a basis ${E_l}$ of $H^2(X,\Z)$ and 
$N_\Gamma(A_i,A_j,A_k)$ are the Gromov-Witten invariants 
\cite{gromov,witten_tsm} of $X$ 
(see \cite{GW} for a review). The sum over Gromov-Witten invariants in the 
right hand side appears through an instanton expansion in the nonlinear sigma 
model, with the powers of $q_l$ characterizing the contributions from 
instantons wrapping various holomorphic 2-cycles. 
In the limit that
the volume of $X$ and all of its rational curves goes to infinity, the 
instanton corrections are suppressed and the correlator 
can be interpreted in terms of the triple intersection form
of $X$:
\be
<{\cal O}_i {\cal O}_j {\cal O}_k> \approx \int_X{A_i \wedge A_j \wedge A_k}~~.
\ee
But as we move away from this limit, 
the  instanton corrections are generally nonzero, yielding the {\em quantum} 
triple 
intersection --- the quantum cohomology ring --- on $H^{1,1}(X,\C)$.

The right hand side of (\ref{threepoint}) is calculationally unwieldy, but as 
first shown in
\cite{greene_plesser} and subsequently evaluated to successful conclusion in
\cite{quintic}, if $X$ has a mirror $Y$ then (\ref{threepoint}) can be
evaluated {\em exactly} by a variation of Hodge
structure calculation on $Y$. This procedure has been employed in a
variety of circumstances, one of the most well-known being the 
enumeration of rational curves on the quintic hypersurface carried out
in \cite{quintic}.  Our interest here, though, is not in enumerating
rational curves. Rather, as implicitly used in \cite{quintic} and
developed in \cite{small_distances1}, we note that at a given point in the
conformal field theory moduli space we can use the instanton expansion
on the right hand side to {\em define} the
`quantum volume' of the rational curves in $X$. Namely, the curves are 
assigned whatever volume is necessary to make the expansion correct. 
More precisely, 
the variables $t_l$
appearing in the instanton expansion are tuned to those values which ensure
that equality (\ref{threepoint}) holds. Since 
$t_l = \int_{e_l}{(B + iJ)}$ (with $(e_l)$ the basis of $H_2(X,\Z)$ 
dual to $(E_l)$), we interpret the absolute value of
$t_l$ as the quantum volume of the holomorphic two-cycles in the class $e_l$.

As strongly motivated in \cite{quintic} and subsequently proved in
\cite{BCOV}, this constraint is met if $t_l$ are related to the complex 
structure of the mirror $Y$ through the mirror map:
\be
\label{mirrormap}
t_l = B_l + iJ_l =\frac{\int_{\gamma_l} \Omega(z)}{\int_{\gamma_0}\Omega(z)}
\ee where $z$ are coordinates on the complex structure moduli space of
$Y$, 
$\gamma_l \in H_3(Y, \Z)$ are suitably chosen three-cycles in $Y$ (to
be discussed in more detail shortly), and $\Omega(z)$ is the
holomorphic three-form of $Y$ at the point $z$.

Given a mirror pair $(X,Y)$, it follows that we can map out the quantum volumes
of rational curves in $X$ as its K\"ahler structure is varied 
by calculating appropriate ratios
of periods of $\Omega$, as functions of the complex structure of $Y$. 
Classically, $B_l + iJ_l$  can take on any
value such that $J_l \geq 0$, but the approach just described generally
yields a different conclusion. For instance, in some examples,
$J_l \ge r_l > 0$, implying a minimum size for certain rational
curves. This departure from the classically allowed two-cycle volumes
motivates the use of the term
``quantum volume'' even though in this case 
$B_l + iJ_l = \int_{e_l}{(B + i J)}$ is a classical formula.
The quantum aspect arises when this formula is examined over
the space of physically realizable models.

This approach for studying the quantum volumes of
two-cycles was developed and applied in a number of
different contexts in \cite{small_distances1} and \cite{small_distances2}. 
For example, the quantum 
volume of two-cycles involved in the
flop transitions of \cite{topchange}
were shown to attain zero value at the center of
the flop (the conifold point of the transition), in keeping
with classical expectations. But in another example in
which one moves
from a smooth Calabi-Yau phase to a phase in which the Calabi-Yau 
manifold acquires a $\Z_3$ orbifold singularity, two-cycle volumes
were shown to be bounded from below, in conflict with geometric
expectations (at an abelian quotient singularity of
this type, an exceptional divisor has been blown down to a point,
which entails vanishing of the geometric volume of all holomorphic 
two-cycles embedded in the divisor).
Thus, there can be significant differences between the quantum and
classical  properties of two-cycle volumes.

While revealing these interesting features of quantum volumes, the
approach of \cite{small_distances1} is deficient for two main reasons: First,
it is limited to the study of two-cycle volumes and second 
(relatedly) it only uses perturbative string probes.
 D-branes, as
emphasized in \cite{quantum_volumes} provide a natural 
tool for going further.

\subsection{Nonperturbative Quantum Volumes}

Since BPS saturated D-brane states wrapped around cycles in a Calabi-Yau
space have a mass that depends on the ``size'' of
the cycles they wrap, we can use these masses to define the quantum volume of
cycles. Since Type IIA string theory has even dimensional branes while
Type IIB string theory has odd dimensional branes, considering both theories 
compactified on the same Calabi-Yau manifold $X$ 
allows us to define a notion of quantum volume for cycles of all 
possible dimensions.

More specifically,  in Type IIB string theory on 
$X$, BPS saturated D3-branes wrap
three-dimensional special Lagrangian cycles of $X$ \cite{Dcycles}
(these are the type A branes of \cite{Ooguri}). While it is a difficult
mathematical problem to identify the special Lagrangian 3-cycles of 
a Calabi-Yau space, if $C$ is such a cycle then the mass of a D3-brane
wrapped on $C$ depends only on the homology class $\gamma = [C]\in H_3(X)$ 
of $C$, and identifying elements $\gamma$ of $H_3(X,\Z)$ is a much easier 
problem. If, for example,
we choose an integral basis $\gamma_i$ of $H_3(X,\Z)$, then we can express
$\gamma = q^i \gamma_i$ with $q^i \in \Z$ (note that $[C]$ is an
integral class since $C$ is a submanifold of $X$) and write the BPS mass
as \cite{BPS_B,quantum_volumes}:
\be
\label{BPSmassesinIIB}
m(\gamma)=
\frac{|\int_{\gamma}{\Omega}|}{(\int_{X}{\Omega\wedge {\overline \Omega}}
)^{1/2}}~~=
\frac{|q^i\int_{\gamma_i}{\Omega}|}{(\int_{X}{\Omega\wedge 
{\overline \Omega}})^{1/2}} ~~.
\ee
The integers $q^i$ can be interpreted as gauge charges with respect to
the $U(1)$-gauge fields resulting from the Kaluza-Klein reduction of the 
Ramond-Ramond four-form along the cycles 
$\gamma_i$, while the periods $\int_{\gamma_i}{\Omega}$ are related to the 
vacuum expectation values of the scalar fields
which parameterize the complex structure moduli space. Hence 
(\ref{BPSmassesinIIB}) takes the standard form of charges
times vevs. Since $C$ is special Lagrangian\footnote{We assume that 
$C$ is special Lagrangian with respect to $\Omega(z)$. This condition depends 
on the point $z$ on the moduli space, as discussed in some detail in 
\cite{Joyce}--such dependence is related to issues of marginal stability, 
which we ignore in this paper.}, we have $m(\gamma)=vol(C)$ with
the proper normalization of $\Omega$. Hence the mass of an A-type $D$-brane
is proportional to the volume of the cycle it wraps, a relation which is 
quantum-mechanically exact. Thus, modulo issues of marginal stability, 
no interesting quantum effects contribute to the  masses of $A$-type D-branes.

In the type IIA theory on $X$, BPS saturated D-branes 
wrap holomorphic cycles (these are the type B-branes of \cite{Ooguri}).
If we choose an integral basis $h_i$ of
$H_{\rm even}(X,\Z) = H_0(X,\Z) \oplus  H_2(X,\Z) \oplus H_4(X,\Z) \oplus
H_6(X,\Z)$, then a D-brane wrapping a holomorphic cycle in the 
class $h = {\tilde q}^i h_i$ has mass
of the form $|\tilde q^i \phi_i|$ where $\phi_i$ are the
vevs of the scalar fields associated with the K\"ahler structure of
$X$. In the large radius limit the mass of the D-brane is 
proportional to the classical volume of the cycle it wraps, but this 
simple relation is destroyed by quantum corrections as one moves away 
from this limit. 

As observed in \cite{quantum_volumes}, one can can turn this argument
around and {\em define} the {\em quantum volume}
of a holomorphic cycle of $X$ to be proportional to
the mass of the type $B$-brane wrapping it. This is a natural generalization 
of the geometric notion of volume inasmuch as its reduces to the former in 
the large radius limit, where the full content of the string theory 
compactified on $X$ (at weak string coupling) can be described in 
terms of standard geometric concepts.

Unfortunately, whereas the integrals  
$\int_{\gamma_i}{\Omega}$
give a simple, closed form for the relevant vevs for the odd-dimensional
branes on $X$ (associated with the complex structure of $X$) there is
no simple, closed form expression for the $\phi_i$. 
Indeed, the semiclassical formula relating 
the mass of such a brane to the volume of the associated holomorphic cycle 
(and the values of the $B$-field and worldvolume gauge field)
is known to receive important quantum corrections, which are difficult to  
analyze directly on $X$. The situation is in many ways similar to that 
encountered for the action of string instantons 
wrapped over rational curves and encoded by  
relations (\ref{threepoint}) and 
(\ref{mirrormap}).
However, we can follow \cite{quantum_volumes} and circumvent this 
obstacle by using mirror symmetry. 

For this, consider the Type IIB theory compactified on the mirror $Y$
of $X$. Conformal and effective 
field theory arguments \cite{Ooguri} show that under mirror symmetry 
$D_{2k}$-branes in the IIA theory on $X$ are mapped to $D3$-branes in the 
IIB-theory on $Y$\footnote{More precisely, they are mapped to $D3$-branes 
wrapping special Lagrangian cycles whose homology classes belong to the 
component ${\cal W}_k$ of the reduced monodromy weight filtration 
associated with the 
large complex structure point of $Y$ (see appendix A).}. This identifies 
the mass of a type B D-brane with the mass of its mirror type A 
brane, which is amenable to direct computation via equation 
(\ref{BPSmassesinIIB}).
Hence invoking both $X$ and its mirror allows us reduce the problem of
computing masses of BPS saturated D-branes
to computing periods of holomorphic three-forms
\footnote{It is believed \cite{Moore} that the moduli space 
of type $B$ branes admits a physically correct compactification related to 
certain moduli spaces of semi-stable simple sheaves on $X$, which should 
form the natural framework for finding an expression of the associated BPS 
charge directly in terms of type $B$ D-brane data. Such an analysis will 
necessarily encode important information about moduli spaces of stable 
sheaves on $X$. By using mirror symmetry, we are somehow `summing up' this 
information and re-expressing it in terms of the data of $Y$--
its complex structure and the collection of its special Lagrangian 3-cycles.
It should be possible to follow this approach in the other direction, 
in order to extract information about moduli spaces of sheaves on $X$.}.

\subsection{Strategy of Calculation}

In view of (\ref{BPSmassesinIIB}), 
our strategy for ``measuring'' quantum volumes of holomorphic
submanifolds of $X$ is in principle as follows:
\vskip.1in
(1) Choose an integral basis of three-cycles $\gamma_i$ in
    $H_3(Y,\Z)$.
\vskip.1in
(2) Calculate the periods $\int_{\gamma_i} \Omega$ with respect to
this basis for all points in the complex structure moduli space of $Y$.
\vskip.1in
(3) Find the explicit mirror map between $\oplus_j H_{2j}(X, \Z)$ and
    $H_3(Y,\Z)$.
\vskip.1in
(4) Determine the quantum volume of any even-dimensional cycle in
$\oplus_j H_{2j}(X, \Z)$ by equating it (up to an overall normalization)
with the mass of a D-3-brane wrapped on the mirror three-cycle in $Y$.
\vskip.1in
In practice, one encounters two 
\footnote{In fact, there is a third -- and much more serious -- problem
involved in the whole subject, which belies many difficulties of 
interpretation and various issues of marginal stability touched upon in 
\cite{Douglas_quintic}. This rests on the following question: given an integral 
class $\gamma \in H_3(Y,\Z)$, does this class support a special 
Lagrangian 3-cycle? If so, what is the moduli space of cycles in this class?
We will mostly ignore this issue in the present 
paper, though we plan to discuss some aspects of this problem elsewhere. 
For recent work on this subject 
we refer the 
reader to \cite{Joyce} (see also \cite{Kachru_Greevy} for a partial physical 
translation of some of those results). A complete understanding of this issue 
will most probably require a systematic development of
 the theory of moduli spaces of 
special Lagrangian submanifolds along the lines of \cite{Hitchin},
\cite{Gross}.}
difficulties in carrying out
this strategy.  First, it is generally a significant challenge to find
explicitly a basis of integer three-cycles, i.e. it is difficult to carry 
out step (1). Second, while the mirror map between $H_2(X,\C)$ and 
(a part
\footnote{This is the component ${\cal W}_1\subset H_3(Y)$ 
of the large complex 
structure reduced monodromy 
weight filtration. The behaviour of the elements of ${\cal W}_1$ 
under mirror symmetry is clearly discussed in 
\cite{morrison_quintic, morrison_cpctfs}.} of) 
$H_3(Y,\C)$ is well understood, and while we certainly know that
this map extends to an isomorphism between $\oplus_j H_{2j}(X, \Z)$ and
$H_3(Y,\Z)$, realizing this map explicitly is generally beyond what as
yet has been worked out, i.e. step (3) nontrivial. How, then, do we
proceed?

Let us deal with the second problem first. In \cite{quintic} the
authors showed that integral two-cycles on $X$ are mirror to integral
three-cycles on $Y$ with prescribed monodromy properties around a
large complex structure point in the moduli space.  Concretely, in a
one-parameter example, the three-cycle $\gamma_l$ in the numerator of
(\ref{mirrormap}) must satisfy $\gamma_l \rightarrow \gamma_l +
\gamma_0$ upon parallel transport about a large complex structure
point. When translated into the language of periods, this implies that
$\int_{\gamma_l} \Omega$ has a logarithmic dependence on the
coordinate $z$ (chosen such that $z = 0$ is the large complex structure point).
In \cite{quintic, morrison_aspects,msh,Morrison_II, SYZ}, this notion was 
formalized and
generalized to the statement that holomorphic cycles of real dimension
$2j$ on $X$ are mirror to three-cycles on $Y$ whose periods have
leading ${\rm log}^j z$ behavior near $z = 0$. Thus, finding a complete
set of periods of $\Omega$ and classifying their leading logarithmic
behavior gives us a means of identifying the dimension of their
even-cycle counterpart on $X$.

Although this is part of the approach we will follow, it is certainly
deficient since it says nothing about the subleading logarithmic
dependence of the periods. For instance, is the six-cycle on $X$
mirror to a three-cycle on $Y$ whose period is purely proportional to
${\rm log}^3(z)$ or is there some nontrivial dependence on ${\rm log}^2(z)$ and
${\rm log}(z)$? This is a question that, at present, we do not know how to
answer completely---it awaits a more thorough understanding of the action of
mirror symmetry on the integral structure of a mirror pair.  Thus, in
this paper when we speak about a cycle on $X$ of real dimension $2j$,
that will often refer to a cycle on $X$ with $2j$ being the maximal
dimensional component, but with the identity of the admixture of lower
cycles left unspecified. In Section 3 we
will quantify how close we can currently get
toward erasing this ambiguity. Gaining greater precision in this
regard is an important problem for future work.

As for the first problem mentioned above---that of ensuring that the
three cycles with which we work are integral---we propose the
following. First, our main focus in this paper is to examine D-branes
wrapping cycles whose mass vanishes at special points in the moduli
space. For this purpose we can weaken the requirement that the
$\gamma_i$ are integral classes, and demand only that they are
{\em proportional} to such classes. That is, if $\Lambda = H_3(Y,\Z)$ is
the integral lattice of 3-cycles in
$H_3(Y,\C)$, then all we require is that each of the $\gamma_i$ we
study is proportional to an element in $\Lambda$, with a {\em common} 
proportionality factor. In this case, given a cycle $\gamma$ which is 
an integral linear combination of $\gamma_i$, the vanishing of 
$\int_{\gamma}{\Omega}$ at some point in the moduli space 
implies the existence of an integral 
cycle (of class proportional to $\gamma$) having the same property. 
We call cycles that are proportional to
integral cycles {\em weakly integral}. Second, given a three-cycle
$\gamma$ which is weakly integral, we can generate other 
weakly integral cycles $\gamma^{(j)}$ in $H_3(Y,\C)$ by acting
on it with monodromy transformations. A cycle $\gamma\in H_3(Y,\C)$ with the 
property that 
the set of weakly integral cycles thus produced is a system of generators 
for the complex vector space $H_3(Y,\C)$ will be called {\em cyclic}; indeed, 
it is a cyclic vector for the monodromy representation of the fundamental 
group of the moduli space by automorphisms of $H_3(Y,\C)$. If $\gamma$ is 
both cyclic and weakly integral, then 
there exists some complex constant $\lambda$  such that 
$\lambda^{-1}\gamma$ is integral. Since the monodromy operators 
preserve the lattice $\Lambda$, it follows that the cycles 
$\lambda^{-1}\gamma^{(j)}$ are integral as well. Then one can easily 
construct the maximal lattice $\Lambda_0\subset H^3(Y,\C)$ having 
the property that it contains all $\gamma^{(j)}$, and it is an 
easy exercise to show 
that $\Lambda':=\lambda^{-1}\Lambda_0$ is a full sublattice
\footnote{That is, a maximal rank sublattice of $\Lambda$, i.e. a sublattice 
such that the $\Lambda/\Lambda_0$ is a finite group.}
of $\Lambda$, though it may fail to coincide with $\Lambda$ itself.

For example, in compact one parameter
models, three-cycles undergo monodromy
transformations not only about the large complex structure point $z = 0$ 
but also about the points $z = 1$ and $z = \infty$ 
(with our choice of coordinate on the moduli space). There always exists a 
{\em fundamental cycle} $\gamma_0\in H_3(Y,\C)$ 
with the property that it is left invariant 
by all monodromy transformations about the large complex structure point
(in the framework of \cite{SYZ}, this is the cycle mirror to a 0-cycle, 
namely the homology class of the fiber of the special Lagrangian 
$T^3$-fibration of $Y$). However, $\gamma_0$ will not generally 
be invariant under $T[\infty]$ and $T[1]$, and it is often the case that 
it is a cyclic vector for the action of these monodromy operators. In fact, 
a particular version of this procedure (involving only the monodromy operator 
$T[\infty]$ and applicable for the case when $z=\infty$ is a Landau-Ginzburg 
point) underlies the method used in \cite{quintic} 
for producing a basis of periods.

We can now recast our strategy for measuring
quantum volumes into the following form:
\vskip.1in
(1) Calculate the period $\int_{\gamma_0} \Omega$ for a weakly
integral three-cycle $\gamma_0$.
\vskip.1in
(2) Calculate the monodromy matrices $T[0]$, $T[1]$, and $T[\infty]$,
and use them to produce a weakly integral basis of three-cycles, and
their corresponding periods
\vskip.1in
(3) Search for places in moduli space where a period associated 
to a weakly integral three-cycle vanishes.
\vskip.1in
(4) Interpret the vanishing period of a weakly integral three-cycle on
$Y$ with leading ${\rm log}^j(z)$ behavior around $z=0$ as the vanishing
quantum volume of a holomorphic cycle on $X$ of real dimension $2j$.
\vskip.1in

As mentioned before, a more precise version of step (4) would
involve identifying the admixture of lower dimensional cycles
on $X$ mirror to  the vanishing cycle on $Y$. This requires
understanding mirror symmetry at the level of integral
structures and in Section 3 we discuss aspects of this issue.
In the next section we consider the problem of efficiently
calculating a complete set of periods over a basis of three-cycles,
together with procedures for identifying the monodromy matrices ---
that is, techniques for carrying out steps (1) through (3).  We
will then use these results to carry out the four steps above in various 
examples.

Before proceeding with this analysis, let us make a few brief remarks on 
the status of results which can be found in the literature. Most previous 
work which involved the calculation of periods was concerned with
deepening
and generalizing the computation of Gromov-Witten invariants first performed 
in \cite{quintic}. Since these invariants can be extracted from the mirror 
map (\ref{mirrormap}), the literature is mainly focused on the 
computation of the fundamental and ${\rm log}^1$ periods. Ironically,
though, some
of the earliest works on applying mirror symmetry to enumerative
geometry did consider the higher periods (e.g. \cite{quintic,2pm1,2pm2,msh}), 
but their physical relevance in those pre-D-brane
studies was, of course, not considered. In this work we present
a method for analyzing all of the periods, throughout the moduli space,
in a scheme that is calculationally easier than previous methods
and which allows their physical content to be directly extracted.

\section{Techniques}

Although we shall consider an example with $h^{1,1} > 1$, we will
follow \cite{small_distances1}\ and focus on judiciously chosen
one-dimensional subspaces in the relevant moduli space.  Hence, in
this section we start by discussing a calculationally efficient technique for
analyzing one-parameter systems. The approach we describe is based on
classical mathematical results due to C.~S.~Meijer and uses the so-called
G-functions (or Meijer functions). As we show below, these
functions are especially well-suited for a systematic computation of a
basis of periods and its monodromies. 
Our use of G-functions grew out of our effort to
find a more effective approach to such computations, and hopefully
represents an improvement over the methods used in \cite{quintic}. 
From an abstract point of view, the approach taken
there consists of the following four steps.  First, one computes the
fundamental period $\omega_0$ near a large complex structure point by
explicitly integrating the holomorphic 3-form $\Omega$ along a
carefully chosen 3-cycle. Second, one analytically continues this
period to a point of finite-order monodromy (assuming that such a
point exists, which is not always the case). Third, assuming that the
order of the monodromy action at that point is sufficiently high and that the
fundamental period is a cyclic vector for the corresponding monodromy
operator, one can produce a basis of periods $\omega_j$ by repeatedly
acting with this operator on $\omega_0$. In practice, this is carried
out by using an explicit realization of the finite-order monodromy as
a local orbifold action on a branched multiple cover of the moduli
space; this procedure then amounts to `rotating the branch cut' used
in the expansion of $\omega_0$ near the small radius point in order 
to obtain the `higher' periods
$\omega_j$.  Finally, one can in principle analytically continue back from the
finite-monodromy point to the large complex structure point by using
carefully chosen integral representations compatible with the
branch-cuts introduced in defining $\omega_j$. In practice, finding 
convenient integral representations of $\omega_j$ for this purpose is 
not always straightforward. 

This approach depends on various assumptions and can be difficult
to carry out in practice, due to the computational complexity
involved and the less than algorithmic nature of some of the steps. 
The alternative we propose is calculationally simpler and
more algorithmic. It consists of the following steps.

First, we
use the theory of Meijer functions to systematically (and rather
easily) produce a special basis of periods adapted to the large complex
structure monodromy-weight filtration. (The reader unfamiliar with this
concept can find a brief description in Appendix A).
These periods (which we will call `Meijer periods') 
are expressed in terms of Meijer functions, which are {\em
defined} by certain integral representations of Mellin
type. Analytic continuation of the Meijer periods is thus straightforward to
perform, with the choice of branch-cuts automatically fixed by the
condition of convergence of the associated contour integral. This
systematically produces the expansions of the periods in all regions
of the moduli space, without requiring the computation of
monodromies. If knowledge of the monodromy matrices is required, 
then they can be computed systematically from the
expansions of the Meijer periods. Once this has been achieved, a
`cyclic basis' of the type considered in
\cite{quintic} can be easily produced, and can be
used to provide some (but generally not complete) information about
the integral lattice $H^3(X,\Z)\subset H^3(X)$.

We begin with a short review of Meijer functions, with the main
purpose of fixing our notation and conventions. Then we illustrate the
technique outlined above in a class of one-parameter examples which
includes the mirror quintic as well as other models considered in the
literature. Finally, we consider a certain `degeneration' of this
class of models, which in particular contains a model we will
encounter again when studying a two-parameter example in Section 6.

\subsection{Meijer functions}

We start by reviewing a few basic facts about a class of higher
transcendental functions introduced by C.~S.~Meijer. These functions can 
be used to generate a fundamental system of solutions of a
generalized hypergeometric equation, and therein lies their
utility. In a sense, they form the `natural' class of functions
associated with hypergeometric systems. As the periods of the
holomorphic three-form satisfy a generalized hypergeometric
equation, we will be able to use these methods to calculate quantum
volumes. (We refer the interested reader to \cite{Meijer_refs} for
details about the general theory of such functions and to the
excellent paper \cite{Norlund} for a review of their applications to
hypergeometric equations.)

The modern definition of Meijer functions proceeds via Mellin-Barnes
integrals. To simplify subsequent formulae, we introduce the notation:
\be \G[x_1~...~x_n,y_1~...~y_m]:=\frac{\Gamma(x_1)~...~\Gamma(x_n)}
{\Gamma(y_1)~...~\Gamma(y_n)} \ee
\noindent for any complex numbers $x_i,y_j$.
Consider the Mellin-Barnes integral in the complex plane: \be
\label{MB}
\I[{a_1~...~a_A},{b_1~...~b_B},{c_1~...~c_C},{d_1~...~d_D}](z)=
\B[\gamma,s][{a_1-s~...~a_A-s}~,~{b_1+s~...~b_B+s}~,~{c_1-s~...~c_C-s}~,~
{d_1+s~...~d_D+s}](z)~,
\ee with complex $a_i,b_j,c_k,d_l$ and $z$.  For generic $a,b,c,d$,
the integrand has two series of poles:

\

\centerline {(A)  $a_i~-~s=-n$}

\

\centerline {(B)  $b_j~+~s=-n$}

\

\noindent with $n$ a nonnegative integer. 
The contour $\gamma$ is taken to extend
from $-i\infty$ to $+i\infty$ (in this order)
in such a way as to separate the A-type poles 
from the B-type poles (see Figure 1). The conditions for convergence of 
(\ref{MB}) are discussed \cite{Meijer_refs}, to which we 
refer the reader for details. Here we only mention that an essential 
requirement is that $|{\rm arg}(z)|<\pi$, which is necessary in 
order to assure rapid decrease of the integrand at imaginary infinity. 
This is an important point implicit in the following sections: 
when considering the 
analytic continuation of periods from large to small radius 
for one-parameter models, the choice of branch-cuts is {\em automatically} 
enforced by the use of Mellin-Barnes representations.

\

\hskip 1.2in\scalebox{0.4}{\begin{picture}(0,0)%
\epsfbox{contour_mellin.pstex}%
\end{picture}%
\setlength{\unitlength}{3947sp}%
\begingroup\makeatletter\ifx\SetFigFont\undefined%
\gdef\SetFigFont#1#2#3#4#5{%
  \reset@font\fontsize{#1}{#2pt}%
  \fontfamily{#3}\fontseries{#4}\fontshape{#5}%
  \selectfont}%
\fi\endgroup%
\begin{picture}(7931,8306)(482,-7680)
\put(7951,-61){\makebox(0,0)[lb]{\smash{\SetFigFont{17}{20.4}{\rmdefault}{\mddefault}{\itdefault} s}}}
\end{picture}
}

\hskip 1in \begin{center}Figure 1. {\footnotesize 
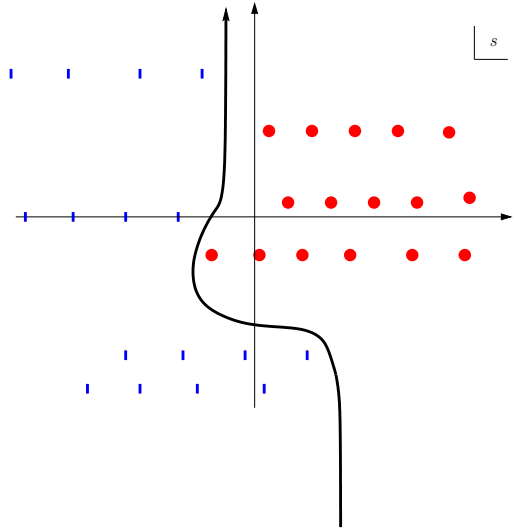
The defining contour for a Mellin-Barnes integral. A-type poles are 
represented by circles and B-type poles by short vertical lines.}
\end{center}

\

The {\em Meijer functions} are defined by:
{\footnotesize \be
\label{Meijer}
\M[\rho_1~...~\rho_r ,\rho'_1~...~\rho'_{r'} , \sigma_1~...~\sigma_s , \sigma'_1~...~
\sigma'_{s'} ](z)=
\I[\sigma_1~...~\sigma_s, 1-\rho_1~...~1-\rho_r,\rho'_1~...~\rho'_{r'},
1-\sigma'_1~...~1-\sigma'_{s'}](z)
\ee}\noindent whenever the Mellin integral appearing on the right hand side 
converges.
We will frequently use ${\bf I}$ rather than ${\bf G}$ since the former 
displays the structure of the Mellin-Barnes integrand more clearly.
The functions ${\bf I}$ and ${\bf G}$ satisfy certain differential equations
which can be obtained by considering the finite difference identities 
obeyed by the associated integrands. This is easiest to write 
down for ${\bf G}$. Indeed, it is not hard to see that the function
{\footnotesize 
$u(z) = \M[{\rho_1 ~...~ \rho_r},{\rho_{r+1} ~...~ \rho_{R}},
{\sigma_1 ~...~ \sigma_s},{\sigma_{s+1} ~...~ \sigma_{S}}](z)$}
satisfies:
\be
\label{Meijer_eq}
\D[\mu,z][{\rho_1~...~\rho_R},{\sigma_1~...~\sigma_S}] u=0~~,
\ee
with the operator:
\be
\label{Meijer_op}
\D[\mu,z][{\rho_1~...~\rho_R},{\sigma_1~...~\sigma_S}]=
\prod_{i=1~...~s}
{(\delta-\sigma_i)}-(-1)^\mu z~\prod_{j=1~...~R}{(\delta-\rho_j+1)}~~,
\ee
where $\delta=z\frac{d}{dz}$ and $\mu=R-r-s~(mod~{\rm 2})=r'-s~(mod~2)$. 
While $u$ depends strongly on the  partitions 
$(\rho_1~...~\rho_R) = (\rho_1~...~\rho_r~;~\rho'_1~...~\rho'_{r'})$ 
 and $(\sigma_1~...~\sigma_S) = 
(\sigma_1~...~\rho_s~;~\sigma'_1~...~\sigma'_{s'})$ 
used for its definition, the 
differential operator (\ref{Meijer_op}) depends on these partitions 
only through the number $\mu \in \Z_2$. Thus, by considering
different partitions of $(\rho_1~...~\rho_R)$
and $(\sigma_1~...~\sigma_S)$ as arguments in $G$, together with
the observation that 
{\footnotesize $\D[\mu,-z][{\rho_1~...~\rho_R},{\sigma_1~...~\sigma_S}]=
\D[\mu+1,z][{\rho_1~...~\rho_R},{\sigma_1~...~\sigma_S}]$}, we learn 
that the functions
{\footnotesize \bea
\begin{array}{c}
\I[a_1~...~a_A, b_1~ ...~ b_B, c_1~ ...~ c_C, d_1~ ...~ d_D](z)~~, \\
\nn\\ 
\I[a_1~..~ 1-d_l~ ..~a_A, b_1~ ...~ b_B, c_1~ ...~ c_C, d_1 .. {\hat d_l} .. d_D](-z) \mbox{~and~}
\I[a_1~ ...~ a_A, b_1~..~{\hat b_j}~..~b_B, c_1~ ..~ 1-b_j~ .. c_C, d_1~ ...~
 d_D](-z)
\end{array}\nn
\eea}\noindent (where a hat indicates removal of the corresponding element 
from the list)
satisfy the {\em same} differential equation. 
Therefore, given a Meijer function 
${\bf I}$, one can produce other solutions of the same equation 
by the simple procedure 
of lifting/lowering the parameters `diagonally' in the symbol of ${\bf I}$,
together with replacing each parameter $\alpha$ thus moved by $1-\alpha$ and 
performing a change of sign of the argument $z$ for each such operation.
This simple observation is the basis of our procedure for obtaining 
all ${\rm log}^j$-monodromy periods starting from the fundamental period, as 
we explain in more detail below.

The Meijer functions are relevant to our problem for the following reason.
The periods of the holomorphic three-form on a Calabi-Yau manifold are
closely related to generalized hypergeometric functions
\footnote{Here $(x)_n:=x(x+1)~...~(x+n-1),~(x)_0:=1$ denotes the Pochhammer 
symbol.},
\bea
\label{HG_general}
\F[p,q][{\alpha_1 ~...~ \alpha_p},{\beta_1 ~...~ \beta_q}](z)=
\G[\beta_1 ~...~ \beta_q,\alpha_1 ~...~ \alpha_p] 
\sum_{n=0}^{\infty}
\G[{\alpha_1 + n ~...~ \alpha_p + n},{\beta_1 + n ~...~ \beta_q + n}] 
\frac{z^n}{n!}
=\sum_{n=0}^{\infty}{\frac{(\alpha_1)_n~...~(\alpha_p)_n}{(\beta_1)_n~ ... 
~(\beta_q)_n}\frac{z^n}{n!}}~~,\nn
\eea
\noindent (with $p=q+1$), which are solutions to
the generalized hypergeometric equation:
\be
\label{HG_eq}
\left[
\delta~\prod_{i=1..q}{(\delta+\beta_i-1)}-
z~\prod_{j=1..p}{(\delta+\alpha_j)}
\right]u=0~~.
\ee 
The generalized hypergeometric functions, in turn, are particular cases of 
Meijer functions 
\footnote{Note that convergence of the integral  
on the right hand side requires $|{\rm arg}(-z)|<\pi$.}:
{\footnotesize 
\bea
\label{HG_Meijer}
\F[p,q][{\alpha_1 ~...~ \alpha_p},{\beta_1 ~...~ \beta_q}](z)=
\G[{\beta_1 ~...~ \beta_q},{\alpha_1 ~...~ \alpha_p}]
\M[{1-\alpha_1 ~...~ 1-\alpha_p},{\cdot},{0},{1-\beta_1 ~...~ 1-\beta_q}](-z)=\nn\\
\G[{\beta_1 ~...~ \beta_q},{\alpha_1 ~...~ \alpha_p}]
\I[{0},{\alpha_1 ~...~ \alpha_p},{\cdot},{\beta_1 ~...~ \beta_q}](-z)\nn
\eea}\noindent (where $\cdot$ indicates that the corresponding group of 
parameters is missing).
Indeed, the hypergeometric operator:
\be
d_z\left(\begin{array}{c}\alpha_1~...~\alpha_p\\\beta_1~...~\beta_q\end{array}
\right)=\delta~\prod_{i=1..q}{(\delta+\beta_i-1)}-
z~\prod_{j=1..p}{(\delta+\alpha_j)}
\ee
is related to the Meijer operator by:
\be
d_z\left(\begin{array}{c}
\alpha_1~...~\alpha_p\\
\beta_1~...~\beta_q\end{array}\right)=
\D[1,-z][{1-\alpha_1~...~1-\alpha_p},{0,~1-\beta_1~...~1-\beta_q}]~~.
\ee

In our Calabi-Yau applications, the hypergeometric equations arise as 
typical Picard-Fuchs equations obeyed by the periods of a 
one-parameter model. In such a model, we will always choose the 
coordinate $z$ on the moduli space such that $z=0$ corresponds to the large 
complex structure (or large radius) point, 
$z=\infty$ to the small radius limit and $z=1$ to 
the other regular singular point (the analogue of the conifold point of 
\cite{quintic}). This assures that the Picard-Fuchs equation will always have 
the standard hypergeometric form (\ref{HG_eq}). We will sometimes call this 
coordinate on the moduli space the `standard' or `hypergeometric' coordinate. 
Then the parameters ${\alpha_i,\beta_j}$ will always be such
that $z=0$ is a point of maximally unipotent monodromy, which implies that 
the only solutions of (\ref{HG_eq}) which are regular at the origin are 
multiples of ~{\footnotesize 
$\F[p,q][{\alpha_1 ~...~ \alpha_p},{\beta_1 ~...~ \beta_q}](z)$}
(the hypergeometric function itself is characterized among such solutions by 
the fact that it has value $1$ at $z=0$). The hypergeometric solution thus 
corresponds to the fundamental period of \cite{Candelas_periods}, and is 
trivial to write down in practice. Writing this solution as a Meijer 
function and making simple operations on its symbol as above will allow us to 
produce a basis of solutions of (\ref{HG_eq}), i.e. a basis of 
periods for the model. Moreover, the `Meijer periods' 
thus produced are automatically adapted to the monodromy weight filtration 
(see Appendix A for an explanation of this concept) 
associated with the large complex structure point. In fact, these periods 
display a rather universal behaviour around $z=0$, as we will see in 
detail below.

\subsection{Monodromies}

The strategy outlined in Section 1 requires that we 
understand the behavior of the
Meijer periods under monodromy transformations. In this section
we set up some general formalism that will allow us to accomplish this goal.
Our remarks will clarify the connection with the general theory of Fuchsian 
systems as explained, for example in \cite{fuchsian} (see also 
\cite{morrison_fuchs}).

\subsubsection{General remarks}

Let us start with a discussion of monodromies for the Meijer equation 
(\ref{Meijer_eq}). For simplicity (and direct relevance
to this paper) we consider only the case $S=R:=p$, which includes the 
hypergeometric equation.
The logarithmic form of the Meijer operator
(\ref{Meijer_op}) is:
\be
\frac{1}{1+(-1)^{\mu +1}z}
\D[\mu,z][{\rho_1~...~\rho_p},{\sigma_1~...~\sigma_p}]=\delta^p+\sum_{k=0}^{p-1}
{B_k(z)\delta^k}~~,
\ee
where we defined:
\be
B_k(z):=\frac{Q_k+(-1)^{\mu+1}zR_k}{1+(-1)^{\mu+1}z}~~,
\ee
with:
\bea
Q_k&:=&(-1)^{p-k}\sum_{1\leq i_1<~...~<i_{p-k}\leq p}
{\sigma_{i_1}~...~\sigma_{i_{p-k}}}~~\\
R_k&:=&(-1)^{p-k}\sum_{1\leq i_1<~...~<i_{p-k}\leq p}
{(\rho_{i_1}-1)~...~(\rho_{i_{p-k}}-1)}~~.\\
\eea\noindent 
The coefficient functions $B_k(z)$ interpolate between 
$B_k(0)=Q_k$ and $B_k(\infty)=R_k$.

The Meijer equation (\ref{Meijer_eq}) can be rewritten as the first order 
system:
\be
\label{Meijer_system}
\delta~w(z)=A(z)~w(z)~~,
\ee
where 
{\footnotesize 
$w(z):=\left[\begin{array}{c}u(z)\\ \delta~u(z)\\ ... \\ \delta^{p-1}u(z)
\end{array}\right]$} and we set:
{\footnotesize \be
A(z):=\left[\begin{array}{cccccc} 0&1&0& ... &0&0\\
0&0&1& ... &0&0\\
...&...&..& ... & ...&...\\
0&0&0& ... &0&1\\
-B_0(z)&-B_1(z)&..& ... &-B_{p-2}(z)&-B_{p-1}(z)\\
\end{array}\right]
\ee} 

Given an arbitrary basis $\{U_j(z)\}_{j=0...p-1}$ of solutions of 
(\ref{Meijer_eq}), we define its monodromy matrices 
$T[0]$ and $T[\infty]$ around $z=0$ and $z=\infty$ by:
\bea
U(e^{2\pi i}z)&=&T[0]U(z)~~\mbox{~~for~~}|z|\ll 1~~,\nn\\
U(e^{2\pi i}z)&=&T[\infty]U(z)~~~\mbox{~~for~~}|z|\gg 1~~,\nn
\eea\noindent 
where {\footnotesize $U(z):=\left[\begin{array}{c}U_0(z)\\ U_1(z)\\ ... \\ 
U_{p-1}(z)
\end{array} \right]$}. 
If $\Phi(z)$ is the associated fundamental matrix of (\ref{Meijer_system}),
i.e. the matrix having 
{\footnotesize $w_j=\left[\begin{array}{c}U_j(z)\\ \delta~U_j(z)\\ ... \\ 
\delta^{p-1}U_j(z)\end{array}\right]~~(j=0~...~p-1)$} as its columns, then the 
differential version of the nilpotent orbit theorem states that there 
exists a regular\footnote{Regular in the vicinity of $z=0$.} 
matrix-valued function $S$ (which, following modern terminology, 
we call the nilpotent orbit of $\Phi$)
and a matrix $R$ such that\footnote{This decomposition has a standard 
geometric interpretation, which we recall in Appendix A.}:
\be
\label{nilpotent}
\Phi(z)=S(z)~z^R~~.\\
\ee
This immediately gives $T=e^{2\pi i R^t}$ and implies that $S(z)$
satisfies the matrix differential equation:
\be
\label{nilpotent_eq}
\delta S(z)=A(z)S(z)-S(z)R~~.
\ee
Expanding $S(z)=\sum_{n=0}^{\infty}{S^{(n)}z^n}$ shows that the matrix 
$S^{(0)}:=S(0)$ 
(which plays the role of initial condition for (\ref{nilpotent_eq}), and thus 
must be invertible if $\Phi(z)$ is to be a fundamental matrix for 
(\ref{Meijer_system})) must satisfy:
\be
S(0)^{-1}A(0)S(0)=R~~.
\ee
Under a change of basis $U(z)\rightarrow M U(z)$, we have 
$\Phi(z)\rightarrow \Phi(z)M^t$ and:
\bea
R ~~~&\rightarrow& M^{-t}RM^t~~\\
S(z)&\rightarrow& S(z)M^t~~
\eea
(where $M^{-t}:=(M^{-1})^t$). In particular, we have $S(0)\rightarrow S(0)M^t$.
Choosing $M:=S(0)^{-t}$ gives a distinguished 
fundamental system $\Phi_{can}(z)$ with the 
properties:
{\footnotesize \bea
\label{can}
R_{can}[0]&:=&A(0)=\left[\begin{array}{cccccc} 0&1&0& ... &0&0\\
0&0&1& ... &0&0\\
...&...&..& ... & ...&...\\
0&0&0& ... &0&1\\
-Q_0&-Q_1&..& ... &-Q_{p-2}&-Q_{p-1}\\
\end{array}\right]~~\\
S_{can}(0)&:=&I~~,
\eea}\noindent whose associated basis $U_{can}$ will be called the 
{\em canonical basis}.

To understand the behaviour at $z=\infty$, note that 
$\delta_{1/z}=-\delta_z$, which gives:
\be
\D[\mu,1/z][{\rho_1~...~\rho_p},{\sigma_1~...~\sigma_p}]=
\frac{(-1)^{\mu+p+1}}{z}
\D[\mu,z][{1-\sigma_1~...~1-\sigma_p},{1-\rho_1~...~1-\rho_p}]~~.
\ee
This is again a Meijer operator, with symbols $\rho'_j:=1-\sigma_j$ 
and $\sigma'_i:=1-\rho_i$. The canonical monodromy is again of the 
form (\ref{can}), but with $Q_k$ replaced by $Q'_k:=(-1)^{p-k}R_k$:
{\footnotesize \be
R_{can}[\infty]:=-\left[\begin{array}{cccccc} 0&1&0& ... &0&0\\
0&0&1& ... &0&0\\
...&...&..& ... & ...&...\\
0&0&0& ... &0&1\\
-Q'_0&-Q'_1&..& ... &-Q'_{p-2}&-Q'_{p-1}\\
\end{array}\right]~~.
\ee}\noindent The minus sign in front of this expression keeps track of 
the fact that the transformation $z\rightarrow e^{2\pi i}z$ corresponds 
to $\frac{1}{z}\rightarrow e^{-2\pi i}\frac{1}{z}$; that is, 
inversion in the complex plane reverses the orientation of a contour 
which surrounds the origin. 
This amounts to inversion of the associated 
monodromy matrix, an observation which will be used again in 
Subsection 4.3.

\subsubsection{The hypergeometric case}

Let us now apply these results to the hypergeometric case, which is the case 
of interest for us. In view of (\ref{HG_Meijer}), this is defined
by $R=S=p=q+1$ and:
\bea
\rho_i~&:=&1-\alpha_i~~(i=1~...~p)\nn\\
\sigma_1&:=&0,~\sigma_{j+1}:=1-\beta_j~~(j=1~...~q)~~.\nn 
\eea
According to the discussion above, if $u(z)$ satisfies the 
generalized hypergeometric equation with (hypergeometric) symbol 
{\footnotesize 
$\HG_symbol[{\alpha_1~...~\alpha_{q+1}},{\beta_1~...~\beta_q}]$}, then 
${\tilde u}(t):=u(1/t)$ satisfies the Meijer equation 
{\footnotesize 
$\D[\mu,t][{1,\beta_1~...~\beta_q},{\alpha_1~...~\alpha_{q+1}}]{\tilde u}=0$}
(note that this equation is {\em not} generally of hypergeometric type). 
Applying the general results derived above shows 
that the canonical forms of the 
hypergeometric monodromies around $z=0$ and $z=\infty$ are:
{\footnotesize \bea
\label{canonical_monodromies}
R_{can}[0]:&=~~&\left[\begin{array}{cccccc} 0&1&0& ... &0&0\\
0&0&1& ... &0&0\\
...&...&..& ... & ...&...\\
0&0&0& ... &0&1\\
0&-Q_1&..& ... &-Q_{q-1}&-Q_{q}\\
\end{array}\right]~~\\
R_{can}[\infty]:&=-&\left[\begin{array}{cccccc} 0&1&0& ... &0&0\\
0&0&1& ... &0&0\\
...&...&..& ... & ...&...\\
0&0&0& ... &0&1\\
(-1)^qR_0&-(-1)^{q-1}R_1&..& ... &-R_{q-1}&R_q\\
\end{array}\right]~~,
\eea}\noindent with:
{\footnotesize \bea
Q_k:&=&\sum_{1\leq i_1<~...~<i_{q-k+1}\leq q}{(\beta_{i_1}-1)...
(\beta_{i_{q-k+1}}-1)}~~\\
R_k:&=&\sum_{1\leq i_1<~...~<i_{q-k+1}\leq q+1}{\alpha_{i_1}...
\alpha_{i_{q-k+1}}}~~,
\eea}\noindent where $k=0~...~q$. 
In this paper we only consider  
models for which all $\beta_i$ are equal to $1$. Then all $Q_k$ are 
zero and the matrix $R_{can}[0]$ is in Jordan form. On the other hand, the 
Jordan form of the matrix $R_{can}[\infty]$ depends on the relative values 
of the parameters $\alpha_k$, as we will illustrate in detail below. 

Before proceeding to apply the theory of Meijer functions to various examples, 
let us mention that the remaining regular singular point of (\ref{HG_eq}) 
(namely the point $z=1$) is somewhat special in the following sense. 
While one can change variables (for example, by the transformation 
$z\rightarrow \frac{1-z}{z}$) in such a way as to move the point $z=1$ to the 
origin of the complex plane, the resulting equation is not of Meijer type. 
This is the root of the technical difficulties encountered when studying the 
behaviour of a system of solutions of (\ref{HG_eq}) around $z=1$, and the main 
reason why a completely uniform treatment of this problem is not yet 
available. We refer the interested reader to \cite{Norlund} for a thorough 
discussion of what is known about this subject. 

On the other hand, the monodromies around $z=1$ can be easily computed for a 
convenient basis of solutions around that point, by making use of 
the general theory of 
\cite{fuchsian}. However, this will not be relevant for us, for the 
following reason. The canonical bases associated with the points $0$ and 
$\infty$ (as well as the more general standard basis associated with the point 
$z=1$) generally correspond to different fundamental systems of solutions 
of (\ref{HG_eq}). What one is interested in when studying one-parameter models
is to compute the monodromies of a {\em given} basis around these three points;
in other words, one considers the analytic continuation of a fundamental 
system throughout the entire complex plane and one asks for the monodromies 
of the basis of solutions thus obtained. If $T[0]$ and $T[\infty]$ are
the monodromies of such a system about $0$ and $\infty$, then 
the monodromy of {\em that} system about $z=1$ is easily obtained as 
$T[1]=T[0]^{-1}T[\infty]$, an equation which follows from the structure of 
the fundamental group of the thrice-punctured Riemann sphere 
$\P^1- \{0,1,\infty\}$. This relation, as  well as our convention 
for the monodromy matrix $T[1]$ are depicted in Figure 2. 
In practice, it will thus suffice to compute 
$T[0]$ and $T[\infty]$ for an analytically continued fundamental system.
As we illustrate in detail below, the most convenient choice 
is to consider the analytic continuation of the Meijer basis, which is
easily obtained from its Mellin-Barnes representation through a computation of 
residues. This basis has the added advantage that it allows for a  
very systematic computation of the monodromies about $0$ and $\infty$. 
The utility of the canonical monodromies computed above is that they will
appear as intermediate steps in this procedure, which essentially consists in 
computing the transition matrices from the Meijer basis to the two canonical 
bases associated with the points $0$ and $\infty$.

\vskip 0.5in

\hskip 1.2in\scalebox{0.5}{\input{mons.pstex_t}}

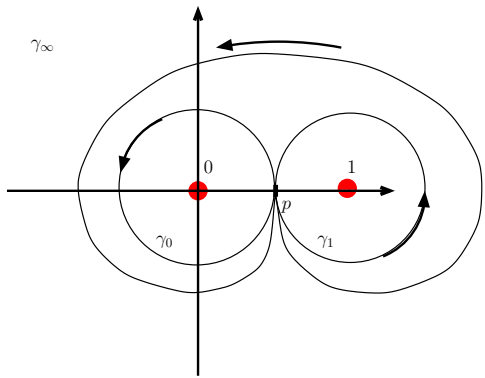
\begin{center}Figure 2. {\footnotesize Generators of the fundamental group 
$\pi_1(\P^1-\{0,1,\infty\},p)$ of the thrice punctured sphere (using the 
base point $p=\frac{1}{2}$). The monodromy operators $T[0],T[1],T[\infty]$ 
used in this paper correspond to the paths $\gamma_0,~\gamma_1,~\gamma_\infty$,
all of which are oriented in counterclockwise manner. The 
relation $\gamma_0\cdot\gamma_1=\gamma_\infty$ induces the equality 
$T[1]=T[0]^{-1}T[\infty]$.}
\end{center}

\section{The Mirror Map and Integral Structure}

Identifying the B-type D-brane state associated with a collapsing 
holomorphic cycle requires a deeper understanding of the map 
induced by mirror symmetry between $H_3(Y,\Z)$ and
$H_{even}(X,\Z)$. In this section we discuss what is known about this 
issue and identify the obstacle that as yet needs to be surmounted. 
While the we believe that the essential point we make below has
not been widely appreciated, most of what follows 
involves manipulating and reinterpreting ideas of
\cite{Candelas_moduli}\ and \cite{quintic} to which we refer the reader
for background and notation.

In deriving the mirror map  $t_l = \frac{\int_{\gamma_l} \Omega}
{\int_{\gamma_0} \Omega}$ for the quintic mirror-quintic
pair $(X,Y)$, 
the authors of \cite{quintic} proceeded in the following way. Choose
an integral symplectic basis $(A_a,B^b)$ of $H_3(Y,\Z)$ with dual 
cohomology basis $(\alpha^a,\beta_b)$, where $a,b = 1,2$. Then 
write the holomorphic three-form $\Omega$ of $Y$ as

\be
\label{Omegaexpanded}
\Omega = (\int_{A_a} \Omega) \alpha^a - (\int_{B^b} \Omega) \beta_b~~,
\ee
and the prepotential $\cal G$ as
\be
\label{Prepotential}
{\cal G} = \frac{1}{2}(\int_{A_a} \Omega)(\int_{B^a} \Omega)~~.
\ee
Defining coordinates $z^a$ in terms of the periods through 
$z^a = \int_{A_a} \Omega$ allows us to write:
\be
\label{PiOmega}
\Omega = \Pi^j  g_j~~,
\ee
with $\Pi^t = (\partial_{z_1} {\cal G}, \partial_{z_2} {\cal G},z_1, z_2)$
and  $g = (-\beta_1, -\beta_2, \alpha^1, \alpha^2)$.

One can follow a similar  procedure for $X$. Choose an integral
basis of cycles $(E_1, E_2, F^1, F^2)$ for $H_0(X,\Z)$,
$H_2(X,\Z)$, $H_4(X,\Z)$, and $H_6(X,\Z)$, respectively.
Let $(\rho^1, \rho^2, \sigma_1, \sigma_2)$ be dual
integral cohomology classes, and following \cite{Candelas_moduli},
let $\lambda$ and $\mu$ be anticommuting constant Grassman 
parameters so that 
$(\theta^1, \theta^2, \theta_1, \theta_2) := (\lambda \rho^1, \lambda\rho^2, \mu \sigma_1, \mu \sigma_2)$
has a symplectic intersection form. 
Then, introduce $\Upsilon 
\in \oplus_{k} H^{k,k}(X,\Z)$, the mirror to $\Omega$,
expressed as:
\be
\label{Moexpanded}
\Upsilon = (\int_{E_a} \Upsilon) \theta^a - (\int_{F^b} \Upsilon) \theta_b~~,
\ee
and the mirror prepotential on $X$:

\be
\label{PrepotentialF}
{\cal F} = \frac{1}{2}(\int_{E_a} \Upsilon)(\int_{F^a} \Upsilon)~~.
\ee

The latter expression is purely formal as we know of
no a priori means of calculating the periods of $\Upsilon$, without
recourse to mirror symmetry. However, following 
\cite{Candelas_moduli}\ and \cite{quintic}, we can write down
an expression for $\cal F$ which is valid up to worldsheet
instanton corrections, by recalling that Yukawa couplings
are given by third derivatives of the prepotential.
Indeed, writing the complexified K\"ahler form on $X$ as
$B + i J = \frac{w^1}{w^2} \rho^2$ for $w^1 \in \C$, we have:
\be
\label{PrepotentialF1}
{\cal F} = -\frac{1}{3!} \int_X (\rho^2)^3 \frac{(w^1)^3}{w^2} + \frac{1}{2}a(w^1)^2 +
bw^1w^2 + \frac{1}{2}c(w^2)^2 + {\cal O}(e^{-w^1/w^2}), 
\ee
where $w^2$ is an auxiallary homogeneous coordinate on the moduli space and 
$t = w^1/w^2$.
Now, if we could actually evaluate (\ref{PrepotentialF1}) 
directly (and exactly), then we could write 
\be
\label{MoIp}
\Upsilon = \amalg^j  f_j~~,
\ee
with $\amalg^t = (\partial_{w_1}{\cal F}, \partial_{w_2}{\cal F}, w_1, w_2)$
and  $f = (-\theta_1, -\theta_2, \theta^1, \theta^2)$. Then, the 
basis-independent equality $\Omega = \Upsilon$ (an equality that follows 
since in the underlying conformal field theory 
$\Omega$ and  $\Upsilon$   correspond to one and the same
operator) can be expressed using (\ref{PiOmega}) and (\ref{MoIp}) as:
\be
\label{components}
\amalg^i = N^i_j \Pi_j
\ee
\be
\label{bases}
g_j = N^i_j f_i,
\ee
where $N$ is a matrix relating the bases $f$ and $g$.

In practice, we cannot carry out this program as written because
we do not know how to evaluate
(\ref{PrepotentialF1}) exactly. But as noticed and used in
\cite{quintic}, if we call  ${\cal F}^{\rm pert}$ the approximation
to $\cal F$ with all instantons corrections dropped,
equating $\Omega$ and $\Upsilon$ in the large complex structure/large
radius limit via
\be
\label{components1}
\amalg^{{\rm pert}~i} = N^i_j \Pi_j
\ee
is enough to determine $N$. This statement relies
on the 
conjecture of Aspinwall and Lutken \cite{int_conj} which asserts
that $N$ should
be an integral symplectic matrix. As such, it can be
determined exactly even if we only impose (\ref{components1})
in the large radius limit.

Once we have determined $N$, we can use it to relate
the integral cohomologies of $X$ and $Y$ via (\ref{bases}),
or the integral homologies via its dual relation:
\be
\label{bases1}
(F^1, F^2, E_1, E_2)^t = N (B^1, B^2, A_1, A_2)^t.
\ee

In principle, this solves the problem of 
identifying the map between $H_3(Y,\Z)$ and
$\oplus_k H_{2k}(X,\Z)$. Although
we have recounted this discussion at the level
of the quintic/mirror quintic example, it
is of course more general. Nevertheless, there is
an important outstanding issue:
Even if we work in the large radius limit and
thereby drop instanton corrections to $\cal F$, from
(\ref{PrepotentialF1}) we see that there are three
numbers, $a,b,c$ which are as yet undetermined.
As pointed out in \cite{quintic}, the real part of these
numbers do not contribute to the Yukawa couplings or
the metric on the moduli space, and the  only imaginary
parts they can contain (which do affect the metric)
arise at four loops in the sigma model, thereby only
contributing to $c$. In the case of the quintic,
$a, b, c$ enter into $N$ in the following manner
\be
\label{Nmatrix}
N=\left[\begin{array}{cccc}
-1+2a'&b'&-a'&0\\
2b'&c'&-b'&-1\\
2&0&-1&0\\
0&1&0&0
\end{array}\right]
\ee
with $a' = a+\frac{11}{2},~b'=b-\frac{25}{12},~c'=c+\frac{25i}{\pi}\zeta(3)$.
In \cite{quintic}, the authors arbitrarily chose
$a' = b' = c' = 0$, noting that other choices differ
by an integral symplectic change of basis that would
not affect the calculations of their paper. In our case,
though, such a change of basis affects the mirror
map on integral homology and hence does affect
our conclusions. For example, in the case of the quintic
with the choice $a' = b' = c' = 0$, we have the map
$A_2 \rightarrow F^2$. $A_2$ is the cycle whose period
vanishes at the conifold point of the mirror quintic 
while $F^2$ is its mirror six-cycle on the quintic itself. Hence one is 
tempted to conclude that the pure six-cycle collapses on
the quintic when its K\"ahler form is tuned to be mirror
to the conifold point \cite{quantum_volumes}. But, for arbitrary 
$a' = b' = c'$, $A_2$ is mapped to a linear combination of
the six, four, two, and zero cycles. The fact that
the coefficient of the six-cycle is nonzero for any choice
of $a' = b' = c'$ tells us that a six-brane is collapsing,
but the lower order contributions dictate a particular binding
of lower brane charges to the six-brane.

Can we calculate $a' = b' = c'$ from first principles? At the
level of the sigma model its hard to see how, as their real
parts seem to have no impact on sigma model observables.
At the level of D-brane field theory, perhaps there is a way
to calculate them, but as yet this is an unsolved problem.
Hence, as discussed in the introduction, we are forced in
this paper to discuss the masses of wrapped $2j$-branes,
leaving their lower brane content unspecified.

Let us add a few remarks about the connection with recent 
work on the relation between D-branes and $K$-theory. It is now well-known 
\cite{Ktheory} that the charges of type B 
D-branes on $X$ should generally be considered to 
take values in certain $K$-groups $K(X)$. While 
this was initially discovered \cite{Ktheory} through anomaly-cancellation 
arguments, it is in fact not surprising given the relation
between D-brane moduli spaces and moduli spaces of holomorphic vector bundles 
proposed in \cite{Moore}. From this point of view, the lattice $H^{even}(X)$
appears as a coarser version of the space of charges, via the natural 
embedding $K(X)\rightarrow H(X)$. Above, we considered the problem of 
identifying the integral structure from a perturbative (conformal 
field-theoretic) point of view, but a probably more powerfull approach to 
this question is to analyze the issue form the point of view of 
the Strominger-Yau-Zaslow conjecture \cite{SYZ,vafa_mirror}. 
In that set-up, mirror symetry is expected to 
map special Lagrangian submanifolds $C$ of $Y$ 
(together with a flat connection on $C$) to holomorphic vector bundles 
on $X$ (or, rather, a certain type of gerbe generalization of this concept). 
The trace of such a map in some cohomology theory should then yield the mirror 
map between (a version of ) $H(Y)$ and $K(X)$. 
Finally, the map 
$H^3(Y)\rightarrow H^{even}(X)$ would  be induced from this through the 
embedding $K(X)\rightarrow H(X)$. Some expectations of these type are 
summarized  by the homological mirror symmetry conjecture of Kontsevich 
(see \cite{kontsevich}), but it is difficult to be more precise 
without first gaining of a proper understanding of the many physical and 
mathematical isssues involved in such an approach.

\section{Examples I: The `generic' family of compact 
one-parameter models}

We are now ready to apply the methods discussed above, and we begin
with the hypergeometric equation satisfied by the periods of a 
class of one-parameter examples:
\be
\label{HG_eq1}
\left[
\delta^4-
z (\delta+\alpha_1)(\delta+\alpha_2)(\delta+\alpha_3)(\delta+\alpha_4)
\right]u=0~~.
\ee
This is associated with the hypergeometric symbol 
{\footnotesize 
$\HG_symbol[{\alpha_1,\alpha_2,\alpha_3,\alpha_4},{1,1,1}]$}, where we 
take  $\alpha_j$ to be {\em rational} numbers\footnote{This seems to always 
be the case in practice, and thus is not a significant reduction in 
generality.}. 
In particular, the mirror quintic of \cite{quintic} is 
of this type. We first 
illustrate our method for computing a fundamental system of 
periods and the associated monodromies for this class of models.

To simplify subsequent discussion, we assume that:

\

(1) None of the differences $\alpha_k-\alpha_l$~(with $k\neq l$) is an integer.

\

(2) None of the number $\alpha_k$ is an integer.

\

(3) $\alpha_k>0$ for all $k=1~...~4$. 

\

\noindent These conditions 
will allow us to avoid certain exceptional cases in the 
discussion of analytic continuation below.

Before proceeding with a detailed analysis, let us make some brief 
comments on the 
role and range of applicability of the results we derive in this 
section. The models we consider are `generic'
\footnote{This is not to say that such models 
are generic (meaning abundant) as a class of one-parameter examples. In fact, 
quite the contrary seems to be the case, at least based on a limited 
investigation of the models constructed in the literature.} 
from the point of view of 
their hypergeometric equation, inasmuch as conditions (1) -- (3) 
are satisfied for generic systems of {\em rational} 
parameters $\alpha_1~...~\alpha_4$.
This class of examples exhibits most of the qualitative 
features familiar from the 
study of the mirror quintic performed in \cite{quintic}; in particular, such 
models admit a Landau-Ginsburg point, due to the fact that the expansion of 
the 
periods near the small radius limit does not contain logarithmic factors
\footnote{This is enforced by condition (1), which, as can be checked 
immediately by 
using the results of subsection 2.2, and will be confirmed by  
explicit computation below, assures that the monodromy around the 
point $z=\infty$ is of finite order}. Of the three conditions above, only 
$(1)$ is essential for assuring this behaviour; imposing the other 
conditions plays mostly the role of avoiding formulae which would be 
too complicated to treat in a general fashion. 

On the other hand, there exist many one-parameter 
examples which do not 
satisfy condition (1) and hence do not entirely fit into this scheme.
Even in such cases, however, the expansions and monodromies we derive 
near the large complex structure point can often be applied.  
Such models can be thought of as `degenerations' of the 
`generic' family considered here. Hence this class 
can be viewed as a `generating' or `universal' family which 
contains other models as various limits. 
This leads to a hierarchy 
of one-parameter models (discussed in more detail in Section 5), 
having the `generic' family at its top.

\subsection{The Meijer periods}

Starting with the fundamental period 
{\footnotesize 
$U_0(z)=\F[4,3][{\alpha_1,\alpha_2,\alpha_3,\alpha_4},{1,1,1}](z)$} and 
applying the procedure discussed in the previous subsection gives the 
following fundamental system of solutions:
{\footnotesize \be
\label{Meijer_periods_4}
U_j(z)=\G[{1,1,1},{\alpha_1,\alpha_2,\alpha_3,\alpha_4}]
\I[{0^{(j+1)}},{\alpha_1,\alpha_2,\alpha_3,\alpha_4} ,{\cdot},{1^{(3-j)}}]((-1)^{j+1}z)~~,
\ee}\noindent where $j=0~...~3$ and the notation $x^{(k)}$ indicates that the 
symbol $x$ is repeated $k$ times.

The expansions of (\ref{Meijer_periods_4}) around the points $z=0$
(large complex structure) and $z=\infty$ (Landau-Ginzburg) can be 
obtained easily by a computation of residues. Indeed, let us introduce the 
following notation for the integrand of the associated Mellin-Barnes 
representations:
{\footnotesize \bea
\label{phi}
\phi_j(s):=\G[{1,1,1},{\alpha_1,\alpha_2,\alpha_3,\alpha_4}]
\G[{(-s)^{(j+1)}~~s+\alpha_1 ~...~ s+\alpha_4},{~~~~~~~~~~ (s+1)^{(3-j)}}]((-1)^{j+1}z)^s=\nn\\
=\frac{1}{\prod_{i=1~...~4}{\Gamma(\alpha_i)}}~
\frac{{\Gamma(-s)}^{j+1}\prod_{i=1~...~4}{\Gamma(s+\alpha_i)}}{{\Gamma(s+1)}^{3-j}}
((-1)^{j+1}z)^s~~. 
\eea}\noindent With this notation, we have:
\be
U_j(z)=\frac{1}{2\pi i}{\int_{\gamma}{ds~\phi_j(s)}}~~.
\ee
The integrand $\phi_j(s)$ has poles at:

\

\centerline {(A) $s = n$}

\

\centerline{(B) $s = -\alpha_i-n$,}

\

\noindent for each nonnegative integer $n$.

\

\hskip 1.2in\scalebox{0.4}
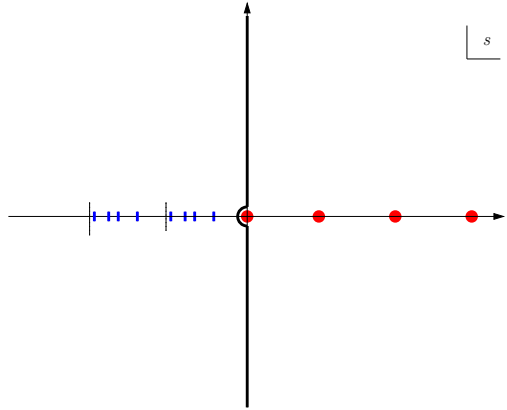
{\begin{picture}(0,0)%
\epsfbox{contour1_0.pstex}%
\end{picture}%
\setlength{\unitlength}{3947sp}%
\begingroup\makeatletter\ifx\SetFigFont\undefined%
\gdef\SetFigFont#1#2#3#4#5{%
  \reset@font\fontsize{#1}{#2pt}%
  \fontfamily{#3}\fontseries{#4}\fontshape{#5}%
  \selectfont}%
\fi\endgroup%
\begin{picture}(7824,6431)(589,-5805)
\put(7951,-61){\makebox(0,0)[lb]{\smash{\SetFigFont{17}{20.4}{\rmdefault}{\mddefault}{\itdefault} s}}}
\end{picture}
}

\begin{center}Figure 3. {\footnotesize 
The 
defining contour for the Mellin-Barnes representation of the Meijer periods.}
\end{center}

\

Consider first the expansions for $|z|<1$. In this case, the asymptotic 
behaviour of the integrand in the complex plane allows us to close the 
contour to the right at $+\infty$ (see Figure 4). 

\

\hskip 1.2in\scalebox{0.4}{\begin{picture}(0,0)%
\epsfbox{contour1_right.pstex}%
\end{picture}%
\setlength{\unitlength}{3947sp}%
\begingroup\makeatletter\ifx\SetFigFont\undefined%
\gdef\SetFigFont#1#2#3#4#5{%
  \reset@font\fontsize{#1}{#2pt}%
  \fontfamily{#3}\fontseries{#4}\fontshape{#5}%
  \selectfont}%
\fi\endgroup%
\begin{picture}(7824,6399)(589,-5773)
\put(7951,-61){\makebox(0,0)[lb]{\smash{\SetFigFont{17}{20.4}{\rmdefault}{\mddefault}{\itdefault} s}}}
\end{picture}
}

\hskip 1in \begin{center}Figure 4. {\footnotesize 
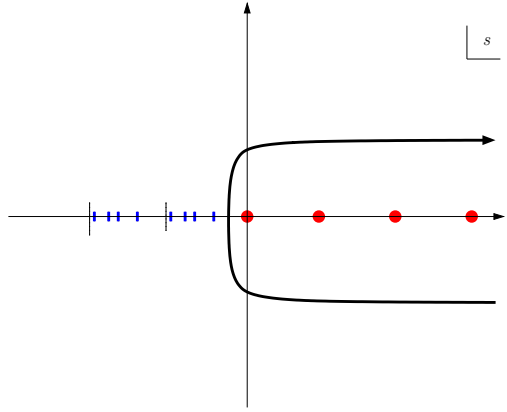
Contour for the expansion around the large complex structure point.}
\end{center}

\

\noindent Only A-type poles contribute to the associated Cauchy expansion, 
and these poles are multiple of order $j+1$. 
The computation of the associated residues proceeds in the following 
standard manner. In order to isolate the poles of the integrand, notice 
that the identity 
$\Gamma(-s)=\frac{(-1)^{n+1}}{s-n}\frac{\Gamma(-s+n+1)}{(s-n+1)_n}$
allows us to write:
{\footnotesize
\be
\phi_j(s):=\frac{(-1)^{(j+1)(n+1)}}{\prod_{i=1~...~4}{\Gamma(\alpha_i)}}~
\frac{f_j(s)}{(s-n)^{j+1}}~,
\ee}\noindent
with:
{\footnotesize
\be
f_j(s)=\frac{{\Gamma(-s+n+1)}^{j+1}\prod_{i=1~...~4}{\Gamma(s+\alpha_i)}}{
{\Gamma(s+1)}^{3-j}\left[(s-n+1)_n\right]^{j+1}}~((-1)^{j+1}z)^s~.
\ee}\noindent
Computing the residues is thereby reduced to calculating the 
derivatives $f_j^{(j)}(n)$ for $j=0~..~3$. 
One can achieve this efficiently by first computing the derivatives of  
$g_j(s)={\rm log}(f_j(s))$, with the result that 
$f_j^{(j)}(n)=f_j(n)\nu_j(n)$, 
where:
{\footnotesize 
\bea
\nu_0&=&1\nn\\
\nu_1(n,z)=g_1'(n,z)&=&\eta_1(n)+{\rm log}(z)\nn\\
\nu_2(n,z)=g_2''(n,z)+[g_2'(n,z)]^2]&=&\eta_2'(n)+(\eta_2(n)+{\rm log}(-z))^2
~~~~~\\
\nu_3(n,z)=g_3'''(n,z)+3g_3''(n,z)g_3'(n,z)+g_3'(n,z)^3\nn &=&
\eta_3''(n)+3\eta_3'(n)(\eta_3(n)+{\rm log}z)+(\eta_3(n)+{\rm log}z)^3~~,
~~~~\nn
\eea
}
with:
{\footnotesize
\be
\eta_j^{(i)}(n)=
\sum_{k=1}^{4}{\psi^{(i)}(n+\alpha_k)}-(3-j)\psi^{(i)}(n+1)
-(-1)^i(j+1)\left[\psi^{(i)}(1)+i!\sum_{l=1}^{n}\frac{1}{l^{i+1}}
\right]~~.
\ee
This gives the expansion of the Meijer periods for $|z|<1$:
\be
\label{LCS_expansions}
U_j(z)=\frac{(-1)^j}{j!}\sum_{n=0}^{\infty}{
\frac{(\alpha_1)_n(\alpha_2)_n(\alpha_3)_n
(\alpha_4)_n}{(n!)^4}\nu_j(n,z)z^n}~~.
\ee}

If $|z|>1$ then one can close the contour to the left (Figure 5). 
In this case, we have contributions  
only from the (B)-type poles of the integrand, which are all simple:
{\footnotesize \bea
{\rm Res}_{-n-\alpha_k}\phi_j(s)=\frac{(-1)^n}{n!}
\frac{{\Gamma(n+\alpha_k)}^{j+1}\prod'_{l=1~...~4}{\Gamma(\alpha_l-\alpha_k-n)}}
{{\Gamma(1-n-\alpha_k)}^{3-j}\prod_{k=1~...~4}{\Gamma(\alpha_k)}}
((-1)^{j+1}z))^{-n-\alpha_k}=~~~~~~~~~~~~~~~~~~~~~~~~~~\nn\\
\left(\frac{\sin(\pi\alpha_k)}{\pi}\right)^{3-j}
\frac{{\Gamma(\alpha_k)}^4}{\prod_{i=1~...~4}{\Gamma(\alpha_i)}}
\left[\prod'_{l=1~...~4}{\Gamma(\alpha_l-\alpha_k)}\right]((-1)^{j+1}z)^{-\alpha_k}
\frac{[(\alpha_k)_n]^4}{n!\prod'_{l=1~...~4}{(1+\alpha_k-\alpha_l)_n}}
z^{-n}~,~~~~~~~~~~~~~\nn
\eea}\noindent where 
{\footnotesize $\prod'_{l=1~...~4}{\{~...~\}}:=\prod_{l=1~...~4,l\neq k}{\{~...~\}}$} 
and the second equality follows upon using the completion formula:
\be
\Gamma(x)\Gamma(1-x)=\frac{\pi}{\sin(\pi x)}~~.
\ee

\

\hskip 1.2in\scalebox{0.4}{\begin{picture}(0,0)%
\epsfbox{contour1_left.pstex}%
\end{picture}%
\setlength{\unitlength}{3947sp}%
\begingroup\makeatletter\ifx\SetFigFont\undefined%
\gdef\SetFigFont#1#2#3#4#5{%
  \reset@font\fontsize{#1}{#2pt}%
  \fontfamily{#3}\fontseries{#4}\fontshape{#5}%
  \selectfont}%
\fi\endgroup%
\begin{picture}(8381,6399)(32,-5773)
\put(7951,-61){\makebox(0,0)[lb]{\smash{\SetFigFont{17}{20.4}{\rmdefault}{\mddefault}{\itdefault} s}}}
\end{picture}
}

\begin{center}Figure 5. {\footnotesize 
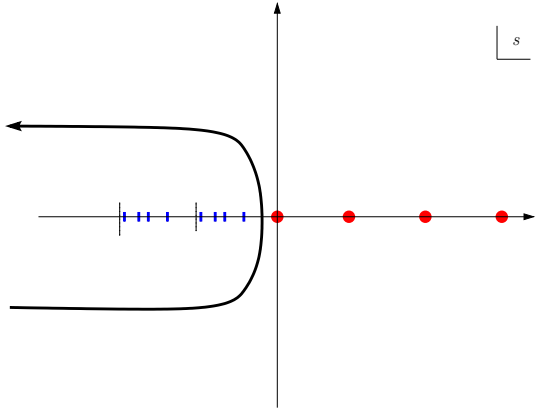
Contour for the expansion around the Landau-Ginzburg point.}
\end{center}

\

\noindent Putting everything together, we obtain the expansion of the Meijer 
periods for $|z|>1$:
{\footnotesize
\bea
\label{LG_expansions}
U_j(z)=\frac{1}{\prod_{i=1~...~4}{\Gamma(\alpha_i)}}
\sum_{k=1}^{4}{
\left(\frac{\sin(\pi\alpha_k)}{\pi}\right)^{3-j}{\Gamma(\alpha_k)}^4
\prod'_{l=1~...~4}{\Gamma(\alpha_l-\alpha_k)}
((-1)^{j+1}z)^{-\alpha_k}~}\times \nn\\
\times~
\F[4,3][{\alpha_k,\alpha_k,\alpha_k,\alpha_k},{1+\alpha_k-\alpha_1,...,{\hat 1},...,1+\alpha_k-\alpha_4}](1/z)~~,
\eea}\noindent where ${\hat 1}$ 
indicates absence of $1=1+\alpha_k-\alpha_k$ in the hypergeometric symbol.

\subsection{Monodromies of the Meijer periods}

The canonical form of the monodromies around the points $z=0$ and $z=\infty$
is easily determined by making use of the general relations 
(\ref{canonical_monodromies}). In our case, one has $S_k=0~(k=1~...~3)$
and $R_k:=\sum_{1\leq i_1<~...~<i_{4-k}\leq 4}{\alpha_{i_1}~...~\alpha_{i_{4-k}}}$.
In particular, it follows that the matrix $R_{can}[0]$ is in Jordan form:
{\footnotesize \be
R_{can}[0]=\left[\begin{array}{cccc} 
0&1&0&0\\
0&0&1&0\\
0&0&0&1\\
0&0&0&0\\
\end{array}\right]~~,
\ee}\noindent while the Jordan form of $R_{can}[\infty]$ is given by:
{\footnotesize \be
R_J[\infty]=\left [\begin {array}{cccc} -\alpha_{{1}}&0&0&0\\0&-
\alpha_{{2}}&0&0\\0&0&-\alpha_{{3}}&0
\\0&0&0&-\alpha_{{4}}\end {array}\right ]~~.
\ee} \noindent Since $\alpha_i$ are all rational, 
it immediately follows that $T_J[\infty]$ 
(and thus $T_{can}[\infty]$) has finite order, so that $z=\infty$ can be 
interpreted as a Landau-Ginzburg point. Following the program outlined above, 
we proceed to compute the monodromies of the Meijer basis around the points 
$z=0,1,\infty$.

\subsubsection{Meijer monodromies around $z=0$}

The Meijer monodromy $T[0]=e^{2\pi i R[0]^t}$  around $z=0$ can 
be determined as follows. Let $U(z)$ be the column vector
with entries $U_0(z)~...~U_3(z)$, $\Phi(z)$ be the associated fundamental 
matrix of the first order system (\ref{Meijer_system}), which has entries 
$\Phi_{ij}(z):=\delta^iU_j(z)$, and $S(z)$ the associated nilpotent orbit, 
which  satisfies (\ref{nilpotent}) and (\ref{nilpotent_eq}).
We can separate the logarithmic factors in $U(z)$ by writing:
\be
U(z)^t=Z(z)q(z)~~,
\ee 
i.e. 
\be
U_j(z)=\sum_{s=0}^{3}{q_{sj}(z)({\rm log}z)^s}~~,
\ee
where $Z(z)$ is the row vector:
\be
Z(z):=\left[\begin{array}{cccc}
1&{\rm log}z& ({\rm log} z)^2&({\rm log} z)^3\end{array}\right]~~
\ee
and $q(z)$ plays the role of the nilpotent orbit of $U(z)$ as discussed 
in subsection 2.2.

Under a change of basis $U(z)\rightarrow U'(z)=M^{-1}U(z)$ 
(with $M$ an invertible matrix), we have $U'(z)^t=Z(z)q'(z)$ with 
$q'(z)=q(z)M^{-t}$ and $\Phi'(z)=\Phi(z)M^{-t}, S'(z)=S(z)M^{-t},
R'=M^{t}RM^{-t}$. 
As discussed earlier, choosing $M$ such that $S'(0)=I$, we have 
$R'(0)=A(0)=R_{can}[0]$. In this case, $\Phi'(z)=S'(z)z^{R_{can}[0]}$, 
and since $R_{can}[0]$ is in Jordan form, we immediately obtain:
{\footnotesize 
\be
z^{R_{can}[0]}=
\left[\begin{array}{cccc} 
1&{\rm log} z&\frac{1}{2}({\rm log} z)^2&\frac{1}{6} ({\rm log} z)^3\\
0&1&{\rm log} z&\frac{1}{2}({\rm log} z)^2\\
...&...&...& ...\\
0&0&0&1\\
\end{array}\right]~~,
\ee}\noindent
i.e. $(z^{R_{can}[0]})_{ij}=\frac{({\rm log} z)^{j-i}}{(j-i)!}$. This 
allows us to write $U'(z)^t=Z(z)q'(z)$ with $q'(z)_{sj}:=
\frac{S'_{0,j-s}(z)}{s!}$. In particular, we obtain:
\be
q'(0)={\rm diag}(1,1,1/2,1/6)~~,
\ee
which allows us to determine $M=q(0)^tq'(0)^{-1}$. Once $M^t$ is known, 
the monodromy of the Meijer basis follows from the relation 
$T[0]=MT_{can}[0]M^{-1}$, where $T_{can}[0]^t=e^{2\pi i R_{can}[0]}$ is 
given by:
{\footnotesize \be
\label{can0}
T_{can}[0]^t=
\left[\begin{array}{cccc} 
1&2\pi i&\frac{1}{2}(2 \pi i )^2&\frac{1}{6} (2 \pi i)^3\\
0&1&2 \pi i&\frac{1}{2}(2 \pi i)^2\\
...&...&...& ...\\
0&0&0&1\\
\end{array}\right]~~.
\ee}

In our class of examples, $q(0)$ can be easily determined from the 
expansions given above. Indeed, we can write:
\be
\label{nu_expansion}
\nu_j(n)=\sum_{s=0}^{j}{v_{sj}(n)({\rm log} z)^s}~~,
\ee
with:
{\footnotesize 
\be
\label{vs}
\begin{array}{ccccccc}
v_{00}&:=&1& &v_{01}(n)&:=&\eta_1(n)\\
v_{11}&:=&1& &v_{02}(n)&:=&\eta_2'(n)+(\eta_2(n)+i\pi)^2\\
v_{12}(n)&:=&2(\eta_2(n)+i\pi) & &v_{22}&:=&1\\
v_{03}(n)&:=&\eta_3''(n)+3\eta_3'(n)\eta_3(n)+\eta_3(n)^3 & &v_{13}(n)&:=&
3\left[\eta_3'(n)+\eta_3(n)^2\right]\\
v_{23}(n)&:=&3\eta_3(n) & &v_{33}&:=&1
\end{array} ~~.
\ee}\noindent
This gives $U_j(z)=\sum_{s=0}^{j}{q_{sj}(z)({\rm log} z)^s}$, with:
\be
q_{sj}(z):=\frac{(-1)^j}{j!}\sum_{n=0}^{\infty}
{\frac{(\alpha_1)_n(\alpha_2)_n(\alpha_3)_n(\alpha_4)_n}{(n!)^4}v_{sj}(n)z^n}
~~.
\ee
In particular, we have $q_{sj}(0)=\frac{(-1)^j}{j!}v_{sj}(0)$.
Substituting the formulae for $\eta_j^{(i)}$ given above, we obtain a 
complicated expression for $q(0)$ in terms of 
polygamma functions which we will not reproduce here. This expression can be 
simplified by the following indirect procedure. Let us introduce the variable 
$w=\kappa z$, where $\kappa={\rm exp}(\sum_{k=1}^{4}{\psi(\alpha_k)}-4\psi(1))$
(in practice, this plays the role of the natural coordinate on the moduli 
space dictated by the monomial-divisor mirror map of \cite{mdmm}). 
One can then consider isolating the factors $({\rm log} w)^s$ instead of 
$({\rm log} z)^s$
in the expansion of $U_j(z)$ around $z=0$. Doing so amounts to writing:
\be
U_j(w)=\sum_{s=0}^{j}{{\tilde q}_{sj}(w)({\rm log} w)^s}~~.
\ee
where the functions ${\tilde q}_{sj}(w)$ are now given by:
\be
{\tilde q}_{sj}(w):=\frac{(-1)^j}{j!}\sum_{n=0}^{\infty}
{\frac{(\alpha_1)_n(\alpha_2)_n(\alpha_3)_n(\alpha_4)_n}{(n!)^4}
{\tilde v}_{sj}(n)\left(\frac{w}{\kappa}\right)^n}~~,
\ee
as can be seen upon expanding:
\be
\nu_s(n,w)=\sum_{s=0}^{j}{{\tilde v}_{sj}(n)({\rm log} w)^s}~~,
\ee
with:
{\footnotesize 
\be
\begin{array}{ccccccc}
{\tilde v}_{00}&:=&1& &{\tilde v}_{01}(n)&:=&\eta_1(n)\\
{\tilde v}_{11}&:=&1& &{\tilde v}_{02}(n)&:=&\eta_2'(n)+(\theta_2(n)+i\pi)^2\\
{\tilde v}_{12}&:=&2(\theta_2(n)+i\pi)& &{\tilde v}_{22}&:=&1\\
{\tilde v}_{03}(n)&:=&\eta_3''(n)+3\eta_3'(n)\theta_3(n)+\theta_3(n)^3 & &
{\tilde v}_{13}(n)&:=&3\left[\eta_3'(n)+\theta_3(n)^2\right]\\
{\tilde v}_{23}(n)&:=&3\theta_3(n)& &{\tilde v}_{33}&:=&1~~.
\end{array}
\ee}\noindent
Here $\theta(n)$ are defined through:
\be
\eta_j(n)+{\rm log}((-1)^{j+1}z)=\theta_j(n)+{\rm log}((-1)^{j+1}w)~~.
\ee
Since ${\rm log}(w)={\rm log}(z)+{\rm log}(\kappa)=
{\rm log}(z)+\sum_{k=1}^{4}{\psi(\alpha_k)}-4\psi(1)$,
we have:
\be
\theta_j(n):=\sum_{k=1}^{4}{\left[\psi(n+\alpha_k)-\psi(\alpha_k)\right]}+
(3-j)\left[\psi(1)-\psi(n+1)\right]-(j+1)\sum_{l=1}^{n}{\frac{1}{l}}~~.
\ee
The new sequences $\theta_j(n)$ have the convenient property $\theta_j(0)=0$,
which allows us to obtain a simple expression for the matrix ${\tilde q}(0)$:
{\footnotesize
\be
{\tilde q}(0):=
\left[\begin{array}{cccc} 
1&0&\frac{1}{2}(\eta_2'(0)-\pi^2)&-\frac{1}{6}\eta_3''(0)\\
0&-1& i\pi &-\frac{1}{2}\eta_3'(0)\\
0&0&\frac{1}{2}&0\\
0&0&0&-\frac{1}{6}\\
\end{array}\right]~~.
\ee}\noindent
In this expression:
{\footnotesize 
\bea
\eta'_2(0)&=&
\psi'(\alpha_1)+\psi'(\alpha_2)+\psi'(\alpha_3)+
\psi'(\alpha_4)+1/3\pi^2\nn\\
\eta'_3(0)&=&
\psi'(\alpha_1)+\psi'(\alpha_2)+\psi'(\alpha_3)+
\psi'(\alpha_4)+2/3\pi^2\\
\eta''_3(0)&=&
\psi''(\alpha_1)+\psi''(\alpha_2)+\psi''(\alpha_3)+
\psi''(\alpha_4)+8\zeta(3)~~.\nn
\eea}

Using ${\rm log}(w)={\rm log}(\kappa)+{\rm log}(z)$, it is easy to see that:
\be
q(0)=c{\tilde q}(0)~~,
\ee
with $c$ an upper triangular matrix whose nonzero entries are given by 
{\footnotesize 
$c_{sp}:=\left(\begin{array}{c}p\\s\end{array}\right)({\rm log}\kappa)^{p-s}$}
for all $s\leq p$. Moreover, it is not hard to see that the matrix 
$q'(0)c^tq'(0)^{-1}$ commutes
with $R_{can}^t$, so that we can use  
${\tilde M}:={\tilde q}(0)^tq'(0)^{-1}$ instead of $M$. In other words, we 
have:
\be
T[0]={\tilde M}T_{can}[0]{\tilde M}^{-1}~~,
\ee
which allows for the computation of $T[0]$ upon using (\ref{can0}):
{\footnotesize
\be
T[0]=\left[\begin {array}{cccc} 
1&0&0&0\\
-2i\pi &1&0&0\\
-4\pi^2&-2i\pi &1&0\\
0&0&-2i\pi &1
\end {array}\right ]~~.
\ee}
\noindent Remarkably, this expression is independent of $\alpha_k$, which
underscores the universal behaviour of the Meijer periods in the large 
complex structure limit, as mentioned in the introduction. In particular, 
these periods are adapted to the large complex structure monodromy weight 
filtration, as can be seen form the form of $T[0]$ or directly from 
the expansions (\ref{LCS_expansions}).

\subsubsection{Meijer monodromies around $z=\infty$}

The monodromy of the Meijer basis around $z=\infty$ is straightforward to 
extract. For this, first notice that:
\be
((-1)^{j+1}z)^{-\alpha_k}=z^{-\alpha_k}\left(\delta_{j,odd}+\delta_{j,even}
e^{-i\pi\alpha_k}\right)
\ee
where $\delta_{j,odd}$ equals $1$ if $j$ is odd, and $0$ if $j$ is even
(and a similar definition holds for $\delta_{j,even}$). 
This allows us to write $U_j(z)=\sum_{k=0}^{3}{a_{jk}u_k(z)}$, where 
\linebreak 
{\footnotesize 
$u_{k-1}(z):=z^{-\alpha_k}~
\F[4,3][{\alpha_k,~\alpha_k,~\alpha_k,~\alpha_k},{1+\alpha_k-\alpha_1~..~
{\hat 1}~..~1+\alpha_k-\alpha_4}](1/z)$} and the transition matrix 
$A:=(a_{jk})_{j,k=0~...~3}$ is given by:
\be
A=\frac{1}{\prod_{i=1~...~4}{\Gamma(\alpha_k)}}CD~~,
\ee
where:
\bea
C_{jk}&:=&\left(\frac{\sin(\pi \alpha_{k+1})}{\pi}\right)^{3-j}
\left(\delta_{j,odd}+\delta_{j,even}e^{-i\pi\alpha_{k+1}}\right)~~\nn\\
D&:=&{\rm diag}(d_0,d_1,d_2,d_3)~~\nn\\
d_k&:=&\Gamma(\alpha_k+1)^4\prod_{l=1~...~4,l\neq k+1}
{\Gamma(\alpha_l-\alpha_{k+1})}~~.\nn
\eea
The monodromy of the basis $u$ has the simple form:
\be
T_u[\infty]={\rm diag}(e^{-2\pi i\alpha_1}~...~e^{-2\pi i \alpha_4})~~,
\ee
which coincides with $T_J=e^{2\pi i R_J[\infty]^t}$, where $R_J[\infty]$ is 
the Jordan form  of the matrix $R_{can}[\infty]$. 
This allows us to compute the monodromy of the Meijer basis via the relation 
$T[\infty]=AT_u[\infty]A^{-1}$. Since both $D$ and $T_u[\infty]$ are diagonal, 
we have $DT_u[\infty]D^{-1}=T_u$, which allows us to reduce this relation to:
\be
T[\infty]=CT_u[\infty]C^{-1}~~.
\ee
The last equality is more convenient for computing $T[\infty]$ 
since it involves only the matrix $C$.

\subsection{The mirror quintic revisited}

Let us now show how the results of \cite{quintic}
can be recovered in our framework. Beyond acting as a check on our 
computations, this will also serve to illustrate the connection between 
vanishing cycles and arithmetic identities in a well-understood example.

The mirror quintic can be obtained as a particular case of the discussion 
above by choosing $\alpha_i=i/5$ for all $i=1~...~4$. 
Let $\sigma:=e^{\frac{2\pi i}{5}}$. In this case, one has:
{\footnotesize 
\bea
\begin{array}{ccc}
{\tilde q}(0)=\left [\begin {array}{cccc} 1&0&5/3\,{\pi }^{2}&40\,\zeta (3)
\\0&-1&i\pi &-7/3\,{\pi }^{2}
\\0&0&1/2&0\\0&0&0&-1/6
\end {array}\right ]&~,~&T_u[\infty]=\left [\begin {array}{cccc} {\sigma}^{-1}&0&0&0\\0&{\sigma}^{-2}&0&0\\0&0&{\sigma}^{-3}&0
\\0&0&0&{\sigma}^{-4}\end {array}\right ]
\end{array}~~\nn,
\eea}\noindent which immediately gives:
{\footnotesize
\bea
\begin{array}{ccc}
T[0]=\left [\begin {array}{cccc} 1&0&0&0\\-2\,i
\pi &1&0&0\\-4\,{\pi }^{2}&-2\,i\pi &1&0
\\0&0&-2\,i\pi &1\end {array}\right ]&~,~&
T[\infty]=\left [\begin {array}{cccc} -4&-5/2\,{\frac {i}{\pi }}&5/4\,{\pi }^{-2}&5/8\,{\frac {i}{{\pi }^{3}}}\\-2
\,i\pi &1&0&0\\-4\,{\pi }^{2}&-2\,i\pi &1&0\\
0&0&-2\,i\pi &1\end {array}
\right ]\end{array}\nn
\eea}\noindent and $T[1]=T[0]^{-1}T[\infty]$. These matrices 
satisfy $(T[0]-I)^4=0,~(T[1]-I)^2=0,~T[\infty]^5-I=0$.
To make contact with the results in \cite{quintic}, first 
note that our convention for monodromies differs from the one used there by 
a change of orientation -- which implies an inversion of the matrices 
$T[0]$ and $T[\infty]$
\footnote{This is due to our use of different coordinates on the moduli 
space, as we explain below.}. 
Then notice that the fundamental period $U_0$ is cyclic for the operator 
$T[\infty]^{-1}$. Thus one can generate a cyclic set by acting with 
$T[\infty]^{-1}$ on $U_0$, which produces $5$ periods $\omega_j$ 
related to the Meijer periods through:
{\footnotesize 
\be
\left[\begin{array}{c}\omega_0\\\omega_1\\\omega_2\\\omega_3\\
\omega_4\end{array}
\right]=
\left [\begin {array}{cccc} 1&0&0&0\\1&5/2\,{\frac {
i}{\pi }}&0&-5/8\,{\frac {i}{{\pi }^{3}}}
\\-4&-5\,{\frac {i}{\pi }}&5/4\,{\pi }^{-2}
&{\frac {15}{8}}\,{\frac {i}{{\pi }^{3}}}\\
6&5\,{\frac {i}{\pi }}&-5/2\,{\pi }^{-2}&-{\frac {15}{8}}\,{
\frac {i}{{\pi }^{3}}}\\-4&-5/2\,{\frac {
i}{\pi }}&5/4\,{\pi }^{-2}&5/8\,{\frac {i}{{\pi }^{3
}}}\end {array}\right ]
\left[\begin{array}{c}U_0\\U_1\\U_2\\U_3\end{array}
\right]~~.
\ee
} 
This corresponds to the procedure used in \cite{quintic} (as well as in 
\cite{Font, Candelas_periods,2pm1,2pm2}) in order to generate a basis of 
periods from the fundamental period. 
Indeed, the cyclic basis used in \cite{quintic} corresponds to 
the periods $\omega_2,\omega_1,
\omega_0$ and $\omega_4$ denoted by the same letters 
in that paper. Up to a rescaling by $-(2 \pi i/5)^3$, 
these form a period vector denoted there by $\omega$. It follows 
that the Meijer period vector  is related to the period vector $\omega$ 
of \cite{quintic} through:
{\footnotesize
\bea
\omega=-\left(\frac{2 \pi i}{5}\right)^3
\left[\begin{array}{c}\omega_2\\\omega_1\\\omega_0\\\omega_4\end{array}\right]=
L U~~\mbox{~with~}~L=
\left [\begin {array}{cccc} -{\frac {32}{125}}\,i{\pi }^{3}&{
\frac {8}{25}}\,{\pi }^{2}&{\frac {2}{25}}\,i\pi &-{\frac {3}
{25}}\\{\frac {8}{125}}\,i{\pi }^{3}&-{
\frac {4}{25}}\,{\pi }^{2}&0&1/25\\{\frac {8}{125}}
\,i{\pi }^{3}&0&0&0\\-{\frac {32}{125}}\,
i{\pi }^{3}&{\frac {4}{25}}\,{\pi }^{2}&{\frac {2}{25}}\,
i\pi &-1/25\end {array}\right ]~~.\nn
\eea}\noindent
In order to make contact with the results of \cite{quintic}, 
one must also take into account the fact 
that the monodromy matrices given there correspond 
to working on a 5-fold cover of the moduli space. This is parameterized in 
\cite{quintic} by 
a variable $\psi$, related to our coordinate by $z=\psi^{-5}$.
It follows that our monodromy matrices $LT[0]L^{-1},~LT[1]L^{-1},~
LT[\infty]L^{-1}$ 
should be compared respectively with 
the matrices $t_\infty^{1/5},t_\infty^{-1/5}a^{-1}$ and $a^{-1}$ 
of \cite{quintic}, and it is easy to see that they agree. 
As a further check, 
one can easily show that the $|\psi|<1$ expansions of $\omega_j$ given in 
Appendix B of \cite{quintic} agree with the expansions of the cyclic basis 
following from (\ref{LG_expansions}). 

A full sublattice of the integral lattice $\Lambda=H_3(Y,\Z)$ can be 
identified as explained in Section 2.
This gives a lattice $\Lambda_0$ characterized by the fact that 
the period vector associated with one of its bases (which we call $E$) 
is (up to a global factor):
{\footnotesize
\be
U_E=E U=\left [\begin {array}{cccc} 1&0&0&0\\-4&-5/2{
\frac {i}{\pi }}&5/4{\pi }^{-2}&5/8{\frac {i}{{
\pi }^{3}}}\\6&5{\frac {i}{\pi }}&-5/2{
\pi }^{-2}&-{\frac {15}{8}}{\frac {i}{{\pi }^{3}}}
\\-4&-5{\frac {i}{\pi }}&5/4{\pi }^{-2}
&{\frac{15}{8}}{\frac {i}{{\pi }^{3}}}\end {array}\right ]U~~.
\ee}\noindent 
The integral basis $P$ of 
$\Lambda$ used in \cite{quintic} is characterized
by a period vector $\Pi$, which is related to the Meijer periods $U$ by
$\Pi=\Theta U$, with:
{\footnotesize
\be
\Theta=\left [\begin {array}{cccc} 0&{\frac {12}{125}}{\pi }^{2}&{\frac {2}
{25}}i\pi &0\\-{\frac {8}{125}}i{\pi }^{3}&0&0&0\\0&{\frac {4}{25}}{\pi }^{2}&{
\frac {4}{25}}i\pi &0\\0&-{\frac {4}{25}}
{\pi }^{2}&0&1/25\end {array}\right ]~~.
\ee}\noindent
Hence the transition matrix from the basis $P$ of $H_3(Y,\Z)$ to 
the  basis $E$ of $\Lambda_0$ is:
{\footnotesize
\be
K=(E \Theta^{-1})^t=\left(\frac{2\pi i}{5}\right)^{-3}
\left [\begin {array}{cccc} 0&5&-20&15\\1&-4&6&-4
\\0&-3&11&-8\\0&1&-3&3\end {array}
\right ]=\left(\frac{2\pi i}{5}\right)^{-3}K_0~~.
\ee}\noindent
Since ${\rm det}(K_0)=5$, we see that 
$\Lambda'_0=\left(\frac{2\pi i}{5}\right)^3\Lambda_0$ 
is an index 5 sublattice of 
$\Lambda$, i.e. the finite group $\Lambda/\Lambda'_0$ has order 5.  

The period \footnote{This equals $z^2$ in the notations of \cite{quintic}.} 
$-(2\pi i/5)^3(\omega_1-\omega_0)$
which vanishes at the conifold point can be expressed in terms of the 
Meijer periods as $\frac{1}{25}(-4\pi^2U_1+U_3)$. Hence collapse of the 
associated 3-cycle at $z=1$ is reflected by the relation:
\be
U_3(1)=4\pi^2 U_1(1)~~,
\ee
which after a few simplifications can be seen to be equivalent to:
{\footnotesize 
\bea
\label{quintic_id}
192\pi^5\sqrt {2}\sqrt {5+\sqrt {5}}\left(5-2\sqrt {5}\right)\Gamma (3/5)^5
~\F[4,3][{1/5,1/5,1/5,1/5},{4/5,2/5,3/5}](1)+\nn\\
+960\pi^{5}\sqrt {2}\sqrt {5+\sqrt {5}}\left(2\sqrt {5}-5\right)\Gamma (4/5)^5
~\F[4,3][{2/5,2/5,2/5,2/5},{4/5,3/5,6/5}](1)+
\nn\\
+3125\left(11-5\sqrt{5}\right)\Gamma(3/5)^5\Gamma(4/5)^{10}
~\F[4,3][{4/5,4/5,4/5,4/5},{8/5,7/5,6/5}](1)+\nn\\
+3750\Gamma (4/5)^{5}
\Gamma (3/5)^{10}~\F[4,3][{3/5,3/5,3/5,3/5},{4/5,7/5,6/5}](1)
=0~~.
\eea}\noindent
One can of course write down a pair identity by using the expansions 
(\ref{LCS_expansions}) 
around the large complex structure point, which can be thought of as 
following from the one above through analytic continuation. It is a 
challenging problem to find direct proofs of these identities.

Figures 6 and 7 compare the classical and quantum volumes of the  
even-dimensional cycles on the quintic. 
In Figure 6, we plot classical volumes as a function
of the classical K\"ahler parameter $s$, defined by $J = se$ with $e$ the
generator of $H^2(X,\Z)$. In figure 7, for ease of comparison,
we plot the quantum volumes as a function of essentially the same
parameter $s$ ---
the so-called `algebraic measure' on the quantum moduli space 
(see \cite{small_distances1} for a discussion of this concept), which
in this case is related to $z$ via $s=\frac{1}{2\pi}{\rm log}
(\frac{5^5}{z})$ (note that $\kappa=5^{-5}$ in this example). In the language
of \cite{small_distances1}, this coordinate is $s={\rm Im}(t_{alg})$, where 
$t_{alg}=\frac{1}{2\pi i}{\rm log}(\frac{z}{5^5})$.

For comparison with the classical situation, 
we consider only the region $s\ge 0$ in Figure 7, 
even though quantum mechanically we can continue $s$ to negative values as 
well.
The point $P$, where $s=\frac{5}{2\pi}{\rm log}5\approx 1.28$ and $z=1$, 
corresponds to the conifold. Figure 7 also displays the values of the 
special coordinate 
$t=\frac{2(\omega_1-\omega_0)+\omega_2-\omega_4}{5\omega_0}$, which measures
the volume of a D2-brane wrapped over the generator of $H_2(X,\Z)$. We 
also plot the absolute value of the integral period $-\frac{2i}{25}U_2$, 
which can be interpreted as the mass of a wrapped D4-brane.  
Note that the  scale used for the $y$-axis is different for the two graphs
and different from the scale used on the $x$-axis (this is needed in order 
to fit the interesting portion of the graphs in the limited space available).
Taking this rescaling into account, one can easily check that 
the curve $m=|t(s)|$  asymptotes to the line $m=s$ (the diagonal of the first 
quadrant) in the the large radius limit $s\rightarrow \infty$.
In this limit, 
$|U_v|$ asymptotes to $1.923-4.134s+1.653s^3$.

These figures capture the essential point, one that we will find
repeated in various forms in subsequent examples. Namely, the classical
relations between 2, 4, and 6 cycle volumes are significantly modified
in the quantum setting. In particular, we see that a 6-cycle collapses
to zero quantum volume at the conifold point, even though 2 and 4 cycle 
volumes stay positive.

\vskip 0.2 in

$\begin{array}{cc}
\scalebox{0.3}{\input{quintic_graph_cl.pstex_t}}&
\scalebox{0.3}{\input{quintic_graph_q.pstex_t}}\\
\begin{array}{c}
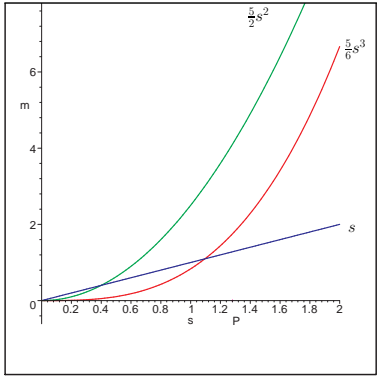
\mbox{Figure 6. {\footnotesize  Graph of the classical volumes}}\\
\mbox{\footnotesize $s,~\frac{5}{2}s^2$ and $\frac{5}{6}s^3$ of a set of 
two,four and six}\\
\mbox{\footnotesize cycles vs. the classical Kahler parameter $s$. }
\end{array}~~~~~~~
&
\begin{array}{c}
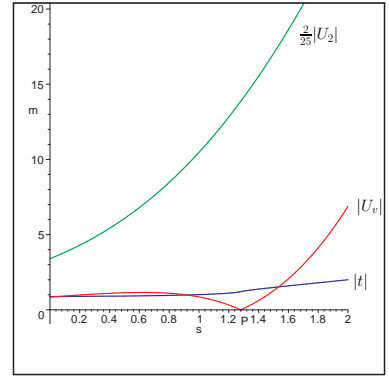
\mbox{Figure 7. {\footnotesize  Graph of $|U_v|$ vs. the classical Kahler }}\\
\mbox{\footnotesize
parameter $s$. For comparison, we also plot the}\\
\mbox{\footnotesize 
absolute value of the special coordinate $t$, which }\\ 
\mbox{\footnotesize measures the mass of a wrapped D2-brane, and the}\\
\mbox{absolute value of the (integral) period $-\frac{2i}{25}U_2$.}
\end{array}~~~~~~~~~
\end{array}$~~~~~~~~~~~~~~~~~~~~~~~~~~~~~~~~~~~~~~~~~~~~~~~~~~~~~~~~~~~~~~~

\section{Examples II: Some `degenerate' models}

Let us now consider the case when the parameters $\alpha_i$ satisfy
conditions (2) and (3), but they violate condition (1).
Namely, consider $0<\alpha_i < 1$ chosen so that
$\alpha_1,\alpha_2,\alpha_3$ are distinct but $\alpha_3=\alpha_4$.
This situation is encountered for some of the  
examples studied in \cite{Candelas_periods} (see Table 3.1. on page 26 of 
that paper). In particular, it is satisfied for a complete intersection 
of a quadric and a quartic in $\P^5$ (the third entry of that table), 
a model which we will encounter as an interesting sub-locus of a 
two-parameter example in Section 4. Another case which 
satisfies this condition is the complete intersection $\P^6[2,2,3]$.

\subsection{The Meijer periods}

Clearly nothing qualitatively new happens in such a model when computing the 
expansions for $|z|<1$, since closing the contour to the right only gives 
contributions form the poles in $\Gamma(-s)$ (the A-type poles) 
in (\ref{phi}), whose nature is independent of the relative 
values of $\alpha_i$. Hence the expansions around the large complex structure
point have the same form as above. 
On the other hand, closing the contour to the 
left gives two types of contributions from the (B) - type poles:

$(B_1)$ Contributions from $s=-n-\alpha_3$ ($n$ a nonnegative integer), which 
are double poles

$(B_2)$ Contributions from $s=-n-\alpha_1$ and $s=-n-\alpha_2$ 
($n$ a nonnegative integer), which are simple poles.

\noindent Correspondingly, we can write the expansions of $U_j(z)$ for 
$|z|>1$ as:
\be
\label{deg_LG_expansion}
U_j(z)=u^{(1)}_j(z)+u^{(2)}_j(z)~~,
\ee
separating the contributions from these two sub-types. 
The $(B_2)$-type contributions are of exactly the same form as above:
{\footnotesize
\bea
\label{deg_LG_expansion1}
u^{(2)}_j(z)=\frac{1}{\Gamma(\alpha_1)\Gamma(\alpha_2)\Gamma(\alpha_3)^2}
\sum_{k=1}^{2}{
\left(\frac{\sin(\pi\alpha_k)}{\pi}\right)^{3-j}{\Gamma(\alpha_k)}^4
\prod'_{l=1~...~4}{\Gamma(\alpha_l-\alpha_k)}
((-1)^{j+1}z)^{-\alpha_k}~}\times \nn\\
\times~
\F[4,3][{\alpha_k,\alpha_k,\alpha_k,\alpha_k},{1+\alpha_k-\alpha_1,~...~,{\hat 1},~...~,1+\alpha_k-\alpha_4}](1/z)~~,
\eea}\noindent but the $B_1$-type poles induce logarithmic terms: 
{\tiny
\bea
\label{deg_LG_expansion2}
u^{(1)}_j(z)=
\frac{1}{\Gamma(\alpha_1)\Gamma(\alpha_2)\Gamma(\alpha_3)^2}
\left(\frac{\sin(\pi \alpha_3)}{\pi}\right)^{3-j}
((-1)^{j+1}z)^{-\alpha_3}
\sum _{n=0}^{\infty }{
\frac{\Gamma (n+\alpha_3)^{4}\Gamma (-n+\alpha_1-\alpha_3)
\Gamma (-n+\alpha_2-\alpha_3)}{n!^2}}]z^{-n}~~~~~~~~~~~~~~~~~~~~~\nn\\
\left[\psi(-n+\alpha_1-\alpha_3)+\psi(-n+\alpha_2-\alpha_3)
-\left (j+1\right )\psi(n+\alpha_3)-\left (3-j\right )\psi(1-n-
\alpha_3)+2\psi(1)+2\sum _{l=1}^{n}{\frac{1}{l}}+
{\rm log}((-1)^{j+1}z)\right]~~.
\eea}

\subsection{Monodromies of the Meijer basis}

The monodromies about $z=0$ follow by substituting $\alpha_4=\alpha_3$ in the 
results of Section 3. In order to obtain the monodromies about $z=\infty$, 
one has to be a bit more careful, as we now explain. 

By using the general results of Section 2, it is easy to compute 
the matrix $R_{can}[\infty]$:
{\footnotesize 
\bea
R_{can}[\infty]=
\left[\begin {array}{cccc} 
0&-1&0&0\\
0&0&-1&0\\
0&0&0&-1\\
\alpha_1\alpha_2\alpha_3^2&-2\alpha_1\alpha_2\alpha_3-
\alpha_1\alpha_3^2-\alpha_2\alpha_3^2&\alpha_1
\alpha_2+2\alpha_1\alpha_3+2\alpha_2\alpha_3+\alpha_3^2&-\alpha_1- 
\alpha_2-2\alpha_3
\end {array}\right ]~~.\nn
\eea}\noindent and its Jordan form:
{\footnotesize 
\bea
\label{J}
R_J[\infty]=\left [\begin {array}{cccc} 
-\alpha_1&0&0&0\\
0&-\alpha_2&0&0\\
0&0&-\alpha_3&1\\
0&0&0&-\alpha_3
\end {array}\right ]~~.
\eea}\noindent These are related by a transition matrix $P\in GL(4,\C)$ 
satisfying:
\be
\label{P}
R_{can}[\infty]=P R_J[\infty] P^{-1}~~.
\ee
While such a transition matrix is not unique, for what follows it suffices to 
pick any $P$ with the property (\ref{P}), for example the matrix:
{\scriptsize 
\bea
P=\left [\begin {array}{cccc} -{\frac {\alpha_{{2}}{\alpha_{{3}}}^{2}}{
\left (\alpha_{{1}}-\alpha_{{3}}\right )^{2}\left (\alpha_{{1}}-\alpha
_{{2}}\right )}}&{\frac {\alpha_{{1}}{\alpha_{{3}}}^{2}}{\left (\alpha
_{{2}}-\alpha_{{3}}\right )^{2}\left (\alpha_{{1}}-\alpha_{{2}}\right 
)}}&{\frac {\alpha_{{1}}\alpha_{{2}}\alpha_{{3}}}{\left (\alpha_{{2}}-
\alpha_{{3}}\right )\left (\alpha_{{1}}-\alpha_{{3}}\right )}}&{\frac 
{\left (-2\alpha_{{1}}\alpha_{{3}}+\alpha_{{1}}\alpha_{{2}}+3{
\alpha_{{3}}}^{2}-2\alpha_{{2}}\alpha_{{3}}\right )\alpha_{{1}}
\alpha_{{2}}}{\left (\alpha_{{2}}-\alpha_{{3}}\right )^{2}\left (
\alpha_{{1}}-\alpha_{{3}}\right )^{2}}}\\-{\frac {
\alpha_{{1}}\alpha_{{2}}{\alpha_{{3}}}^{2}}{\left (\alpha_{{1}}-\alpha
_{{3}}\right )^{2}\left (\alpha_{{1}}-\alpha_{{2}}\right )}}&{\frac {
\alpha_{{1}}\alpha_{{2}}{\alpha_{{3}}}^{2}}{\left (\alpha_{{2}}-\alpha
_{{3}}\right )^{2}\left (\alpha_{{1}}-\alpha_{{2}}\right )}}&{\frac {
\alpha_{{1}}\alpha_{{2}}{\alpha_{{3}}}^{2}}{\left (\alpha_{{2}}-\alpha
_{{3}}\right )\left (\alpha_{{1}}-\alpha_{{3}}\right )}}&-{\frac {
\alpha_{{1}}\alpha_{{2}}{\alpha_{{3}}}^{2}\left (\alpha_{{1}}-2
\alpha_{{3}}+\alpha_{{2}}\right )}{\left (\alpha_{{2}}-\alpha_{{3}}
\right )^{2}\left (\alpha_{{1}}-\alpha_{{3}}\right )^{2}}}
\\-{\frac {{\alpha_{{1}}}^{2}{\alpha_{{3}}}^{2}
\alpha_{{2}}}{\left (\alpha_{{1}}-\alpha_{{3}}\right )^{2}\left (
\alpha_{{1}}-\alpha_{{2}}\right )}}&{\frac {\alpha_{{1}}{\alpha_{{2}}}
^{2}{\alpha_{{3}}}^{2}}{\left (\alpha_{{2}}-\alpha_{{3}}\right )^{2}
\left (\alpha_{{1}}-\alpha_{{2}}\right )}}&{\frac {\alpha_{{1}}\alpha_
{{2}}{\alpha_{{3}}}^{3}}{\left (\alpha_{{2}}-\alpha_{{3}}\right )
\left (\alpha_{{1}}-\alpha_{{3}}\right )}}&-{\frac {\alpha_{{1}}\alpha
_{{2}}{\alpha_{{3}}}^{2}\left (\alpha_{{1}}\alpha_{{2}}-{\alpha_{{3}}}
^{2}\right )}{\left (\alpha_{{2}}-\alpha_{{3}}\right )^{2}\left (
\alpha_{{1}}-\alpha_{{3}}\right )^{2}}}\\-{\frac {{
\alpha_{{1}}}^{3}\alpha_{{2}}{\alpha_{{3}}}^{2}}{\left (\alpha_{{1}}-
\alpha_{{3}}\right )^{2}\left (\alpha_{{1}}-\alpha_{{2}}\right )}}&{
\frac {\alpha_{{1}}{\alpha_{{2}}}^{3}{\alpha_{{3}}}^{2}}{\left (\alpha
_{{2}}-\alpha_{{3}}\right )^{2}\left (\alpha_{{1}}-\alpha_{{2}}\right 
)}}&{\frac {\alpha_{{1}}\alpha_{{2}}{\alpha_{{3}}}^{4}}{\left (\alpha_
{{2}}-\alpha_{{3}}\right )\left (\alpha_{{1}}-\alpha_{{3}}\right )}}&-
{\frac {\alpha_{{1}}\alpha_{{2}}{\alpha_{{3}}}^{3}\left (-\alpha_{{1}}
\alpha_{{3}}+2\alpha_{{1}}\alpha_{{2}}-\alpha_{{2}}\alpha_{{3}}
\right )}{\left (\alpha_{{2}}-\alpha_{{3}}\right )^{2}\left (\alpha_{{
1}}-\alpha_{{3}}\right )^{2}}}\end {array}\right ]~~.\nn
\eea}\noindent Assuming that we made such a choice, 
we can define a {\em Jordan basis} of periods
(which depends on the choice of $P$)
through $U_J(z)=P^{-t} U_{can}(z)$. In terms of fundamental 
matrices for the associated Picard-Fuchs system, this relation reads:
\be
\Phi_{can}(z)=\Phi(z)P^{-1}~~,
\ee
which implies that the monodromy of the basis $U_J$ about $z=\infty$ is 
given by the matrix $R_J[\infty]$. The associated nilpotent orbits 
are related through $S_{can}(z)=S_J(z)P^{-1}$, which, together with the 
defining property $S_{can}[\infty]=I$, allows us to compute:
\be
\label{SJinf}
S_J[\infty]=P~~.
\ee
The Jordan basis is a useful device for extracting the monodromy of the 
Meijer periods. To see this, note that equation (\ref{J}) assures us that 
we can extract the nontrivial behaviour of periods about $z=\infty$ by 
writing:
\be
\label{qs}
U(z)=Z(z)q(z)~~,~~U_J(z)=Z(z)q_J(z)~~,
\ee
where $q(z),q_J(z)$ are one-valued and regular at $z=\infty$ while the 
row vector $Z(z)$ is defined by\footnote{Note that this differs from the 
vector $Z(z)$ used in Section 3.}:
\be
Z(z)=\left[\begin{array}{cccc}z^{-\alpha_1}&z^{-\alpha_2}&z^{-\alpha_3}
&z^{-\alpha_3} {\rm log} z\end{array}\right]~~.
\ee
Since the Jordan monodromy
{\scriptsize 
\be
T_J[\infty]=e^{2\pi i R_J[\infty]^t}=
\left [\begin {array}{cccc} 
e^{-2\pi i\alpha_1}&0&0&0\\
0&e^{-2\pi i\alpha_2}&0&0\\
0&0&e^{-2\pi i\alpha_3}&0\\
0&0&2\pi ie^{-2\pi i\alpha_3} &e^{-2\pi i\alpha_3}
\end {array}\right ]
\ee}\noindent is trivially known, it follows that, in 
order to compute the monodromy of $U(z)$, it suffices to identify the 
transition matrix $M\in GL(4,\C)$ such that $U(z)=MU_J(z)$. The latter can be 
determined form knowledge of $q(\infty)$ and $q_J(\infty)$ 
through the relation:
\be
M=q(\infty)^tq_J(\infty)^{-t}~~,
\ee
provided that we can compute the two matrices appearing in this equation. 
The matrix $q_J(\infty)$ is straightforward to extract by noting that 
$U_J(z)^t$ coincides with the first row of $\Phi_J(z)=S(z)z^{R_J}$ and 
by rewriting $U_j(z)=S_i(z)(z^{R_J})_{ij}$ in the form (\ref{qs}):
{\footnotesize \be
q_J(z)=\left [\begin {array}{cccc} 
S_{1,1}(z)&0&0&0\\
0&S_{1,2}(z)&0&0\\
0&0&S_{1,3}(z)&S_{{1,4}}(z)
\\0&0&0&S_{1,3}(z)
\end {array}\right ]~~,
\ee}
\noindent which together with (\ref{SJinf}) gives:
{\footnotesize \be
q_J(\infty)=\left [\begin {array}{cccc} 
P_{1,1}&0&0&0\\
0&P_{1,2}&0&0\\
0&0&P_{1,3}&P_{1,4}
\\0&0&0&P_{1,3}
\end {array}\right ]~~.
\ee}
\noindent On the other hand, the matrix $q(z)$ can be determined via its 
defining 
property (\ref{qs}) by using the expansions 
(\ref{deg_LG_expansion1},\ref{deg_LG_expansion2}). This amounts to writing:
\be
U_j(z)=\sum_{i=0}^{4}{q_{ij}(z)Z_i(z)}~~,
\ee
which immediately gives:
\be
q(\infty)=\frac{1}{\Gamma(\alpha_1)\Gamma(\alpha_2)\Gamma(\alpha_3)^2}D C^t~~,
\ee
with $C$ and $D$ given by:
{\footnotesize 
\bea
C_{jk}&=&\left(\frac{\sin\pi \alpha_k}{\pi}\right)^{3-j}
\left(\delta_{j,odd}+\delta_{j,even}e^{-i\pi \alpha_k}\right)~~\mbox{for~}
k=0,1,3~~~~~~~~~~~~~\nn\\
C_{j2}&=&\left(\frac{\sin\pi \alpha_3}{\pi}\right)^{3-j}
\left(\delta_{j,odd}+\delta_{j,even}e^{-i\pi \alpha_3}\right)
(\psi(\alpha_1-\alpha_3)+\psi(\alpha_2-\alpha_3)-\nn\\
&-&(j+1)\psi(\alpha_3)-(3-j)\psi(1-\alpha_3)+2\psi(1)+
\delta_{j,even}i\pi~)\nn~~.\\
\eea}\noindent with $j=0~...~3$ and:
{\footnotesize
\bea
D&=&{\rm diag}(d_1,d_2,d_3,d_4)~~\nn\\
d_k&=&\Gamma(\alpha_k)^4\prod'_{l=1~...~4}\Gamma(\alpha_l-\alpha_k)~
\mbox{~~~for~~}k=1,2~~\nn\\
d_3&=&d_4~~=~~\Gamma(\alpha_3)^4\Gamma(\alpha_1-\alpha_3)
\Gamma(\alpha_2-\alpha_3)~~.\nn
\eea}\noindent These relations allow us to determine 
$M=\frac{1}{\Gamma(\alpha_1)\Gamma(\alpha_2)\Gamma(\alpha_3)^2}
CDq_J(\infty)^{-t}$ and hence 
the Meijer monodromy $T[\infty]=MT_J[\infty]M^{-1}$. Moreover, it is possible 
to show that the matrices $D$ and $q_J(\infty)^{-t}T_J[\infty]q_J(\infty)^t$ 
commute, so that we can use $N:=Cq_J(\infty)^{-t}$ instead of $M$. 
We conclude that the Meijer monodromy about the small radius point is given by:
\be
T[\infty]=NT_J[\infty]N^{-1}~~.
\ee

The reader will easily recognize that the procedure employed above is a 
natural generalization of the approach used in Section 3 for extracting the 
Meijer monodromy around $z=0$. In that case, the canonical matrix $R_{can}[0]$
was already in Jordan form, so that the matrix $P$ was simply the identity
(while $q_J(0)$ was denoted there by $q'(0)$). 
The procedure presented above can be easily generalized for an arbitrary 
Jordan form, thus allowing for a uniform treatment of all possible types of 
boundary points of the moduli space.

\subsection{The model $\P^5[2,4]$}

The mirror of this model can be described by an orbifold of 
the complete intersection
$Y=\{p_1(u)=p_2(u)=0\}$, where $u=[u_0~...~u_5] \in \P^5$ and:
\bea
p_1(u)=u_0^2+u_1^2+u_2^2+u_3^2-2\psi u_4 u_5~~\\
p_2(u)=u_4^4+u_5^4-4\psi u_0 u_1 u_2 u_3~~.
\eea
The fundamental period of this example was determined in 
\cite{Candelas_periods} (see also \cite{Teitelbaum} for other  
results on this model). 
By using our techniques, we can go beyond these results  
and determine a full basis of periods. We will also find a (weakly) integral 
period vanishing at $x=1$, which can once again be interpreted as a massless 
$6$-brane state.

The complex variable $\psi$ gives a coordinate on a $6$-fold
cover of the moduli space, which is a copy of the Riemann sphere $\P^1$.
As before, we prefer to work with the `hypergeometric' parameter 
$x=\frac{4}{\psi^6}$, which gives a coordinate on the moduli space itself
\footnote{In this subsection we denote the hypergeometric coordinate by 
$x$ instead of $z$, for agreement with Section 7 below.}. 
The normalization of $x$ is fixed by the requirement that $x=0$ corresponds 
to the large complex structure point and $x=1$ corresponds to the `conifold'
point. Then $x=\infty$ corresponds to the small radius limit of the model.
As mentioned above, logarithmic behaviour of the periods for 
$|x|>1$ prevents us from interpreting $x=\infty$  as Landau-Ginzburg point;
rather, it is a more general type of non-geometric point (in the terminology 
of last paper in reference \cite{topchange}, it is a hybrid point).

The hypergeometric symbol of this model is easily computed by the techniques 
of \cite{batyrev_straten,GKZ} 
with the result that it is given by 
{\footnotesize 
$\HG_symbol[{\alpha_1,\alpha_2,\alpha_3,\alpha_4},{1,1,1}]$} with
$\alpha_1=1/4,\alpha_2=3/4,\alpha_3=\alpha_4=1/2$. 
Hence the model fits into the scheme of the 
previous subsection and we can directly apply the results derived above by 
simply substituting these particular values of $\alpha_i$ in relations
(\ref{LCS_expansions}) and (\ref{deg_LG_expansion}), (\ref{deg_LG_expansion1}),
(\ref{deg_LG_expansion2})\footnote{The reader can easily verify
that the fundamental period {\scriptsize $U_0= 
\F[4,3][{1/2,1/2,1/4,3/4},{1,1,1}](x)$} 
coincides with the period $\omega_0=\sum_{n=0}^{\infty}
{\frac{(2n)!(4n)!}{(n!)^6(2^8\psi^6)^n}}$ given in Table 3.1 of 
\cite{Candelas_periods}. This follows from the identity  
$\frac{(2n)!(4n)!}{(n!)^6}=2^{10n}\frac{(1/2)_n^2(1/4)_n(3/4)_n}{(n!)^4}$.}
.

Computation of the Meijer monodromies proceeds as explained above. 
The monodromy around $x=0$ has the same form as in Section 3, while for 
the point $x=\infty$ we have:
{\footnotesize \be
R_{can}[\infty]=\left 
[\begin {array}{cccc} 0&-1&0&0\\0&0&-1&0
\\0&0&0&-1\\{\frac {3}{64}}&-{
\frac {7}{16}}&{\frac {23}{16}}&-2\end {array}\right ]~~,~~
R_J[\infty]=\left [\begin {array}{cccc} -1/4&0&0&0\\0&-3/4&0&0
\\0&0&-1/2&1\\0&0&0&-1/2
\end {array}\right]~~.
\ee}
\noindent A choice for the matrix $P$ is:
{\footnotesize \be
P=\left [\begin {array}{cccc} 6&-2&-3/2&-3\\3/2&-3/2&-
3/4&0\\3/8&-{\frac {9}{8}}&-3/8&3/4
\\{\frac {3}{32}}&-{\frac {27}{32}}&-3/16&3/4
\end {array}\right ]~~.
\ee}\noindent We can now compute:
{\footnotesize \be
N=\left [\begin {array}{cccc} {\frac {1/24-1/24i}{{\pi }^{3}}
}&{\frac {1/8+1/8i}{{\pi }^{3}}}&2/3{\frac {2i
+2i{\rm log} (2)-\pi }{{\pi }^{3}}}&2/3{\frac {i}{{
\pi }^{3}}}\\1/12{\pi }^{-2}&-1/4{\pi }^{-2}&-4/
3{\frac {1+{\rm log} (2)}{{\pi }^{2}}}&-2/3{\pi }^{-2}
\\{\frac {1/12-1/12i}{\pi }}&{\frac {1/4+
1/4i}{\pi }}&2/3{\frac {2i+2i{\rm log} (2
)-\pi }{\pi }}&2/3{\frac {i}{\pi }}\\1/6&
-1/2&-4/3-4/3{\rm log} (2)&-2/3\end {array}\right ]~~,
\ee}\noindent thus obtaining the Meijer monodromies:
{\footnotesize
\bea
T[0]=\left [\begin {array}{cccc} 1&0&0&0\\-2i
\pi &1&0&0\\-4{\pi }^{2}&-2i\pi &1&0
\\0&0&-2i\pi &1\end {array}\right ]~,~
T[\infty]=\left [\begin {array}{cccc} -5&-3{\frac {i}{\pi }}&
2{\pi 
}^{-2}&{\frac {i}{{\pi }^{3}}}\\-2i\pi &1&0&0\\-4{\pi }^{2}&-2i\pi &1&0
\\0&0&-2i\pi &1\end {array}\right ]~~,~~~~~~\nn
\eea}\noindent and $T[1]=T[0]^{-1}T[\infty]$. The monodromy matrices satisfy:
\bea
(T[0]-I)^4=0~~,~~(T[1]-I)^2=0~~,~~(T[\infty]^4-I)^2=0~~. \nn
\eea
\noindent Hence the monodromy about $x=\infty$ is neither 
unipotent, nor of finite order. This confirms our expectation that the small 
radius limit of the model has a `hybrid' character (i.e. is not a  
Landau-Ginzburg orbifold). 

A set of periods associated with a basis of a  full sublattice of 
$H_3(Y,\Z)$ (up to a {\em common} factor) can be obtained by the procedure 
discussed above:
{\footnotesize \be
U_E(x)=EU(x)~~\mbox{~, with~~} 
E=\left[\begin {array}{cccc} 1&0&0&0\\
-5&-3\frac {i}{\pi}&2\frac{1}{\pi^2}&\frac {i}{\pi^3}\\
11&8{\frac {i}{\pi }}&-\frac{6}{\pi^2}&-
4\frac {i}{\pi^3}\\-15&-13{\frac {i}{\pi }}&
\frac{8}{\pi^2}&7\frac {i}{\pi^3}\end {array}\right]~~.
\ee}
Moreover, it is not hard to check that the
period $U_v=\frac{1}{\pi^3}[U_3-3\pi^2 U_1]$ 
vanishes at $x=1$; it is also easy to see 
that $U_v$ is weakly integral: 
\be
U_v(x)=-i[3,2,2,1]~U_E(x)~~.
\ee
We will see evidence of a different kind for the integrality of this period 
in Section 6. Collapse of the associated cycle once again 
rests on two arithmetic identities, of which we mention only the one 
induced by using the expansion of the periods for $|x|>1$:
{\scriptsize
\bea
\label{5_24id}
\sum _{n=0}^{\infty }{4^{n+1}\sqrt{2}~\frac 
{\Gamma(n+\frac{1}{2})^4
\Gamma (-2n-\frac{1}{2})}{n!^2}
\left(\psi(-n-\frac{1}{4})+\psi(-n+\frac{1}{4})-\psi(n+\frac{1}{2})-
3\psi(1/2-n)+2\psi(n+1)\right)}~~+~~~~~~~~~~~~~~~~~~~~~~~~~~~~\nn\\
\frac{1}{2}\Gamma(1/4)^{6}
~\F[4,3][{1/4,1/4,1/4,1/4},{1/2,3/4,3/4}](1)-
16\Gamma(3/4)^6
~\F[4,3][{3/4,3/4,3/4,3/4},{5/4,5/4,3/2}](1)~~~~~~~~~~~~~~~~~~~~~~~~~~ =0~~~~~~~~~~~~~~~~~~~~~~~~~~~~~.
\eea}\noindent 
Figure 8 plots the values of the vanishing period versus the coordinate 
$s=-\frac{1}{2\pi}{\rm log}(\kappa x)$ which defines the algebraic measure on 
the (uncomplexified) Kahler moduli space. In this case 
$\kappa=e^{\sum_{k=1~...~4}{\psi(\alpha_k)}-4\psi(1)}=2^{-10}$. For comparison, 
we also 
display the values of the weakly integral periods $\frac{2}{\pi^2}U_2$ and 
$-\frac{4i}{\pi}U_1$. Up to a common factor, these 
periods correspond to the the integral of $\Omega$ over cycles mirror to a 
$6,4$ and $2$-cycle respectively.

\

\

\

\hskip 1.4in\scalebox{0.4}{\input{5_24.pstex_t}}

\hskip 1in \begin{center}Figure 8. {\footnotesize  
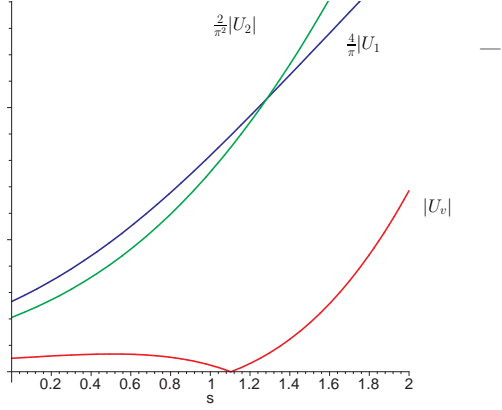
Graph of $|U_v|$ versus the imaginary part $s$ of the algebraic coordinate 
$t_{alg}=\frac{1}{2\pi i}{\rm log}(\kappa x)$ for $s \in [0,2]$. 
The point $x=1$ corresponds to $s=\frac{5{\rm log} 2}{\pi}
\approx 1.103$. We also display the weakly 
integral periods $-\frac{4i}{\pi}U_1$ and $\frac{2}{\pi^2}U_2$.}
\end{center}

\

\subsection{The model $\P^6[2,2,3]$}

This mirror of this 
model is realized as an orbifold of 
the complete intersection $Y=\{p_1=p_2=p_3=0\}$ of two quartics and a cubic
in $\P^6$:
\bea
p_1=x_1^2+x_2^2+x_3^2-2\psi x_6 x_7~~\nn\\
p_2=x_4^2+x_5^2-2\psi x_1 x_2~~\\
p_3=x_6^3+x_7^3-3\psi x_3 x_4 x_5\nn~~.
\eea
\noindent The hypergeometric coordinate on the moduli space is given by 
$z=\frac{3}{2\psi^7} $, while the hypergeometric symbol is 
{\footnotesize 
$\HG_symbol[{1/2,1/2,1/3,2/3},{1,1,1}]$}. 
The large and small radius expansions 
are obtained from  (\ref{LCS_expansions}) and 
(\ref{deg_LG_expansion}), (\ref{deg_LG_expansion1}),
(\ref{deg_LG_expansion2}) by substituting $\alpha_1=1/3,\alpha_2=2/3$ and 
$\alpha_3=\alpha_4=1/2$. 

Computation of the Meijer monodromies proceeds as above, so we only list 
the results: 
{\tiny
\bea
R_{can}[\infty]&=&
\left [\begin {array}{cccc} 0&-1&0&0\\0&0&-1&0
\\0&0&0&-1\\1/18&-{\frac {17}{36}}
&{\frac {53}{36}}&-2\end {array}\right]~,~
R_J[\infty]=
\left [\begin {array}{cccc} -2/3&0&0&0\\0&-1/3&0&0
\\0&0&-1/2&1\\0&0&0&-1/2
\end {array}\right ]~~~~~~~~~~~~~~~~~~~~~~~~~~~~~~~~~~~~~~~\nn\\
P&=&\left [\begin {array}{cccc} -9&18&-4&-8\\-6&6&-2&0
\\-4&2&-1&2\\-8/3&2/3&-1/2&2
\end {array}\right]~,~
N=\left [\begin {array}{cccc} 1/48{\frac {\sqrt {3}+3i}{{
\pi }^{3}}}&-{\frac {1}{96}}{\frac {-\sqrt {3}+3i}{{\pi }
^{3}}}&-1/4{\frac {\pi -4i-4i{\rm log} (2)+3i{\rm log} (3)}{{\pi }^{3}}}
&1/4{\frac {i}{{\pi }^{3}}}
\\-1/12{\pi }^{-2}&1/24{\pi }^{-2}&1/4{\frac {
-4-4{\rm log} (2)+3{\rm log} (3)}{{\pi }^{2}}}&-1/4{\pi }^{-2}
\\1/36{\frac {\sqrt {3}+3i}{\pi }}&-{
\frac {1}{72}}{\frac {-\sqrt {3}+3i}{\pi }}&-1/4{\frac 
{\pi -4i-4i{\rm log} (2)+3i{\rm log} (3)}{\pi }}&
1/4{\frac {i}{\pi }}\\-1/9&1/18&-1-{\rm log} (2
)+3/4{\rm log} (3)&-1/4\end {array}\right ]~~~~~~~~~~~~~~~~~~~~~~~~~~~~~~~~~~~~~~~~~~~\nn\\
T[0]&=&\left [\begin {array}{cccc} 1&0&0&0\\-2i
\pi &1&0&0\\-4{\pi }^{2}&-2i\pi &1&0
\\0&0&-2i\pi &1\end {array}\right ]~,~
T[\infty]=\left [\begin {array}{cccc} -6&-7/2{\frac {i}{\pi }}&3{
\pi }^{-2}&3/2{\frac {i}{{\pi }^{3}}}\\-2
i\pi &1&0&0\\-4{\pi }^{2}&-2i\pi &1&0\\0&0&-2i\pi &1\end {array}
\right]~~,\nn
\eea}\noindent and $T[1]=T[0]^{-1}T[\infty]$. The monodromy matrices satisfy:
\bea
(T[0]-I)^4=0~~,~~(T[1]-I)^2=0~~,~~(T[\infty]^6-I)^2=0~~. \nn
\eea
\noindent A set of periods associated with a basis of a  full sublattice of 
$H_3(Y,\Z)$ is (up to a common factor):
{\footnotesize 
\be
U_E(z)=EU(z)~~\mbox{~, with~~} 
E=\left [\begin {array}{cccc} 1&0&0&0\\-6&-7/2\frac {i}{\pi}&
\frac{3}{\pi^2}&3/2\frac {i}{\pi^3}\\
17&23/2\frac {i}{\pi}&-\frac{12}{\pi^2}&-15/2\frac {i}{\pi^3}
\\-31&-24\frac {i}{\pi}&\frac{24}{\pi^2}&
18\frac {i}{\pi^3}\end {array}\right]~~.
\ee}\noindent 
One can check that the period $U_v=\frac{1}{\pi^3}(3 U_3-7\pi^2 U_1)$, 
vanishes at $z=1$. This can be expressed as:
\be
U_v=-2i~[4,4,3,1]~U_E~~,
\ee
which shows that $U_v$ is weakly integral. 
The behaviour of this model is 
qualitatively very similar to that of the previous example, as we illustrate 
by drawing the graph of $U_v$ versus $s=-\frac{1}{2\pi}{\rm log}(\kappa z)$
with 
$\kappa=e^{\sum_{k=1~...~4}{\psi(\alpha_k)}-4\psi(1)}=
\frac{1}{432}=2^{-4}3^{-3}$.

\vskip 1.0 in
\hskip 1.4in\scalebox{0.4}{\begin{picture}(0,0)%
\epsfbox{6_223.pstex}%
\end{picture}%
\setlength{\unitlength}{3947sp}%
\begingroup\makeatletter\ifx\SetFigFont\undefined%
\gdef\SetFigFont#1#2#3#4#5{%
  \reset@font\fontsize{#1}{#2pt}%
  \fontfamily{#3}\fontseries{#4}\fontshape{#5}%
  \selectfont}%
\fi\endgroup%
\begin{picture}(9024,6774)(1189,-6598)
\end{picture}
}
\begin{center} 
Figure 9. {\footnotesize  
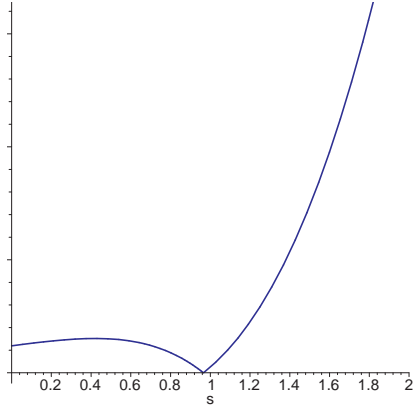
Graph of $|U_v|$ versus $s$ for $s \in [0,2]$. 
The point $z=1$ corresponds to $s=\frac{1}{2\pi}
(4 {\rm log} 2+3{\rm log}3)\approx 0.9658$.}
\end{center}

\subsection{A hypergeometric hierarchy}

Before leaving the subject of compact one-parameter models, let us make 
a few methodological remarks. As illustrated above, starting with the `generic'
family of Section 3 and taking the limit where some of of the parameters 
coincide produces new families of qualitatively different models. 
In this section, we studied only the first layer of such `degenerate' 
models, namely the case where two of the parameters $\alpha$ coincide. 
The procedure clearly generalizes, leading to a hierarchy of models 
on five levels, which are characterized (up to permutations of the parameters)
by the conditions:

(0) all $\alpha_i$ are distinct

(1) three of the parameters $\alpha_i$ are distinct

(2) $\alpha_1=\alpha_2$ and $\alpha_3=\alpha_4$ but $\alpha_1\neq \alpha_3$

(3) $\alpha_1=\alpha_3=\alpha_3=\alpha_4$.

(4) $\alpha_1=\alpha_2=\alpha_3\neq \alpha_4$

For example, it 
is easy to see that all of the complete intersection models in projective 
spaces investigated in \cite{Candelas_periods} fit into one of these classes
\footnote{Class (4) is not realized through compact one-parameter complete 
intersections in toric varieties, though it could be realized through 
different constructions.}.
In this section, we considered the two models of \cite{Candelas_periods} 
which are of type (1), but it should be clear that a very similar approach 
can be followed for the other classes. In fact, one of the 
major advantages of using Meijer functions is that it allows for 
a very systematic treatment of entire families of models, as we have 
illustrated in some detail above. In the next section, we study 
a {\em non-compact} one-parameter model, which displays some new features.
As we will show, our techniques are well adapted for this type of model as 
well.

\section{Examples III: An orbifold example}

In this section we consider a {\em non-compact} one-parameter 
example, namely the $\C^3/\Z_3$ orbifold. While non-compact and hence 
slightly unphysical, global orbifolds can be realized as local singularities 
of Calabi-Yau spaces, or as decompactification limits of singular Calabi-Yau 
varieties. In particular, the orbifold we consider can be realized as a limit 
of the 5-parameter model studied in \cite{small_distances1}. 
For an overview of toric and mirror-symmetry 
techniques in the study of the moduli space of conformal field theories 
associated with partial resolutions of orbifolds we refer the reader to 
\cite{aspinwall_orbifolds}.

\subsection{Classical geometry of the orbifold}

We begin by reviewing the classical geometry of the model. Our orbifold 
$X_0$ can be realized as the quotient of $\C^3$ by the action:
\be
(z_1,z_2,z_3)\rightarrow (\zeta z_1,\zeta z_2,\zeta z_3)~~
\ee
where $\zeta=e^{\frac{2\pi i}{3}}$. 
This has a straightforward toric description as the quotient:
\be
X_0=\C^4/\C^*~~
\ee
of $\C^4$ by the torus action:
\be
(x_0,x_1,x_2,x_3)\rightarrow (\lambda^{-3}x_0,\lambda x_1,\lambda x_2,
\lambda x_3)~~,
\ee
where $\lambda \in \C^*$, which corresponds to the fan in $\R^3$ consisting 
of a single simplicial cone with generators:
\bea
\nu_1&=&(3,-1,-1)\\
\nu_2&=&(0,1,0)\\
\nu_3&=&(0,0,1)~~.
\eea
Note that the interior of this cone contains the integral vector:
\be
\nu_0=(1,0,0)~~.
\ee

The resolution $X$ of $X_0$ can be achieved by a toric blow-up, which 
amounts to replacing $X_0$ by:
\be
X=(\C^4-Z)/\C^*~~,
\ee
with the exceptional set $Z=\{x\in \C^4|x_1,x_2,x_3=0\}$. This corresponds 
to adding $\nu_0$ to our list of toric generators and taking the fan to 
consist of the $3$ cones spanned by $(\nu_0,\nu_i,\nu_j)$ 
($1\leq i\leq j\leq 3$), thereby performing a star subdivision of the original
cone (see Figure 10). Geometrically, this amounts to blowing-up the origin
in the space $\C^3/\Z_3$, thus producing an exceptional divisor
$D_0$ associated with the generator $\nu_0$. The resolved space contains $4$ 
toric divisors $D_\rho$, associated with the four toric generators $\nu_\rho$ 
($\rho=0~...~3$).

\hskip 1.2in\scalebox{0.4}{\input{ orbf_fan.pstex_t}}

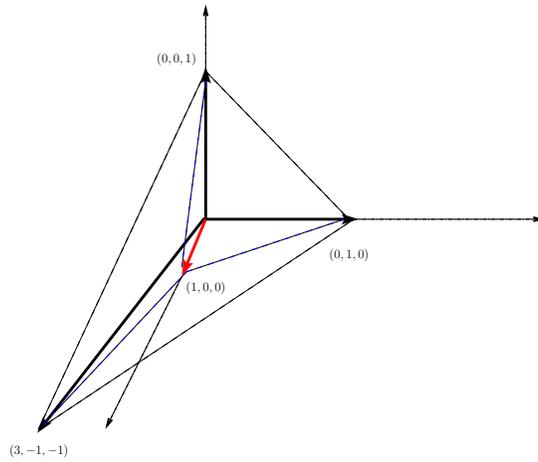
\hskip 1in \begin{center}Figure 10. {\footnotesize Fan for the $\C^3/\Z_3$ 
orbifold.}
\end{center}

\

\

The group ${\rm Div}(X)\approx {\rm Pic}(X)$ of divisor classes on 
$X$ modulo linear equivalence 
is easy to compute by the general technology of \cite{toric} or by simple 
manipulations with the associated linear systems ${\cal O}(D_\rho)$, with the 
result that it is a free group generated by either of $D_1$, $D_2$ or $D_3$.
In particular, we have the relation:
\be
D_0=-3D_1=-3D_2=-3D_3~~,
\ee
where we denote the linear equivalence class of a divisor $D$ by the same 
letter. This corresponds to the fact that 
${\cal O}(-D_0)\approx {\cal O}(D_1)^3\approx {\cal O}(D_2)^3\approx 
{\cal O}(D_3)^3$, as can be seen by considering the charges of $x_\rho$.

It follows that our model has a 1-dimensional K\"ahler moduli space, which 
can be parameterized by writing:
\be
k=(B_0+iJ_0)e_0~~,
\ee
where $k$ is the complexified K\"ahler class, 
and $e_0$ is the Poincare dual of $D_0$.
In particular, the classical volume of $D_0$ 
is proportional to $-\int_{D_0}J^2=27J_0^2$, where we used the fact that 
$D_1^3=1$. Hence the orbifold point corresponds classically 
to $J_0\rightarrow 0$, in which limit the volume of $D_0$ becomes zero. 

\subsection{The quantum moduli space and massless $D$-branes}

As discussed for example in \cite{aspinwall_orbifolds}, the moduli space 
can be obtained without the need of an explicit construction of a mirror
family. For this, it suffices to view the star-subdivision considered above 
as a perestroika associated with changing the phase in a mirror 
secondary fan, which in this case consist of two opposing semi-axes on the 
real line. It follows that the model has two phases, associated with two 
limits which correspond to blowing up the singularity until the exceptional 
divisor acquires very large area (the large radius point), respectively to 
blowing down the divisor to zero area (the deep interior of the orbifold 
phase). Hence the compactified moduli space can be identified 
with a copy of the Riemann sphere. Analyzing the associated Picard-Fuchs 
equation (see next subsection), which we write in terms of a convenient 
coordinate $z$ gives two regular singular points at $z=0,\infty$, which 
with our parameterization correspond to the 
large radius and deep orbifold points, respectively, as well as a regular 
singular point at $z=1$, which can be thought of as defining the `boundary' 
between the two regions 
(in sense which is explained precisely in \cite{small_distances1,topchange}). 
The special coordinate $t=B_0+iJ_0$ on the 
quantum-corrected K\"ahler 
moduli space is then given by the mirror map as a function 
$t=t(z)$ which can be determined from the solutions of the Picard-Fuchs 
equation.

In the limit $z\rightarrow \infty$, the conformal field theory 
is an orbifold theory and hence exactly solvable.
Based on the classical picture, one may expect that the D2-brane wrapping 
a cycle in the class $e_1$, as well as the D4-brane wrapping the exceptional 
divisor would become massless at the orbifold point. As we discuss 
below, this picture is drastically modified by quantum effects. 
First, based on the computations of ${\rm log}^0$ and ${\rm log}^1$-monodromy
periods already performed in \cite{small_distances1} 
(and which will be subsumed by our more detailed results below), one can 
immediately see that the value of $J_0$ at the orbifold point is zero, but the 
value of $B_0$ equals $-1/2$ (in units where $2\pi \alpha'=1$), so that a 
$D2$-brane wrapping a rational curve in $X$ has nonzero mass at $z=\infty$.
This matches the fact that the conformal field theory is perfectly 
well-behaved there; an analogous observation was subsequently made
in \cite{asp_2} in the context of K3 compactifications to explain the physical
smoothness of orbifold points. On the other hand, we will argue that there 
exists a $D4$-brane which becomes massless at the point $z=1$, though no 
$D2$-brane acquires vanishing mass there. Following our earlier
discussion, we are unable to uniquely fix the lower brane charge
on this D4-brane, so it might well be interpretable as a (D0,D4)-bound state.

\subsection{Quantum volumes}

\subsubsection{The hypergeometric equation and the Meijer periods}

The periods of this model satisfy the hypergeometric equation:
\be
\label{HG_eq_orb}
\left[\delta^3-z \delta(\delta+1/3)(\delta+2/3)\right]u=0~,
\ee
which corresponds to the hypergeometric symbol 
{\footnotesize $\HG_symbol[{0,1/3,2/3},{1,1}]$}. 
The Meijer periods can be obtained by the procedure discussed in Section 2
\footnote{
The reader may worry that our expression for $U_0(z)$ is identically zero 
due to presence of the singular quantity   
$\Gamma(0)$ in the prefactor. However, this is a removable singularity. 
One can take $U_0(z)$ to be defined as a 
limit in which $\Gamma(0)$ in the denominator of 
{\scriptsize $\G[{1,1},{0,1/3,2/3}]$} 
and $\Gamma(-s),~\Gamma(s)$ in the numerator of the integrand are replaced by 
$\Gamma(\epsilon)$ and $\Gamma(-s+\frac{\epsilon}{2}),~
\Gamma(s+\frac{\epsilon}{2})$ respectively, 
with a small positive $\epsilon$ which is
taken to zero after performing the integral. 
When closing the contour to right or left,
the factor $\Gamma(\epsilon)$ in the denominator will force us to keep only 
the residue corresponding to the pole at $s=0$. 
Indeed, we have $\Gamma(\epsilon)\approx \frac{1}{\epsilon}$ and 
the quantity $\epsilon~{\rm Res}_{s_0}[\Gamma(-s+\frac{\epsilon}{2})
\Gamma(s+\frac{\epsilon}{2})]$ (with $s_0$ a
pole of type (A) or (B) of the regularized integrand) 
has a nonzero limit as $\epsilon$ tends to zero only if 
$s_0=\frac{\epsilon}{2}$ or $s_0=-\frac{\epsilon}{2}$; this 
limit is  $\pm 1$, which forces the limiting value of the regularized 
integral to be identically $1$
(remembering that the contours about $\pm \infty$ have opposite 
orientations).
One can of course see directly that $U_0=1$ is a solution of the 
hypergeometric equation, but the argument just presented shows how this 
example fits into the general theory of Section 2.
The solution $U_1$ is obtained from $U_0$ by performing {\em two}
diagonal operations on the Meijer symbol. Lifting the parameter $1$ along the 
second diagonal is the usual procedure for trying to make the 
$(A)$-type poles of the integrand 
become double, which assures that the solution thus produced has 
$\log^2(z)$ monodormy about $z=0$. A further operation is performed along 
the first diagonal, by lowering the parameter $0$. This replaces 
the factor $\Gamma(s)$ 
in the numerator by a factor
$\Gamma(1-s)=-s\Gamma(-s)$ in the denominator, which cancels a
$\Gamma(-s)$ introduced in the numerator by the first operation and leaves 
a factor of $-\frac{\Gamma(-s)}{s}$; 
hence all $(A)$-type poles 
remain simple except for the pole at $s=0$.  
We also replace the prefactor
{\scriptsize $\G[{1,1},{0,1/3,2/3}]$} by  
{\scriptsize $-\G[{1,1},{1/3,2/3}]$} in order to avoid problems from  
$\Gamma(0)$. Such a rescaling is always allowed  
since the choice of normalization for our solutions is arbitrary. 
This slightly nonstandard behaviour is
due to non-compactness of the model.}: 
{\footnotesize \bea
\label{Meijer_periods_3}
U_0(z)&=&\G[{1,1},{0,1/3,2/3}]
\I[{0},{0,1/3,2/3} ,{\cdot},{1,1}](-z)=1~~,\\
U_1(z)&=&-\G[{1,1},{~1/3,2/3~}]
\I[{0~,~0},{1/3,2/3} ,{1},{1}](-z)~~,\\
U_2(z)&=&-\G[{1,1},{1/3,2/3}]
\I[{0,0,0},{~1/3,2/3~} ,{1},{\cdot}](+z)~~.
\eea}
\noindent For clarity, let us write down the integral representations for 
the nontrivial Meijer periods:
\bea
U_1(z)&=&\frac{1}{2\pi i\Gamma(1/3)\Gamma(2/3)}\int_{\gamma}{ds~
\frac{\Gamma(s+1/3)\Gamma(s+2/3)\Gamma(-s)}{s\Gamma(s+1)}~(-z)^s}~~\\
U_2(z)&=&\frac{1}{2\pi i\Gamma(1/3)\Gamma(2/3)}\int_{\gamma}{ds~
\frac{\Gamma(s+1/3)\Gamma(s+2/3){\Gamma(-s)}^2}{s}~z^s}~~.
\eea
\noindent The reader can verify directly 
that $U_j$ satisfy equation (\ref{HG_eq_orb}). 

The expansions of the periods can be obtained as before. Instead of 
presenting details of the calculation, let us point out that closing the
contour to the right (which is possible for $|z|<1$) leads to $j$--th order  
poles at $s=n$ ($n$ a strictly positive integer) from the factors 
of $\Gamma(-s)^j$ in the integrand and to a $j+1$ --th order pole at $s=0$
from $\frac{\Gamma(-s)^j}{s}$. Computing the associated residues leads to
the expansions for $|z|<1$:
\bea
U_1(z)={\rm log}(-z)+\left[\psi(1/3)+\psi(2/3)-2\psi(1) \right]+
\sum_{n=1}^{\infty}{\frac{(1/3)_n(2/3)_n}{n!^2n}z^n}~~~~~~~~~~~~~~~~~~~~\nn\\
U_2(z)=-\frac{1}{2}\left[
(\psi'(1/3)+\psi'(2/3)+2\psi'(1))+(\psi(1/3)+\psi(2/3)-2\psi(1)+
{\rm log}z)^2\right]-\nn\\
\sum_{n=1}^{\infty}{(~{\rm log}(z)+\psi(n+1/3)+\psi(n+2/3)-2\psi(1))-
\frac{3}{n}-
2\sum_{k=1}^{n-1}{\frac{1}{k}}~)~\frac{(1/3)_n(2/3)_n}{n!^2n}z^n}~~.\nn
\eea
\noindent The expression for $U_1(z)$ can be simplified by 
noticing that:
{\footnotesize \be
\sum_{n=1}^{\infty}\frac{(1/3)_n(2/3)_n}{n!^2n}z^n=\frac{2}{9}z~
\F[4,3][{4/3,5/3,1,1},{2,2,2}](z)
\ee}\noindent (to see this, one must first shift the summation variable 
according to 
$n\rightarrow n-1$ and then perform some basic manipulations on the 
Pochhammer symbols).
Moreover, applying the identities (\ref{identities1}) of Appendix B
for $n=3$ gives the relations:
\bea
\psi(1/3)+\psi(2/3)-2\psi(1)~~&=&-3{\rm log}(3)~~\\
\psi'(1/3)+\psi'(2/3)+2\psi'(1)&=&10\psi'(1)=\frac{5\pi^2}{3}~~.
\eea
which allow one to further simplify $U_1(z)$ and $U_2(z)$.

The expansions of the Meijer periods for $|z|>1$ can be obtained similarly. 
In this case, one can close the contour to the left, obtaining a Cauchy 
expansion over the simple poles $s=-n-1/3$ and $s=-n-2/3$, with $n$ a 
nonnegative integer. The resulting residues can be simplified by making use 
of the completion formula and the identity:
\be
(x+1)_n=(x)_n\frac{x+n}{x}
\ee 
which allows for the replacement of $\frac{1}{n+x}$ with 
$\frac{1}{x}\frac{(x)_n}{(x+1)_n}$. In this way, one obtains:
{\scriptsize 
\bea
\label{orbf_LG_expansions}
U_1(z)&=&-3\frac{\Gamma(1/3)}{\Gamma(2/3)^2}(-z)^{-1/3}~
\F[3,2][{1/3,1/3,1/3},{2/3,4/3}](1/z)+
\frac{3}{2}\frac{\Gamma(2/3)}{\Gamma(1/3)\Gamma(4/3)}(-z)^{-2/3}~
\F[3,2][{2/3,2/3,2/3},{4/3,5/3}](1/z)~~\nn\\
U_2(z)&=&-3\frac{\Gamma(1/3)^2}{\Gamma(2/3)}z^{-1/3}~
\F[3,2][{1/3,1/3,1/3},{2/3,4/3}](1/z)+
\frac{3}{2}\frac{\Gamma(2/3)^2}{\Gamma(4/3)}z^{-2/3}~
\F[3,2][{2/3,2/3,2/3},{4/3,5/3}](1/z)
\eea
}

\subsubsection{Monodromies of the Meijer basis}

The monodromies of the Meijer basis follow from arguments similar to those 
presented in Section 3. In our case, the matrix ${\tilde q}(0)$ is:
{\footnotesize \be
{\tilde q}(0):=
\left[\begin{array}{ccc}
1&i\pi&-\frac{5\pi^2}{6}\\
0&1&0\\
0&0&-\frac{1}{2}
\end{array}\right]~~,
\ee}\noindent 
where we used $\psi'(1/3)+\psi'(2/3)+2\psi'(1)=\frac{5\pi^2}{3}$ and the 
expansions derived above for 
$|z|<1$.  
The monomial-divisor mirror map gives the coordinate $w=\frac{z}{27}$ 
(in this example $\kappa=\psi(1/3)+\psi(2/3)-2\psi(1)=3^{-3}$), 
while the matrix 
$q'(0)$ has the form $q'(0):={\rm diag}(1,1,1/2)$. This information 
allows us to compute ${\tilde M}={\tilde q}(0)^tq'(0)^{-1}$. 
The matrix $R_{can}[0]$ is again in Jordan form:
{\footnotesize \be
R_{can}[0]=
\left[\begin{array}{ccc}
0&1&0\\
0&0&1\\
0&0&0
\end{array}\right]~~,
\ee}\noindent 
leading to the following expression for the 
Meijer monodromy around $z=0$:
{\footnotesize \be
T[0]:={\tilde M}T_u[0]{\tilde M}^{-1}=
\left [\begin {array}{ccc} 
1&0&0\\
2 i\pi &1&0\\
0&-2 i\pi &1
\end {array}\right ]
\ee}\noindent (here $T_u[0]=e^{2\pi i R_{can}[0]^t}$). 
The monodromy around $z=\infty$ follows 
easily from the expansions (\ref{orbf_LG_expansions}): 
{\footnotesize \be
T[\infty]=\left [\begin {array}{ccc} 
1&0&0\\
0&-2&-\frac{3}{2}\frac {i}\pi\\
0&-2i\pi &1
\end {array}\right]~~.
\ee}\noindent 
Finally, the monodromy around $z=1$ is given by:
{\footnotesize \be
T[1]=T[0]^{-1}T[\infty]=
\left [\begin {array}{ccc} 1&0&0\\-2i\pi 
&-2&-\frac {3i}{2\pi}\\4\pi^2&
-6i\pi &4\end {array}\right ]~~.
\ee}\noindent 
These monodromy matrices satisfy:
\be
(T[0]-I)^3=0~~,~~
(T[1]-I)^2=0~~,~~
T[\infty]^3-I=0~~.
\ee

\subsubsection{Integral structure}

Following the general method described in Section 2, we now look for a 
cyclic period for this model. In this example we cannot use  
$U_0$ as a cyclic period, which is yet another effect of 
non-compactness. Indeed, $U_0$ is a constant function 
over the moduli space, which means that all monodromy operators
$T[0],~T[1]~,T[\infty]$ leave it invariant. In order to be able to 
carry out our program, it is thus necessary to identify another cyclic 
and weakly integral period. Here we present a possible approach to this 
problem, which makes use of the integrality conjecture of Aspinwall and 
Lutken \cite{int_conj}. It was argued in \cite{int_conj} 
that the cohomology class $\Omega\in H^3(Y,\C)$ should align with a 
vector of the lattice $H^3(Y,\Z)$ in the large complex structure 
limit. Taking this conjecture at face value
\footnote{In fact, the status of the arguments of
\cite{int_conj} is not completely clear to us,
for the following reason. In \cite{int_conj}, it is argued that the
Hodge subspaces $H^{3-p,p}(Y)$ $(p=0..3)$ should become rational subspaces
of $H^3(Y)$ (with respect to $H^3(Y,\Z)$) in the large complex structure limit.
On the other hand, the standard approach to mirror symmetry computations
does not actually use the Hodge subspaces as such but, rather, the subspaces
${\cal H}^{3-p,p}(Y)={\cal F}^{p}_Y\cap W^{3-p}$ where ${\cal F}$ is the
Hodge filtration and $W$ the monodromy weight filtration
(see \cite{morrison_aspects} for a clear explanation of this point). Doing so
is desirable because  ${\cal H}^{3-p,p}(Y)$ vary holomorphically
over the moduli space, while the true Hodge subspaces do not. This
makes the associated periods amenable for direct computation. 
Working with ${\cal H}^{3-p,p}(Y)$ is
permitted at the level of the {\em closed} nonlinear sigma model due to
arguments of \cite{BCOV} and \cite{msh} which show that working with elements
of ${\cal H}^{3-p,p}$ instead of $H^{3-p,p}$ does not affect the
closed conformal field theory correlators. However, the status of this claim
needs to be reexamined when one includes boundary states thereby
working at the level of open conformal field theory. While we have
some preliminary calculations that are relevant to this issue, we will
not consider it further here. We emphasize, though, that whereas the
reasoning of subsection 6.3.3 involves the conjecture of \cite{int_conj},
we will provide independent evidence
for integrality of the vanishing period $U_v$  below
(see equation (\ref{foo}) for an independent argument to that effect),
so that our conclusion that $U_v$ is weakly integral is
independent of the considerations of this subsection.}, it is possible to 
obtain a candidate period with the desired properties, as we now explain.

Consider the asymptotic form of the period vector in the 
large complex structure limit $t_{alg}\rightarrow i\infty$, where 
$t_{alg}=\frac{1}{2\pi i}{\rm log}(\frac{z}{27})$ 
is the algebraic coordinate on 
the moduli space:
{\footnotesize 
\be
U(s)\approx\left[\begin{array}{c}
0\\0\\2\pi^2
\end{array}\right]~
t_{alg}^2+O(t_{alg},e^{it_{alg}})~~.
\ee}\noindent 
According to \cite{int_conj}, the leading term in this equation should be 
proportional to a vector of the lattice $\Lambda=H_3(Y,\Z)$. 
It follows that the Meijer period $U_2$ should coincide with the period 
of $\Omega$ over an integral cycle, up a {\em constant} complex factor. 
Assuming this is the case, we can use that cycle in order to generate 
a sublattice of $\Lambda$. In fact, it is easy to check that 
$U_2$ is cyclic for the action of all monodromies. Hence
repeatedly acting on it with 
$T[0],T[1]$ and $T[\infty]$ produces a basis of periods which 
are associated (up to a {\em common} factor) with a basis of a full 
sublattice $\Lambda_0$ of $\Lambda$. One can take $U_E(z)=EU(z)$ as the 
period vector associated to such a basis, with the matrix $E$ given by:
{\footnotesize 
\be
E:=\left[\begin {array}{ccc} 
0&0&1\\
0&-2 i\pi &1\\
4\pi^2&-4i\pi &1
\end {array}\right]~~.
\ee}\noindent 
The monodromies of the basis $E$ are given by 
integral matrices, as they should:
{\footnotesize
\bea
T_E[0]=\left[\begin{array}{ccc}
0&1&0\\
0&0&1\\
1&-3&3
\end {array}\right]~,~
T_E[1]=\left [\begin {array}{ccc} 2&1&1\\0&1&0
\\-1&-1&0\end {array}\right ]~,~
T_E[\infty]=\left[\begin{array}{ccc}
0&1&0\\
-1&-1&0\\
-1&-5&1
\end{array}\right]~~.\nn
\eea}

As a further check, let us consider the period 
$U_\beta(z)=\frac{1}{2\pi i}U_1(z)-\frac{1}{2}$, which is associated with the 
vector:
\be
\beta:=\left[\begin{array}{ccc}
-\frac{1}{2}&\frac{1}{2\pi i}&0\end{array}\right]~~,
\ee
Since this period asymptotes to $U_\beta\approx s$ in the large complex 
structure limit, it is guaranteed to correspond to an integral cycle 
up to an unknown complex factor (see, for example, \cite{morrison_quintic}).
It is easy to express this period in the basis $U_E$, with the result
that $U_\beta=\beta_{int}U_E$, where:
{\footnotesize \be
\beta_{int}=\frac{1}{8\pi^2}
\left[\begin{array}{ccc}-3&4&-1\end{array}\right]~~,
\ee}
\noindent which is indeed an element of $\Lambda_0$ after rescaling by the 
factor $8\pi^2$.

\subsubsection{A vanishing period and some arithmetic identities}

Now consider the period 
$U_v:=\frac{1}{8\pi^2}(4\pi^2U_0+3U_2) $, which corresponds to the vector:
{\footnotesize \be
\gamma:=
\left[\begin{array}{ccc}\frac{1}{2}&0&\frac{3}{8\pi^2}\end{array}\right]~~.
\ee}
\noindent Then we have $U_v=\gamma_{int}U_E$, with:
{\footnotesize \be
\gamma_{int}=\frac{1}{8\pi^2}
\left[\begin{array}{ccc}4&-2&1\end{array}\right]~~.
\ee}
\noindent Hence $U_v$ is again a weakly integral period. This can be 
checked independently by acting with $T[\infty]$ and $T[\infty]^2$ on 
$U_\beta$, thus 
producing a weakly integral basis in which $U_v$ has rational coefficients:
\be
\label{foo}
U_v=-\frac{5}{6}U_\beta-\frac{1}{3}T_U[\infty]U_\beta+\frac{1}{6}T_U[\infty]^2
U_\beta~~.
\ee
We claim that the  period $U_v$ vanishes for $z=1$:
\be
U_v(1)=0 \Longleftrightarrow U_2(1)=-\frac{4\pi^2}{3}~~.
\ee
\noindent 
Using the expansions of $U_2(z)$ for $|z|>1$ and $|z|<1$, this is equivalent 
to the following slightly nontrivial arithmetic identities:
{\footnotesize \be
\label{orbf_id1}
-3~\F[3,2][{1/3,1/3,1/3},{2/3,4/3}](1)+\frac{27}{16\pi^3}
\sqrt{3}~\F[3,2][{2/3,2/3,2/3},{5/3,4/3}](1)~\Gamma(2/3)^6+
\Gamma (2/3)^3=0~~
\ee}\noindent and:
{\footnotesize 
\bea
\label{orbf_id2}
-\frac{1}{2}\left(\psi'(1/3)+\psi'(2/3)+\frac{1}{3}\pi^2
+9{\rm log}(3)^2\right)
-\sum_{n=1}^{\infty }{\frac{\left(\psi(n+\frac{1}{3})+\psi(n+\frac{2}{3})-
\frac{3}{n}-2\psi(n)\right)(1/3)_n(2/3)_n}{n!^2n}}\nn\\
=-\frac{4\pi^2}{3}~~.
\eea
}
We checked both identities numerically, but we lack an 
analytic proof. Note that, given one of the two identities above, 
analytic continuation automatically produces the other. Figure 11  
displays the values of $|U_v|$ as a function of the 
algebraic coordinate $s=-\frac{1}{2\pi}{\rm log}{\left(
\frac{z}{27}\right)}$. We also display the values of the special coordinate 
$t=U_\beta=\frac{1}{2\pi i}U_1(z)-1frac{1}{2}$, which measures the mass of a 
$D2$-brane. 

\vskip 0.8 in 
\hskip 1.4in\scalebox{0.3}{\begin{picture}(0,0)%
\epsfbox{orb_graph.pstex}%
\end{picture}%
\setlength{\unitlength}{3947sp}%
\begingroup\makeatletter\ifx\SetFigFont\undefined%
\gdef\SetFigFont#1#2#3#4#5{%
  \reset@font\fontsize{#1}{#2pt}%
  \fontfamily{#3}\fontseries{#4}\fontshape{#5}%
  \selectfont}%
\fi\endgroup%
\begin{picture}(10828,7845)(1168,-7723)
\put(7426,-361){\makebox(0,0)[lb]{\smash{\SetFigFont{20}{24.0}{\rmdefault}{\bfdefault}{\updefault}$|U_v|$}}}
\put(8251,-1111){\makebox(0,0)[lb]{\smash{\SetFigFont{20}{24.0}{\rmdefault}{\bfdefault}{\updefault}$|t|$}}}
\end{picture}
}

\begin{center} 
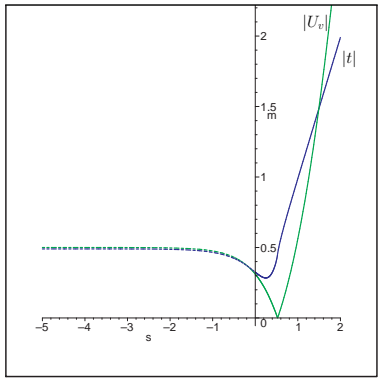
Figure 11. {\footnotesize  Graph of the vanishing period  
and special coordinate (in absolute value)
versus the real Kahler parameter $s=-\frac{1}{2\pi}{\rm log}{\left(
\frac{z}{27}\right)}$, for $z$ belonging to the real axis. 
The point $z=1$ corresponds to 
$s=\frac{3}{2\pi}{\rm log}3$. We also plot the values of $|U_v(s)|$ and $|t|$
for negative $s$. The region $s\rightarrow -\infty$ corresponds to 
the small radius limit, which has no classical analogue. In this limit, 
both $|t|$ and $|U_v|$ asymptote to the value $1/2$.}
\end{center}

\subsection{Interpretation}

Our results confirm the picture announced above. In particular, we obtained 
a vanishing period which has ${\rm log}^2$ monodromy around $z=0$ and thus 
can be 
interpreted as a D4-brane state in a $IIA$ compactification on $X$. This 
object becomes massless at $z=1$, and not at $z=0$ as one would have expected  
based on classical arguments. Moreover, when it becomes massless, the
D2-brane states remain massive, something quite at odds with classical
expectations.

Admittedly, our discussion of integral structure has been less 
than complete, essentially because we avoided working with an explicit 
construction of the mirror, which would necessitate a more careful 
discussion of the effects of non-compactness on the interpretation of 
standard mirror symmetry constructions.
In the next section, we consider a compact 
model in which the integral structure is fully understood. 
In order to make the discussion more realistic, we focus on a two-parameter 
example, some features of which were discussed in \cite{2pm1}.

\section{Examples IV: A two-parameter example}

In this section, we consider a special locus in the moduli space of a 
two-parameter example. We will show that there is a D4-brane state (possibly 
bound with a D0-brane) which becomes massless at every point on this locus. 
Comparison with classical geometry leads to some surprising conclusions.
Two-parameter examples were studied in the papers \cite{2pm1,2pm2,Yau}. 
We will focus on a model discussed in \cite{2pm1}, to which we refer the 
reader for background.

\subsection{The model and its mirror}

One of the two models studied in \cite{2pm1} corresponds to the 
K\"ahler moduli space of a degree $8$ hypersurface in the weighted projective 
space $WP^{1,1,2,2,2}$. This can be described as the space of partial 
resolutions of the Fermat hypersurface:
\be
X_0:~~x_1^8+x_2^8+x_3^4+x_4^4+x_5^4=0~~,
\ee
which has a curve $\Sigma$ of singularities of arithmetic genus $3$, 
given by:
\be
\Sigma:~~{x_1=x_2=0,~x_3^4+x_4^4+x_5^4=0}~~.
\ee
Partial resolutions $X$ are obtained by blowing up $X_0$ along $\Sigma$, which 
produces an exceptional divisor $E$ (a ruled surface over $\Sigma$). 
The Picard group of the resolved variety is generated by two linear systems
$L$ (a pencil of $K3$ surfaces) and $H$, in terms 
of which the (linear equivalence class of the) 
exceptional divisor can be expressed as:
\be
E=H-2L~~.
\ee
In order to describe the full cohomology ring, consider the two classes $l$ 
and $h$ in $H_2(X,\Z)$ defined by the fiber of the ruling of $E$, respectively 
by the intersection of $H$ with $L$, respectively. 
Then the relations defining the intersection ring, 
whose derivation can be found in \cite{2pm1}, are:
\bea
\begin{array}{ccccc}
L^2&=&0~~~~~~,~~~~~H^2&=&4l+8h~~\\
H\cdot L&=&4h~~~~~~,~~~~~H^3&=&8~~\\
~~~~~&~&H^2\cdot L=4~~~~~&~&
\end{array}~~.
\eea
In particular, we have:
\bea
L\cdot l&=&H\cdot h=1~~\\
H\cdot l&=&L\cdot h=0~~,
\eea
hence the bases $(h,l)$ of $H_2(X,\Z)$ and $(H,L)$ of $H_4(X,\Z)$ are dual 
with respect to the intersection form. The K\"ahler cone is then easily seen 
to be given by $J=\tau_1 {\check H}+\tau_2 {\check L}$ with 
$\alpha, \beta \geq 0$, where ${\check A}$ denotes the Poincare dual of a 
homology class $A$.

The mirror manifold $Y$ can be easily constructed via the methods of 
\cite{batyrev_construction} or \cite{greene_plesser}. In the second 
approach $Y$ can be described as the finite quotient ${\tilde Y}/G$
were ${\tilde Y}$ is the following hypersurface in $WP^{1,1,2,2,2}$:
\be
{\tilde Y}:~x_1^8+x_2^8+x_4^4+x_5^4-8\psi x_1 x_2 x_3 x_4 x_5 -2\phi x_1^4 
x_2^4~=~0~~,
\ee
and the group $G$ is a copy of $\Z_4^3$ acting in a way described in detail 
in \cite{2pm1}. The moduli space is the quotient $\C^ 2(\psi,\phi)/\Z_8$~~,
with the generator of $\Z_8$ acting by:
\be
(\psi,\phi)\rightarrow (\alpha\psi,-\phi)~~,
\ee
where $\alpha:=e^{\frac{2\pi i}{8}}$. Hence the parameters $(\psi,\phi)$ 
used in \cite{2pm1} give an eight-fold cover of the complex structure 
moduli space of $Y$. The discriminant locus has three components at 
finite-distance in the variables $(\psi,\phi)$, which following \cite{2pm1}
we denote as follows:

(1) $C_{con}:~(\phi+8\psi^4)^2=1$, where $Y$ acquires a conifold singularity

(2) $C_1:~\phi^2=1$, where $Y$ acquires an isolated singularity not of the 
conifold type

(3) $C_0:~\psi=0$, where the moduli space 
acquires orbifold singularities.

\noindent Beyond these, one must add at least the divisor:

(4) $C_{\infty}:~\psi,\phi\rightarrow \infty$

\noindent in order to form a reasonable compactification of the moduli space. 
The detailed description of this divisor depends on the precise 
compactification under consideration. 

The natural compactification for the purpose of mirror symmetry is a toric
compactification defined by the secondary fan \cite{oda_park,mdmm}. 
In the case under consideration, 
this turns out to be a certain partial resolution of the
weighted projective space $WP^{1,1,2}$, whose toric diagram we draw 
in Figure 12. The model has four phases, the deep interior points of which are 
denoted in the figure by $A$, $B$, $C$ and $D$. Point $A$ corresponds to the 
smooth Calabi-Yau phase, point $D$ is a compact singular phase in which the 
exceptional divisor $E$ has been blown down, while points $B$ and $C$ 
correspond to non-geometric phases. The correct singularity structure of the 
moduli space is obtained upon replacing the generator $(0,1)$ with $(0,4)$, 
which leads to orbifold singularities at the points 
$B$,$C$ and $D$. Point $C$ is a 
Landau-Ginzburg point; at that point, the moduli space has a $\Z_8$ quotient
singularity. In Figure 12, we indicate the divisors associated with the 
generators in square brackets. The divisors $E$, $H$ and 
$L$ are as before, while the generator $(0,1)$ does not correspond to a 
divisor on $X$ but must be added in order to build the moduli space of its 
mirror (this is sometimes called the `non-compact generator'). Figure 12 
provides a parameterization of the moduli space by 
means of algebraic coordinates \cite{small_distances1}. 
The dotted lines in the figure represent the compactification divisors,
which connect the four deep interior points; these are the 
`rational curves at infinity' of \cite{small_distances1}. 
There exists a common point of intersection among 
$C_0,C_1$ and $C_{con}$, denoted by $S$ in the figure. This point 
corresponds to $\psi=0, \phi=1$. The curve $C_1$ also intersects the divisor
$D_{(0,-1)}$ in a point $T$, which corresponds to $\phi=1,\psi=\infty$. 
We also indicate the point $U$ of tangency between $C_{con}$ and 
$C_{\infty}$, which corresponds to $\phi\rightarrow \infty, 
8\phi^{-1}\psi^4\rightarrow -1$. Note that the K\"ahler cone can be identified 
with the cone spanned by the generators $(1,0)$ and $(0,-1)$; in particular,
point $A$ corresponds to the large radius limit of our two-parameter 
model\footnote{Further resolutions--some of them non-toric-- are needed in 
order to produce a compactification which is smooth and 
such that the various boundary 
divisors intersect with normal crossings~\cite{2pm1}. 
This is useful for an exhaustive study of monodromies, 
but will not be important for us.}.

\vskip 0.5in

\begin{center}
\scalebox{0.4}{\input{sec_fan.pstex_t}}
\end{center}

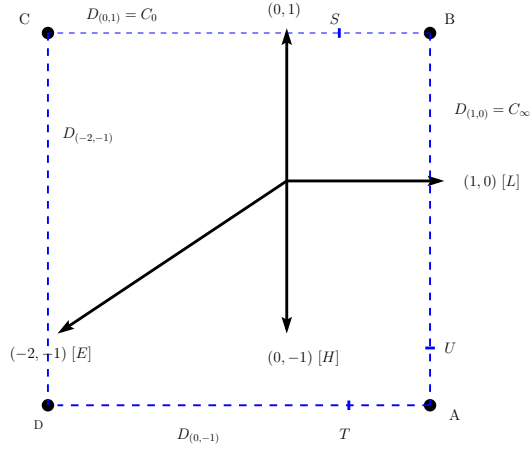
\hskip 1in \begin{center}Figure 12. {\footnotesize The secondary fan}
\end{center}

\vskip 0.4in

Figure 13 is a diagram of the curves $C_1$ and $C_{con}$, as well as of the 
compactification divisors $D_{(0,1)}=C_0$, $D_{(1,0)}=C_\infty$ and 
$D_{(0,-1)}$ (we do not draw $D_{(-2,-1)}$, since it will not be 
important below). The curves 
$C_1$ and $C_{con}$ intersect in two points, one of which is the point $S$ 
already discussed above. The other is the point $R$, which corresponds to 
$8\psi^2=-2,\phi=1$. These points will play an important role below.

\

\

\

\hskip 1.6in\scalebox{0.4}{\input{discrim.pstex_t}}

\hskip 1in \begin{center}Figure 13. {\footnotesize 
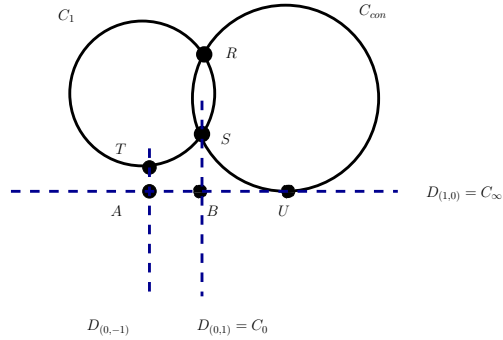
Components of the discriminant locus}
\end{center}

\

\

In order to understand the role of $C_{con}$ and $C_1$ in defining the phase 
structure of the model, it is useful to determine the image of these
components of the discriminant locus in the (uncomplexified) 
K\"ahler moduli space, with respect with the algebraic coordinates defined 
in \cite{small_distances1}. 
This can be easily achieved by using the asymptotic 
form of the mirror map given on pages 40 -- 41 of \cite{2pm1} (or by  
direct consideration of the monomial-divisor mirror map of \cite{mdmm}).
The result is plotted in Figure 14. We see that the asymptotic branches of 
$C_{con}$ separate phases $A$ and $B$, $B$ and $C$ 
as well as phase $C$ from $A$ and $D$, 
but they do not separate phase $A$ from $D$. 
This last phase boundary is induced by the curve $C_1$, 
which also contributes to the separation between phases $B$ and $C$.

\vskip 0.8in
\hskip 1.2in\scalebox{0.6}{\begin{picture}(0,0)%
\epsfbox{disc_graph.pstex}%
\end{picture}%
\setlength{\unitlength}{3947sp}%
\begingroup\makeatletter\ifx\SetFigFont\undefined%
\gdef\SetFigFont#1#2#3#4#5{%
  \reset@font\fontsize{#1}{#2pt}%
  \fontfamily{#3}\fontseries{#4}\fontshape{#5}%
  \selectfont}%
\fi\endgroup%
\begin{picture}(7217,6667)(1179,-6481)
\end{picture}
}

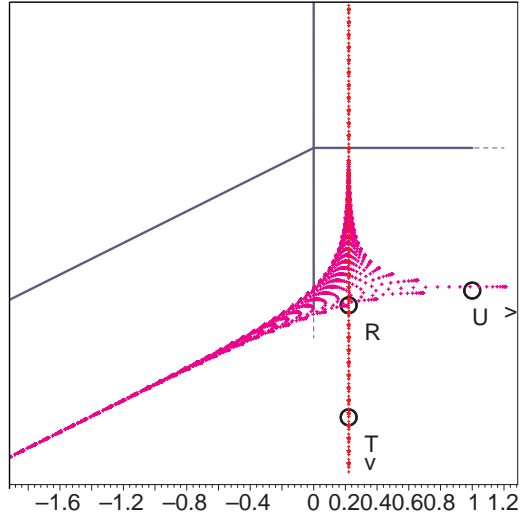
\hskip 1in \begin{center}Figure 14. {\footnotesize The image of the 
discriminant components $C_1$ and $C_{con}$ with respect to the algebraic 
measure. The image of $C_{con}$ is the three-pronged structure at the center 
of the diagram, extending towards infinity in directions parallel with the 
generators $(0,1)$, $(1,0)$ and $(-2,-1)$. The image of $C_1$ is the 
vertical line parallel with the generator $(0,1)$, located at a distance 
given by $\frac{1}{2\pi}{\rm log} 4\approx 0.2206$ in the positive 
direction along 
this generator. We also display the image of the points $R$,$S$,$T$ and $U$. 
The image of $R$ has coordinates 
$(\frac{1}{2\pi}{\rm log} 4 ,-\frac{9}{2\pi}{\rm log 2})\approx 
(.2206,-.9928)$, while the images of $S,T$ lie at infinity along the 
image of $C_1$. The image of $U$ lies at infinity on the horizontal branch of 
the image of $C_{con}$. This branch asymptotes to a line parallel 
with the generator $(1,0)$ and placed at a distance 
$\frac{8}{2\pi}{\rm log} 2 \approx 0.882$ below it.}
\end{center}

\

\

In what follows, we will only be interested in the locus $C_1$, 
leaving a discussion of other interesting loci to future work. 
As observed in \cite{2pm1}, this locus has the interesting 
property that for $(\psi,\phi) \in C_1$, the family $Y$ is birationally 
equivalent to the mirror of a complete intersection of a quadric and a 
quartic in $\P^5$, one of the models we studied in Section 5. 
Naively, one expects $C_1$ to be mirror to the
locus ${\check C}_1$ where the fiber of the ruling of the 
exceptional divisor $E$ of $X$ is blown-down to zero size, since, as 
mentioned in \cite{2pm1}, $X$ becomes birational to a $(2,4)$ complete 
intersection on this locus. Therefore, one naively expects    
that a D2-brane wrapping a rational curve in the class $l$ 
becomes massless everywhere on the locus  ${\check C}_1$. 
Note that it is possible to keep the area of the $\P^1$ fiber $l$ of 
the ruling equal to zero while varying the area of the base $\Sigma$, which 
is why we obtain a one-parameter locus; the classical 
volume of $E$ is of course identically zero on ${\check C}_1$. 
The entire locus ${\check C}_1$ corresponds to 
the wall in the K\"ahler moduli space where $\int_{l}{J}=0$, i.e. 
$\tau_2=0,~\tau_1\geq 0$.

As we show below by direct computation, this picture receives important 
modifications due quantum corrections. Indeed, we will see that the 
D2-brane wrapping $l$ does {\em not} acquire vanishing mass on the mirror 
locus to $C_1$ due to the familiar fact that there is a 
nonzero $B$-field flux through cycles in $l$ at generic points on this locus.
Nevertheless, the classical conclusion that $\int_{l}{J}=0$ is still 
valid after quantum 
corrections are taken into account.
The D-brane which becomes massless everywhere on 
(the mirror to the locus) $C_1$, is a $D4$-brane, as we will
see explicitly.
In the last subsection, we also consider the restricted theory on the 
locus $C_1$ and show explicitly that it is equivalent to the model 
$\P^5[2,4]$. In particular, this allows us to identify the collapsing 6-brane
found in Section 5 with a certain 6-brane state of the 
`ambient' two-parameter model.

\subsection{Restriction of periods to the locus $C_1$}

A basis of periods for our model was obtained in \cite{2pm1}. 
Such a basis consists of the 6 periods denoted there by 
$\omega_0~...~\omega_5$, whose expansions were computed in \cite{2pm1} in some 
regions of the moduli space. The
precise form of $\omega_i$ depends on whether $i$ is even 
or odd, so we think of our collection of periods as being divided into 
two subsets, $\{\omega_{2j}\}_{j=0..2}$ and $\{\omega_{2j+1}\}_{j=1..2}$.

We start with the 
$\psi$-expansions of $\omega_i$ obtained in \cite{2pm1}:
{\footnotesize
\bea
\label{periods_candelas}
\omega_{2j}(\psi,\phi)&=&-\frac{1}{4\pi^3}\sum_{r=1}^{3}(-1)^r
\sin^3(\frac{\pi r}{4}) \alpha^{2jr}\xi_r(\psi,\phi)~~\\
\omega_{2j+1}(\psi,\phi)&=&-\frac{1}{4\pi^3}\sum_{r=1}^{3}(-1)^r
\sin^3(\frac{\pi r}{4}) \alpha^{(2j+1)r}\eta_r(\psi,\phi)~~\nn
\eea}\noindent
where:
{\footnotesize
\bea
\xi_r(\psi,\phi)=\frac{1}{2\pi i}\int_{\gamma}{d\nu}{f_r(\nu,\psi,\phi)}
~~~~,~~~~
\eta_r(\psi,\phi)=-\frac{1}{2\pi i}\int_{\gamma}{d\nu}{e_r(\nu,\psi,\phi)}
~~,\nn
\eea}\noindent
with the integrands:
{\footnotesize
\bea
f_r(\nu,\psi,\phi)&=&\frac{\pi}{\sin\pi(\nu+r/4)}\frac{\Gamma^4(-\nu)}
{\Gamma(-4\nu)}(2^{12}\psi^4)^{-\nu}u_\nu(\phi)\nn\\
e_r(\nu,\psi,\phi)&=&\frac{\pi}{\sin\pi(\nu+\frac{r}{4})}\frac{\Gamma^4(-\nu)}
{\Gamma(-4\nu)}(2^{12}\psi^4)^{-\nu}\frac{u_\nu(\phi)\sin\pi(\nu+\frac{r}{4})-
u_\nu(-\phi)\sin(\frac{\pi r}{4})}{\sin(\pi \nu)}\nn~~.
\eea}\noindent 
The contour $\gamma$ is shown in Figure 15.

\

\

\hskip 1.4in\scalebox{0.6}{\begin{picture}(0,0)%
\epsfbox{2pm_contour.pstex}%
\end{picture}%
\setlength{\unitlength}{3947sp}%
\begingroup\makeatletter\ifx\SetFigFont\undefined%
\gdef\SetFigFont#1#2#3#4#5{%
  \reset@font\fontsize{#1}{#2pt}%
  \fontfamily{#3}\fontseries{#4}\fontshape{#5}%
  \selectfont}%
\fi\endgroup%
\begin{picture}(4844,4544)(2454,-4583)
\put(6751,-436){\makebox(0,0)[lb]{\smash{\SetFigFont{14}{16.8}{\rmdefault}{\bfdefault}{\updefault}$\nu$}}}
\put(3601,-3511){\makebox(0,0)[lb]{\smash{\SetFigFont{14}{16.8}{\rmdefault}{\bfdefault}{\updefault}$\gamma$}}}
\end{picture}
}

\hskip 1in \begin{center}Figure 15. {\footnotesize 
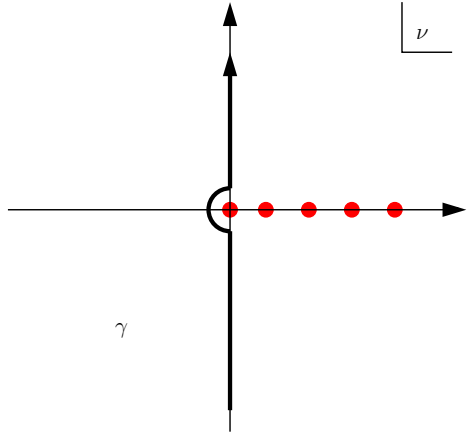
Contour of integration for $\xi_r$ and $\eta_r$. We also display the poles 
of the integrands at $\nu=n$ ($n$ a nonnegative integer) 
on the positive real axis. The integrands 
also have poles on the negative real axis, which are irrelevant for us.}
\end{center}

\

\

\noindent The function $u_\nu(\phi)$ in the expressions above is defined by:
{\footnotesize
\be
u_\nu(\phi):=
(2\phi)^\nu~\F[2,1][{-\frac{\nu}{2}~~,~~\frac{1-\nu}{2}},{1}](\frac{1}{\phi^2})
~~.
\ee}\noindent
We refer the reader to \cite{2pm1} for a discussion of the basic properties
of the functions $u_\nu(\phi)$, some of which we will use below.

The identity (\ref{basic}) of Appendix B, 
together with the completion formula allows us to write:
{\footnotesize
\bea
f_r(\nu,\psi,\phi)&=&8\pi^2\frac{\Gamma(\nu+\frac{r}{4})\Gamma(-\nu)^3}
{\prod'_{k=1~...~3}{\Gamma(-\nu+\frac{k}{4})}}z^{-\nu}u_\nu(\phi)~~\nn\\
e_r(\nu,\psi,\phi)&=&-8\pi\frac{\Gamma(\nu+\frac{r}{4})\Gamma(\nu+1)
\Gamma(-\nu)^4}{\prod'_{k=1~...~3}{\Gamma(-\nu+\frac{k}{4})}}z^{-\nu}
\left[u_\nu(\phi)\sin\pi(\nu+\frac{r}{4})+
u_\nu(-\phi)\sin\frac{\pi r}{4}\right]\nn~~,
\eea}\noindent
where 
{\footnotesize 
$\prod'_{k=1~...~3}{\{~...~\}}:=\prod_{k=1~...~3,k\neq 4-r}{\{~...~\}}$} 
and we defined $z:=(2\psi)^4$.

We are interested in the behaviour of the periods near the curve 
$C_1$, i.e. in the limit of (\ref{periods_candelas}) for $\phi\rightarrow 1$.
The expansions of these periods in the region $|\frac{8\psi^4}{\phi\pm 1}|<1$ 
are given in \cite{2pm1}, but this is not what we need, since $C_1$ lies in 
the complement of that region. In order to extract the information we desire, 
we must obtain the expansions of (\ref{periods_candelas}) for large $\psi$, 
i.e. for $|z|\gg 1$. This can be achieved once again by a computation 
of residues. 

Let us first discuss the expansion of $\xi_r$ in this region. Since
$|z|\gg 1$, we can close the defining contour to the right, towards
$+\infty$. The resulting Cauchy expansion receives contributions only from the 
poles of $f_r$ located at $\nu=n$, with $n$ a nonnegative integer; these poles 
are all of order $3$. Computation of the associated residues proceeds as 
before, giving:
{\footnotesize
\be
\xi_r(\psi,\phi)=4\pi^2\sum_{n=0}^{\infty}
\frac{\Gamma(n+\frac{r}{4})}{n!^3\prod'_{k=1~...~3}{\Gamma(-n+\frac{k}{4})}}
(-z)^{-n}H_r(n,\psi,\phi)u_n(\phi)~~.
\ee}\noindent A similar computation for $\eta_r$ (which receives contributions 
from poles of order $4$) gives:
{\footnotesize 
\be
\eta_r(\psi,\phi)=-\frac{8\pi}{3}\sum_{n=0}^{\infty}
\frac{\Gamma(n+\frac{r}{4})}{n!^3\prod'_{k=1~...~3}{\Gamma(-n+\frac{k}{4})}}
\sin(\frac{\pi r}{4})(-z)^{-n}K_r(n,\psi,\phi)u_n(\phi)~~.
\ee}\noindent 
To arrive at the last formula, we made use of the property:
\be
u_n(-\phi)=(-1)^n~u_n(\phi)~~,
\ee
which can be easily deduced from the Euler representation of $u_n$
(see \cite{2pm1}).

\noindent In these expansions, the quantities $H_r,K_r$ are defined by:
{\footnotesize
\bea
H_r(n,\psi,\phi):=(p_1(n,r)+\alpha_1(n,\phi)-{\rm log}(z))^2+p_2(n,r)+
\alpha_2(n,\phi)~~~~~~~~~~~~~~~~~~~~~~~~~~~~~~~~~~~~~~~~~~~~\nn\\
K_r(n,\psi,\phi):=(q_1(n,r)+\beta_1(n,r,\phi)-{\rm log}(z))^3+
3(q_1(n,r)+\beta_1(n,r,\phi)-{\rm log}(z))(q_2(n,r)+\beta_2(n,r,\phi))+\nn\\
+q_3(n,r)+\beta_3(n,r,\phi)~~,\nn
\eea}\noindent where:
{\footnotesize
\bea
p_i(n,r):=\psi^{(i-1)}(n+\frac{r}{4})-(-1)^i\left[
\sum'_{k=1~...~3}{\psi(-n+\frac{k}{4})}-3(i-1)!\sum_{k=1}^{n}{\frac{1}{k^i}}-
3\psi^{(i-1)}(1)\right]~~~~~~~~~~~~~~~~~~~~~~~~~~~~~~~~\\
q_i(n,r):=\psi^{(i-1)}(n+\frac{r}{4})+\psi^{(i-1)}(n+1)+
(-1)^{i-1}\left[
\sum'_{k=1~...~3}{\psi^{(i-1)}(-n+\frac{k}{4})}-4\psi^{(i-1)}(1)-4(i-1)!
\sum_{k=1}^{n}{\frac{1}{k^i}}\right]~~\nn
\eea
}
and:
{\footnotesize
\bea
\alpha_i(n,\phi)~~~&:=&\left[{\rm \frac{d^i}{d\nu^i}}
{\rm log}(u_\nu(\phi))\right]_{\nu=n}~~\nn\\
\beta_i(n,r,\phi)&:=&\left[{\rm \frac{d^i}{d\nu^i}}
{\rm log}\left(u_\nu(\phi)\sin\pi(\nu+\frac{r}{4})+
u_\nu(-\phi)\sin(\frac{\pi r}{4})\right)\right]_{\nu=n}~~.\nn
\eea}

One can rearrange the result in a form which displays $\omega_i$ 
as expansions in the basis of functions given by 
${z^{-n}~({\rm log}~z)^s}$. Write:
\bea
H_r(n,\psi,\phi)&:=&\sum_{s=0}^{2}{\rho_s(r,n,\phi)({\rm log}~z)^s}~~\nn\\
K_r(n,\psi,\phi)&:=&\sum_{s=0}^{3}{\lambda_s(r,n,\phi)({\rm log}~z)^s}~~.\nn
\eea
with the coefficients:
{\footnotesize
\bea
\rho_0:=p_2+\alpha_2+(p_1+\alpha_1)^2~~,~~
\rho_1:=-2(p_1+\alpha_1)~~,~~
\rho_2:=1\nn
\eea}\noindent
and:
{\footnotesize
\bea
\begin{array}{ccccccc}
\lambda_0&:=&(q_1+\beta_1)^3+3(q_1+\beta_1)(q_2+\beta_2)+(q_3+\beta_3)
&~~~~~&\lambda_2&:=&3(q_1+\beta_1)\nn\\
\lambda_1&:=&-3\left[(q_1+\beta_1)^2+(q_2+\beta_2)\right]&~~~~~
& \lambda_3&:=&-1
\end{array}~~.\nn
\eea}\noindent
We also define $\rho_3:=0$ in order to give a concise form to subsequent 
formulae.
Then we have:
\be
\label{log_expansions}
\omega_i(\psi,\phi):=\sum_{s=0}^3\sum_{n=0}^{\infty}{
z^{-n}({\rm log}~z)^sM_i(s,n,\phi)}~~,
\ee
with the expansion coefficients:
\be
M_i(s,n,\phi)=
\frac{u_n(\phi)}{n!^3\prod_{k=1}^{3}{\Gamma(-n+\frac{k}{4})}}L_i(s,n,\phi)~~,
\ee
where:
\bea
L_{2j}(s,n,\phi)~~~&:=&-\sum_{r=1}^{3}{(-1)^r\alpha^{2jr}
\sin^2(\frac{\pi r}{4})\rho_s(n,r,\phi)}~~\\
L_{2j+1}(s,n,\phi)&:=&\frac{2}{3\pi}\sum_{r=1}^{3}{(-1)^r\alpha^{(2j+1)r}
\sin^3(\frac{\pi r}{4})\lambda_s(n,r,\phi)}~~.\\
\eea
\noindent In order to arrive at these expressions, we made use of the identity:
\be
\sin(\frac{\pi r}{4})\frac{\Gamma(n+\frac{r}{4})}{\prod'_{k=1~...~3}{
\Gamma(-n+\frac{k}{4})}}=\pi (-1)^n
\frac{1}{\prod_{i=1~...~3}{\Gamma(-n+\frac{k}{4})}}~~,
\ee
which follows by applying the completion formula.

It is now straightforward to extract the behaviour of the periods in the 
limit $\phi\rightarrow 1$. Since $u_n(\phi)$ is a polynomial for integer $n$,
it has a finite limit at the point $\phi=1$, which is easily computed 
(\cite{2pm1}):
\be
\label{u1}
u_n(1)=\frac{(2n)!}{n!^2}~~.
\ee
Moreover, we have:
\be
u_\nu(1)=\frac{4^\nu\Gamma (1/2+\nu)}{\sqrt {\pi }\Gamma (1+\nu)}~~,
\ee
which allows us to compute the quantities $\alpha_i(n,1)$ and $\beta_i(n,r,1)$:
{\footnotesize
\bea
\alpha_1(n,1)&=&2{\rm log} (2)-\psi(n)-\frac{1}{n}+\psi(n+1/2)\nn\\
\alpha_2(n,1)&=&-\psi'(n)+\frac{1}{n^2}+\psi'(n+1/2)~~,
\eea
}
and:
{\footnotesize
\bea
\beta_1(n,1,1)&=&2{\rm log}(2)-\psi(n)-\frac{1}{n}+\psi(n+1/2)+\frac{1}{2}\pi +
\frac{1}{2}i\pi=\alpha_1(n,1)+\frac{1}{2}\pi +\frac{1}{2}i\pi\nn\\
\beta_1(n,2,1)&=&2{\rm log} (2)+i\pi -\psi(n)-\frac{1}{n}
+\psi(n+1/2)=\alpha_1(n,1)+i\pi\nn\\
\beta_1(n,3,1)&=&2{\rm log} (2)-\psi(n)-\frac{1}{n}+\psi(n+1/2)-
\frac{1}{2}\pi +\frac{1}{2}i\pi=
\alpha_1(n,1)-\frac{1}{2}\pi +\frac{1}{2}i\pi\nn\\
\beta_2(n,1,1)&=&\frac{1}{n^2}+\psi'(n+1/2)-\pi^2-\psi'(n)-
\frac{1}{2}i\pi^2=\alpha_2(n,1))-\pi^2-\frac{1}{2}i\pi^2\nn\\
\beta_2(n,2,1)&=&-\psi'(n)+\frac{1}{n^2}+\psi'(n+1/2)=\alpha_2(n,1)\\
\beta_2(n,3,1)&=&-\psi'(n)+\frac{1}{n^2}+\psi'(n+1/2)-\pi^{2}
+\frac{1}{2}i\pi^2=\alpha_2(n,1)-\pi^{2}+\frac{1}{2}i\pi^2\nn\\
\beta_3(n,1,1)&=&\frac{3}{2}i{\pi }^{3}+\psi''(n+1/2)+\frac{1}{2}
\pi^3-\psi''(n)-2\frac{1}{n^3}\nn\\
\beta_3(n,2,1)&=&\psi''(n+1/2)-\psi''(n)-2\frac{1}{n^3}\nn\\
\beta_3(n,3,1)&=&\frac{3}{2}i\pi^3 -\psi''(n)+\psi''(n+1/2)-
2\frac{1}{n^3}-\frac{1}{2}\pi^3~~.\nn
\eea}\noindent The relations given above are valid for $n\neq0$. While  
$\alpha,\beta$ are apparently singular at $n=0$, 
the associated analytic extensions $\alpha_i(\nu,1),\beta_i(\nu,r,1)$ 
have finite limits for $\nu\rightarrow 0$, which we list in Appendix B.
It follows that each of the periods $\omega_i$ has a finite limit for 
$\phi\rightarrow 1$, {\em as long as} $|z|>4$
\footnote{For $\phi=1$, our expansions converge if $|z|>4$.}. 
These limits are obtained 
by substituting $u_n(1),\alpha_i(n,1)$ and $\beta_i(n,r,1)$ 
in the expansions given above.

\subsection{Integral structure and a vanishing period}

A symplectic 
basis of the lattice $H_3(Y,\Z)$ was computed in \cite{2pm1}. 
If {\footnotesize 
$\Pi=\left[\begin{array}{c}\Pi_0\\ \Pi_1\\ \Pi_2\\ \Pi_3\\\Pi_4\\\Pi_5
\end{array}\right]=\left[\begin{array}{c}{\cal G}_0\\ {\cal G}_1\\ 
{\cal G}_2\\z^0\\z^1\\z^2\end{array}\right]$} is the period vector of 
$\Omega$ in this basis, then $\Pi$ and 
{\footnotesize 
$\omega:=\left[\begin{array}{c}\omega_0\\\omega_1\\\omega_2\\\omega_3
\\\omega_4\\\omega_5\end{array}\right]$} are related by $\Pi=m\omega$,
with:
{\footnotesize
\be
m=\left[\begin {array}{cccccc} 
-1&1&0&0&0&0\\1&0&1&-1&0&-1\\3/2&0&0&0&-1/2&0\\1&0&0&0&0
&0\\-1/4&0&1/2&0&1/4&0\\1/4&3/4&-1
/2&1/2&-1/4&1/4\end {array}\right ]~~.
\ee
}\noindent In particular, the periods $\Pi_4=z^1,\Pi_5=z^2$, which correspond 
to the integral of 
$\Omega$ along 3-cycles which are mirror to members of the classes
$h$ and $l$ respectively, are given by the following linear combinations of 
the the periods $\omega$:
{\footnotesize
\bea
c_1&=&\left [\begin {array}{cccccc} -1/4&0&1/2&0&1/4&0\end {array}\right]\\ 
c_2&=&\left [\begin {array}{cccccc} 1/4&3/4&-1/2&1/2&-1/4&1/4\end {array}
\right ]~~.
\eea}
Moreover, the period $\Pi_3=z^0$ coincides with the fundamental period 
$\omega_0$, which corresponds to the integral of $\Omega$ 
over a 3-cycle mirror to the homology class of a point.

Now let us look for an integral cycle with the property that the associated 
period of $\Omega$ vanishes identically on the locus $C_1$. Such a period 
is of the form:
\be
\omega_v=\beta\omega=\sum_{i=0~...~5}{\beta_i\omega_i}~~,
\ee 
where the row vector $\beta$ is such that $\beta_{int}=\beta m^{-1}$ has 
integral entries. By virtue of (\ref{log_expansions}), 
the condition that $\Omega_v|_{C_1}$ is identically 
zero implies that the following linear combination of the coefficients 
$L$ must vanish for all nonnegative integers $n$ and all $s=0,1,2,3$:
\be
\label{lc}
\sum_{i=0~...~5}{\beta_iL_i(s,n,1)}=0~~.
\ee
View $L_i=L_i(s,n,1)$~$(i=0~...~5)$ as sequences defined over the 
countable set $\{0,1,2,3\}\times \{0,1,2.....\}$.Then we are looking for 
a nontrivial linear relation among these five sequences. 
The existence of such a 
relation requires that all truncated determinants associated with the infinite 
linear system (\ref{lc}) vanish, a condition which is easily seen to hold 
numerically. Then a solution of (\ref{lc}) can be found by doing a numerical 
search for small integer values of $\beta_i$, with the result that the 
vector:
{\footnotesize
\be
\beta_v:=\left [\begin {array}{cccccc} 0&0&1&-1&1&-1\end {array}\right ]
\ee}\noindent 
satisfies (\ref{lc}). We conclude that the period:
\be
\omega_v(\psi,\phi)=\omega_2(\psi,\phi)-\omega_3(\psi,\phi)+
\omega_4(\psi,\phi)-\omega_5(\psi,\phi)
\ee
vanishes when restricted to $C_1$. To check integrality of the associated 
3-cycle $\gamma_v$, it suffices to compute the row vector:
{\footnotesize
\be
\beta_{int}=\beta m^{-1}=\left [\begin {array}{cccccc} 0&1&-2&2&0&0
\end {array}\right ]~~,
\ee}\noindent
which is integral indeed. 

The nature of the cycle ${\tilde \gamma}_v$ mirror to $\gamma_v$ can be 
determined by finding the position of the periods $\omega_v$ in the 
monodromy weight filtration  of $H_3(Y)$ 
associated with the large complex structure point.
By using the monodromy matrices 
computed in \cite{2pm1}, one can easily see that 
this filtration can be represented in the basis $\omega$ by
\footnote{This is the {\rm reduced} form of the filtration -- see Appendix A.}:
{\scriptsize
\bea
{\cal W}=[{\cal W}_0,{\cal W}_1,{\cal W}_2,{\cal W}_3]=
[\left [\begin {array}{c} 1\\0\\0
\\0\\0\\0
\end {array}\right ],\left [\begin {array}{ccc} 1&0&0
\\0&3&0\\0&0&2\\0
&2&0\\0&0&1\\0&1&0\end {array}
\right ],\left [\begin {array}{ccccc} 1&0&0&0&0\\0&0
&-3&3&0\\0&1&0&0&0\\0&0&0&1&0
\\0&0&0&0&1\\0&0&1&0&0\end {array}
\right ],\left [\begin {array}{cccccc} 0&0&0&0&0&1\\0
&0&0&0&1&0\\0&0&0&1&0&0\\0&0&1&0&0
&0\\0&1&0&0&0&0\\1&0&0&0&0&0
\end {array}\right ]]\nn
\eea}\noindent where the periods associated with the linear combination of 
$\omega_0~...~\omega_5$ given by the columns of each matrix span the 
corresponding weight subspace. It is easy to see that $\beta_v$
is a linear combination of the columns of ${\cal W}_2$, so that the mirror 
cycle ${\tilde \gamma}_v$ is a 4-cycle.

Numerical analysis also shows that the periods 
$\Pi_1={\cal G}_1,\Pi_2={\cal G}_2,\Pi_3=z^0,\Pi_5=z^2$ have vanishing 
imaginary part along the locus $C_1$, though none of them vanishes identically
on this locus. In particular, none of 
the special coordinates on the moduli space, 
which are given by:
\be
t_1=\frac{\Pi_4}{\Pi_3}=\frac{z^1}{z^0}~~,~~
t_2=\frac{\Pi_5}{\Pi_3}=\frac{z^2}{z^0}
\ee
vanishes identically on $C_1$. The coordinate $t_2$ is {\em real} along 
the locus $C_1$, while the coordinate $t_1$ is neither purely real, nor purely 
imaginary. Since the mirror K\"ahler class is given by:
\be
k=B+iJ=t_1{\check H}+t_2{\check L}~~,
\ee
it follows that:
\bea
\begin{array}{ccccccc}
\int_{l}{k}&=&t_2&~~,~~ &\int_{h}{k}&=&t_1
\end{array}~~,
\eea
and since $J={\rm Im}(k)={\rm Im}(t_1){\check H}$, this implies:
\bea
\begin{array}{ccccccc}
\int_{l}{J}&=&0 &~,~ &\int_{h}{J}&=&{\rm Im}(t_1)
\end{array}~~.
\eea

As reviewed above, the mirror of $C_1$ is classically a 
locus in the moduli space of $X$ where the exceptional divisor $E$ 
has been blown down to an curve of genus $3$. 
The result $\int_{l}{J}=0$ on 
this locus confirms that part of this classical geometric 
intuition is preserved in the quantum setting: 
the geometric volume of the fiber of the ruling remains zero.
However, the result $|\int_{l}{k}|=|t_2|\neq 0$ shows that
the B-field modifies this conclusion for the quantum volume 
of $l$, ensuring that 
the mass of the $D2$-brane wrapping $l$ does not vanish identically 
on $C_1$. In fact, we have some numerical evidence that
no $D2$-brane becomes identically massless on this locus.
Rather, as shown above, the vanishing period we found:
\be
\omega_v=\Pi_1-2\Pi_2+2\Pi_3={\cal G}_1-2{\cal G}_2+2z^0~~
\ee
corresponds to a $4$ - cycle. Hence it is the exceptional
divisor $E$ --- possibly mixed together with lower dimensional
cycles --- which has zero quantum volume on the locus $C_1$.

In this example it proves instructive to compare these exact results with 
semiclassical calculations. To do so, consider the quantities:
\bea
\begin{array}{ccccccc}
\int_{H}{k^2}&=&8t_1(t_1+t_2) &~~,~~ &\int_{L}{k^2}&=&4(t_1)^2
\end{array}\\
\int_{E}{k^2}=\int_{H}{k^2}-2\int_{L}{k^2}= 8t_1t_2~~,~~~~~~~~~~~\nn~~
\eea
and:
\bea
\begin{array}{ccccccc}
\int_{H}{J^2}&=&8({\rm Im}(t_1))^2 &~,~ &\int_{L}{J^2}&=&4({\rm Im}(t_1))^2
\end{array}\\
\int_{E}{J^2}=0\nn~~.~~~~~~~~~~~~~~~~~~~~~~~~~~
\eea\noindent
This amounts to using the {\em classical} relations for the volume, but 
applied to the quantum-mechanically correct values of $J$ and $k$ on the 
locus mirror to $C_1$. The result $\int_{E}{J^2}=0$ agrees with classical 
reasoning, showing that the mirror of $C_1$ is indeed the locus where  
$E$ collapses to zero classical volume. 
Thus, whereas in all
previous examples the location in
the K\"ahler moduli space where we expect to
find collapsed cycles based on classical reasoning is shifted
by quantum effects, in this
model the classical and quantum loci agree. (This is somewhat
akin to what happens in flop transitions where the flopping curve
has zero quantum volume on the classical flop wall.)
However, the D4-brane state which 
becomes massless at least naively seems to carry some $D0$-brane charge 
(though this statement is, of course, affected by the ambiguities 
discussed above). Moreover, the D2-branes wrapping holomorphic 
subcycles of $E$ remain massive on $C_1$, a result in stark contrast 
with classical geometric reasoning.

\subsection{Postscript: Quantum geometry of the restricted theory}

As mentioned above, it is believed that the restriction of our two-parameter 
model to the locus $C_1$ is equivalent to the one-parameter model 
$\P^5[2,4]$ discussed in Section 4. The argument given in \cite{2pm1} 
for this equivalence is based on classical geometry, namely on the fact 
that the restrictions of both $Y$ and $X$ to the locus $C_1$ and its 
naive mirror locus are birationally equivalent to the one-parameter mirror 
families $Y'$, respectively $X'$ describing the model of subsection 4.1. 

Having computed the required analytic continuations of all periods, we are 
in a position to perform a detailed check of this statement, which will 
also serve to complete the picture given above of what happens 
quantum-mechanically when one restricts to $C_1$. For this purpose, we must 
understand precisely how the one-parameter model $\P^5[2,4]$ 
is `embedded' in the two-parameter model at the quantum level. We 
start with the following natural geometric proposal for the embedding of 
$Y'$ into $Y$. Consider a period $U$ 
of the two-parameter model $Y$ as one approaches the locus $C_1$. If $U$ 
has nontrivial monodromy around $C_1$, then the process of restricting $U$ to 
$C_1$ is in general ill-defined. Therefore, it is natural to postulate 
that only those periods $U$ which have trivial monodromy around $C_1$ 
have a counterpart in the restricted model $Y$. 
\footnote{The alert reader may have 
noticed that this intuitive argument is not as clear-cut as it may seem. 
Indeed, it is perfectly possible to have periods with 
nontrivial monodromy around some 
component of the discriminant locus, which however do have a well-defined 
limit along that component. A familiar example in lower dimension is provided 
by a generic period on the mirror quintic in the vicinity of the conifold 
point. As shown in \cite{quintic}, such a period has the form:
\bea
U={\rm constant}~(z-1){\rm log}(z-1)+g(z)~~,\nn
\eea 
with $g$ a function regular at $z=1$; in that case, $U$ has a 
well-defined limit at $z=1$ even though it has nontrivial monodromy 
around the conifold. 
Hence the argument above is based on something more than the existence of 
a well-defined limit on $C_1$; essentially, it is an argument about the 
monodromy weight filtration defined by the degeneration associated with $C_1$.
In any case, correctness of 
this ansatz is justified by the 
consistency of the results we will obtain. }

We thus identify $H_3(Y')$ with the subspace of $H_3(Y)$ consisting of those 
classes which have trivial monodromy around $C_1$. The latter can be easily 
determined as the subspace of period vectors fixed by the associated monodromy 
operator (denoted by $B$ in \cite{2pm1}), with the result that it is spanned 
by the linear combinations of $\omega_0~...~\omega_5$ defined by the rows of 
the following matrix:
{\footnotesize 
\be
\label{A_matrix}
A:=\left [\begin {array}{cccccc} 0&0&0&0&1&1\\0&1&0&-3&
-3&0\\1&0&0&3&3&0\\0&0&1&1&0&0
\end {array}\right ]~~.
\ee}\noindent
Denote the associated periods (restricted to $C_1$) by:
\be
\omega'_j(z):=\sum_{i=0~...~5}{A_{ji}\omega_i(z,1)}
~~\mbox{~for~all~}j=0~...~3~~.
\ee
In order to understand the precise relation between the two models, we must
compute the transition matrix $P$ from the Meijer basis $U_j$ of $\P^5[2,4]$ 
to the basis $\omega'_j$. The existence of a {\em constant} such 
matrix on the locus $C_1$ amounts to a proof of the quantum equivalence 
between the restriction of $Y$ to $C_1$ and this model. 

In order to check this statement and obtain the matrix $P$, it is convenient 
to use a matrix formalism. Considering the column vectors
{\footnotesize $\omega':=\left[\begin{array}{c}
\omega'_0\\\omega'_1\\\omega'_2\\\omega'_3\end{array}\right]$} and  
{\footnotesize $U:=\left[\begin{array}{c}
U_0\\U_1\\U_2\\U_3\end{array}\right]$}, \linebreak 
the row vector {\footnotesize $Z(z)=\left[\begin{array}{cccc}
1&{\rm log} z&({\rm log} z)^2&({\rm log} z)^3\end{array}\right]$} as well as 
the matrix 
$M(n)=(M_{si}(n))_{s=0~...~3,i=0~...~5}$ with entries $M_{si}(n):=M_i(s,n)$ (where 
$M_i(s,n)$ are given in (\ref{log_expansions})), we can write :
\be
\omega'(z)^t=\sum_{n=0}^\infty{z^{-n}Z(z)M'(n)}~~,
\ee
with $M'(n)=M(n)A^t$.
Using the results of Section 4, one can similarly write:
\be
U(x)^t=\sum_{n=0}^\infty{x^nZ(x)l(n)}~~,
\ee
where $x$ is the coordinate on $C_1$ defined by the property that 
{\scriptsize $U_0(x)=\F[4,3][{1/2,1/2,1/4,3/4},{1,1,1}](x)$} and 
$l(n)=(l_{sj}(n))_{s,j=0~...~3}$ is a matrix with entries:
\be
l_{sj}(n)=\frac{(-1)^j}{j!}\frac{(1/2)_n(1/2)_n(1/4)_n(3/4)_n}{n!^4}
v_{sj}(n)~~.
\ee
Here $v_{sj}(n):=v_j(s,n)$ with $v_j(s,n)$ given in (\ref{nu_expansion}) and
(\ref{vs}). 

It is important to notice that the `hypergeometric' 
coordinate $x$ on $C_1$ does not coincide with $z$.
In fact, these two coordinates 
are expected to differ by the rescaling:
\be
z=-\frac{4}{x}~~,
\ee
as can be seen by comparing the expression for $\omega_0$ 
given on page 27  of \cite{2pm1} and the expression of the
$\P^5[2,4]$ fundamental period given in Table 3.1. of 
\cite{Candelas_periods}.
In other words, we expect the relation:
\be
\omega'(z)=PU(x)~~.
\ee
Instead of testing this relation directly, it proves more convenient to 
proceed as follows. Define a new coordinate $y$ on $C_1$ by 
$y=\frac{1}{w}=\frac{1}{\kappa x}$,
where $w=\kappa x$ is the natural variable on the moduli space of $\P^5[2,4]$
required by the monomial-divisor mirror map; thus $\kappa>0$ is given by:
\be
{\rm log} \kappa =2\psi(1/2)+\psi(1/4)+\psi(3/4)-4\psi(1)=-5{\rm log}(4)~~.
\ee
Hence $\kappa=4^{-5}$ and $y=\frac{4^5}{x}$. 
Then we expect $z=-\frac{4}{x}=\frac{1}{\mu}y$ with 
$\mu=-4^4$ and we must test the existence of a matrix $P$ such that
$\omega'(\frac{y}{\mu})=P U(\frac{1}{\kappa y})$. It is now easy to see that 
the associated rescalings have the following effect on the matrix function 
$Z$:
\be
Z(\frac{y}{\mu})=Z(y)C(\mu)~~,~~
Z(\frac{1}{\kappa y})=Z(y)c(\kappa)~~.\\
\ee
Here $C(\mu),c(\kappa)$ are 
two upper triangular matrices whose nonzero entries are given by:
\be
C_{sp}(\mu)=(-1)^{p-s}\left(\begin{array}{c}p\\s\end{array}\right)
({\rm log} \mu)^{p-s}~~,~~
c_{sp}(\kappa)=(-1)^{p}\left(\begin{array}{c}p\\s\end{array}\right)
({\rm log} \kappa)^{p-s}
\ee
for all $0\leq s\leq p\leq 3$. 
It follows that there should exist an invertible matrix $P\in {\bf GL}(4,\C)$
such that:
\be
\label{cond}
\mu^nC(\mu)M(n)A^t=\kappa^{-n}c(\kappa)l(n)P^t~~
\mbox{~for~all~integers~} n\geq 0~.
\ee
To determine a candidate for $P$, it suffices to consider this equation for 
$n=0$, which gives:
\be
P^t=l(0)^{-1}c(\kappa)^{-1}C(\mu)M(0)A^t~~.
\ee
The reason why we prefer to work with the variable $y$ can now be
made clear. Substituting the formulae for the expansion 
coefficients of $\omega_i(z)$ and $U_j(x)$ gives very complicated 
expressions for $M(0)$ and $l(0)$, which are difficult to simplify directly. 
However, by virtue of our experience in Section 3, it can be expected 
that rescaling to the variable $y$ has the effect of dramatically simplifying 
the expressions involved. This is because it is much easier in practice to 
simplify the linear combinations of the elements of $l(0)$ which form the 
matrix $c(\kappa)l(0)$ than the elements of $l(0)$ themselves; 
the same turns out to be the case for $C(\mu)M(0)$. 
Going through this exercise leads to the following remarkably simple 
expressions:
{\footnotesize
\bea
C(\mu)M(0)A^t &=&
\sqrt{2\pi}\left [\begin {array}{cccc} 
-2/3\frac{-7\pi^3+99i\zeta (3)}{\pi^3}&
\frac{61\pi^3-1320i\zeta(3)}{6\pi^3}
&\frac{-13\pi^3+396i\zeta(3)}{2\pi^3}&
3{\frac{-\pi^3+22i\zeta (3)}{\pi^3}}\\
\frac {17i}{2\pi}&\frac {29i}{3\pi}&-\frac {15i}{2\pi}&-\frac {9i}{2\pi}\\
-\frac{4}{\pi^2}&-\frac{11}{2\pi^2}&\frac{9}{2\pi^2}&\frac{3}{\pi^2}\\
-\frac {i}{2\pi^3}&-\frac {5i}{3\pi^3}&\frac {3i}{2\pi^3}&
\frac {i}{2\pi^3}\end {array}\right ]~~\nn\\
c(\kappa)l(0)~~~~~~&=&~~~~~~
\left [\begin {array}{cccc} 1&0&7/6{\pi }^{2}&22\zeta (3)
\\0&1&-i\pi &{\frac {11}{6}}{\pi }^{2}
\\0&0&1/2&0\\0&0&0&1/6\end {array}\right ]~~,
\eea}\noindent
which immediately give:
{\footnotesize
\be
\label{embedding}
P=\sqrt{2\pi}\left [\begin {array}{cccc} 14&6\frac {i}{\pi }&
-8\pi^{-2}&-3{\frac {i}{{\pi }^{3}}}\\23&17
{\frac {i}{\pi }}&-11{\pi }^{-2}&-10{\frac {i}{{
\pi }^{3}}}\\-17&-15{\frac {i}{\pi }}&9
{\pi }^{-2}&9{\frac {i}{{\pi }^{3}}}\\-10
&-4{\frac {i}{\pi }}&6{\pi }^{-2}&3{\frac {i}{
{\pi }^{3}}}\end {array}\right ]~~.
\ee}\noindent
In order to arrive at these results, we used relations (\ref{identities2})
of Appendix B.
Using this value of $P$, one can now check numerically that (\ref{cond}) is
indeed satisfied. We did this up to $n=40$ and with $60$ significant digits, 
but we were not able to simplify the resulting expressions in order to give 
an analytic proof to this claim. 

The matrices (\ref{A_matrix}) and (\ref{embedding}), give a complete 
characterization of the 
{\em quantum} 
relation between the restricted model $\P^5[2,4]$ and the ambient 
two-parameter model. It should be noted that we obtained considerably 
more detailed information about this relation than could have been extracted 
by less direct methods such as those of \cite{small_distances1}. 
For example, following 
\cite{small_distances1}, one can
consider the restriction of the Picard-Fuchs equations of the two-parameter 
model to the locus $C_1$ and check that the resulting ordinary equation is 
equivalent to the hypergeometric equation of the model  $\P^5[2,4]$ after 
performing an appropriate change of variable. While this procedure suffices 
to show that the restricted model is equivalent to $\P^5[2,4]$
at the quantum level, it does not fix the precise isomorphism between the two.
Doing so requires knowledge of the relation between two bases of 
periods of these models, which is precisely the information encoded by the 
matrices $P$ and $A$ computed above.

Now let us consider the vanishing period $U_v=U_3-3\pi^2 U_1$ which 
we found for the restricted model in Section 5. Since we know the integral 
structure of 
the 2-parameter model, we can obtain the integral structure of  
$\P^5[2,4]$ by restriction. In particular, since $\Pi=m \omega$ is the period 
vector in an integral basis, and since $\omega'=A\omega=Am^{-1}\Pi$, while 
$U_v=\beta U$, with $\beta:=\left(\frac{1}{\pi^3}\right)
\left[\begin{array}{cccc}0&-3\pi^2&0&1
\end{array}\right]$ and $U=P^{-1}\omega'$, 
it follows that the vanishing period $U_v$ which we found in Section 5 
can be expressed as:
\be
U_v=\beta P^{-1} A m^{-1} \Pi=\beta_{int} \Pi ~~,
\ee
with:
\be
\beta_{int}=-86 i\sqrt{2\pi}
\left [\begin {array}{cccccc} 86&-79&252&-190&172&-156\end {array}
\right]~~,
\ee
i.e. we have:
\be
\label{large_integers}
U_v=-86 i\sqrt{2\pi}
(86{\cal G}^0-79~{\cal G}^1+252~{\cal G}^2 -190z^0+172~z^1-156~z^2)
\ee
In particular, we see that $U_v$ is an integral linear combination of integral
periods up to a global factor; hence it is indeed associated with 
an integral 3-cycle which collapses at $x=1$. The reader can easily check that 
the large integers appearing in (\ref{large_integers}) are coprime. This 
underscores the rather nontrivial character of this vanishing period, when 
viewed from the perspective of the ambient two-parameter model. 
It would be  a difficult task to identify $U_v$ directly 
from the data of the two-parameter model, without making use of independent 
knowledge of the vanishing period for the model $\P^5[2,4]$.

It is instructive to re-interpret the result from 
this `ambient' perspective. The point $x=1$ 
corresponds to $z=-4$; this is the point $R$, one of the two points where 
$C_1$ intersects 
$C_{con}$ (the other point $S$ corresponds to $z=0\iff x=\infty$, 
which is the small radius limit of the model $\P^5[2,4]$.). 
As shown in \cite{2pm1}, there exists an integral cycle vanishing identically 
on the locus $C_{con}$, namely the cycle corresponding to the period 
$\omega_1-\omega_0={\cal G}^0$; however, it is easy to see that this 
period has nontrivial monodromy around $C_1$ and hence it does not 
restrict to a period of the model $\P^5[2,4]$. Clearly both of our cycles 
belong to the highest component ${\cal W}_3$ of the (reduced) large radius 
monodromy weight 
filtration of the two-parameter model; hence both can be interpreted 
as $6$-brane states in the mirror theory (the $IIA$ theory compactified on 
$X$). It follows that a partial picture of massless D-branes in the type IIA 
compactification on $X$ is as follows. First, there is a 6-brane state which 
becomes massless everywhere on the conifold 
locus $C_{con}$; second, there is another 6-brane 
state which becomes massless at one of the intersection points (the 
point $R$) between $C_1$ and $C_{con}$; when restricting to the locus $C_1$, 
this state induces the vanishing 6-brane of the resulting one-parameter 
model. Finally, there is a $4$-brane state which becomes massless 
everywhere on the locus $C_1$. 

\section{Summary and Conclusions}

One of the essential lessons of recent years is that much interesting
physics can be learned from studying systems at points in their
moduli space which naively appear to be singular. In a wide range of
examples, these singularities signal the appearance of new massless 
degrees of freedom that are responsible for yielding non-singular physics.
In the context of compactified string theory, these singularities are often
associated with geometrical degenerations of one form or another,
and the new massless degrees of freedom often 
arise from D-branes wrapped around collapsing cycles.

In this paper, we have studied in some detail the D-branes that
become massless along various loci in the moduli space of
Calabi-Yau examples --- compact and noncompact --- and used this to
extract properties of quantum volume in string theory. The features
we find confirm and extend previous results, and highlight significant
qualitative and quantitative departures from classical expectations.
Figures 6, 7, 8, 9 and 11  illustrate some of the new features
of quantum volume and can be thought of as a visual summary of
our results.  Although we have not overly stressed this point, 
we also showed that  the physical arguments which ensure that some 
D-brane mass vanishes within
a phase boundary can be used to generate interesting and seemingly novel
arithmetic identities. Equations (\ref{quintic_id}), (\ref{5_24id}),
(\ref{orbf_id1}) and (\ref{orbf_id2}) give some representative
examples. In fact, one can reverse the argument and use degenerations 
of Calabi-Yau manifolds in order to produce such identities in a systematic 
manner. It would be interesting to gain a better understanding of the 
mathematics behind this process, as well as of the connection between these 
results and the considerations of \cite{Moore_arithmetics}.  

From the physical point of view, there are, perhaps, two main  issues that
remain unresolved. The first is to determine the precise action of
mirror symmetry on the integral structures of a mirror pair $(X,Y)$. 
As we have seen, our inability
to resolve this issue prevents the precise identification of
the even dimensional D-branes on $X$ mirror to specified odd dimensional
D-branes on $Y$. Nevertheless, it is only the lower brane charges that
remain ambiguous as we can use the leading logarithmic behaviour of periods
on $Y$ to identify the maximal dimensional component of mirror D-branes on
$X$. Certainly, though, filling this gap in our understanding is an 
important subject for the future. The second issue is that of marginal 
stability. While we have studied D-branes at geometrical degenerations, 
there are other loci in moduli space where a given D-brane state might 
be massive, but unstable to decay. Perhaps the techniques
of this paper, as well as the important contributions of 
\cite{Douglas_quintic, Moore_arithmetics}, will allow
us to understand the marginal stability loci in detail for Calabi-Yau
compactifications. We hope to return to this issue in future work.

\appendix

\section{The monodromy weight filtration}

In this appendix, we give a very brief account of role played by monodromies 
in the study of moduli spaces of Calabi-Yau threefolds. 
This topic is intimately connected with the theory of variations of 
Hodge structure. Basic references for 
this subject (which we will not attempt to present in detail) 
are \cite{deligne} for the general theory of local systems and 
the geometric formulation of Fuchsian equations, \cite{deligne_hodge} for 
the abstract theory of Hodge structures, \cite{mwf} as general references
and \cite{mon_theorem} for the monodromy theorem. An excellent introduction 
to the mathematical structures underlying mirror symmetry (up to date at the 
level of 1997) can be found in \cite{morrison_aspects}.

For our purpose, the set-up is that of a family of Calabi-Yau threefolds 
(understood as complex manifolds of vanishing first Chern class), i.e. 
a (proper) holomorphic map $\pi:{\cal X}\rightarrow S$ over a connected base 
$S$ whose fibers 
$X_s=\pi^{-1}(s)$ are all diffeomorphic with a given Calabi-Yau manifold 
$X$. The complex manifold $S$ plays the role of a parameter space for 
the complex structure on $X_s$. For simplicity, we assume that ${\rm dim} 
S:=r$ 
is equal to $h^{2,1}(X)$, so that our family captures the full space of 
deformations of complex structure of $X$.
In general, one must allow for bases $S$ 
which are non-contractible; this appears because one is usually interested 
not only in the family ${\cal X}$, but in a compactification 
${\overline \pi}:{\overline {\cal X}}\rightarrow {\overline S}$ 
given by a (proper) map ${\overline \pi}$ which has singularities 
above the locus ${\overline S}-S$; in other words, we wish to allow the 
complex structure of $X$ to degenerate as we vary the parameters $s\in S$.
In order to describe the middle 
cohomology of the various fibers $X_s$, one considers 
the derived sheaf $R^3\pi_*\C_{\cal X}$, which is intuitively the sheaf 
having $H^3(X_s,\C)$ as its stalk above each point $s$. 
The sections of this sheaf are simply the topologically constant families 
of cohomology classes. The important subtlety is that, for a general basis $S$,
such sections will have nontrivial monodromies around the components 
of ${\overline S} -S$; these are described by actions 
$\rho_s:\pi_1(S,s)\rightarrow Aut(H^3(X_s))$ of the fundamental 
group of $S$ on the spaces $H^3(X_s)$. In other words, if 
$g(s)\in H^3(X)$  is such a section, then upon following a closed path 
$l:[0,1]\rightarrow {\cal S}$ starting and ending at a given point $s_0$,
the cohomology class $g(s)$ does not necessarily return to its original 
value $g(s_0)$ but rather to the value $\rho_{s_0}([l])g(s_0)$, where $[l]$ 
is the homotopy class of $l$ in $\pi_1(S,s_0)$. An important device for 
studying this situation, discussed at length in \cite{deligne} is to 
represent the local system $R^3\pi_*\C_{\cal X}$ 
as the sheaf of flat holomorphic sections of a 
pair $({\cal H},\nabla)$, where 
${\cal H}={\cal O}_S\otimes R^3\pi_*\C_{\cal X}$ is a holomorphic vector 
bundle and $\nabla$ is a connection on ${\cal H}$ which, in our context, is 
called the {\em Gauss-Manin} connection. In other words, each 
topologically-constant cohomology class in $H^3(X,\C)$ can be viewed as
a holomorphic section of ${\cal H}$ which is covariantly-constant with 
respect to $\nabla$. The existence of nontrivial monodromies around 
the components of 
${\overline S}-S$ is then reflected by the fact that the connection 
coefficients of $\nabla$ in a basis of sections which extend to smooth 
sections over a component of ${\overline S}-S$ will have poles along that 
locus (this is made more precise by the nilpotent orbit theorem, as we 
explain below). 

In order to understand the nature of the monodromies around a 
compactification point $P\in {\overline S}-S$, we consider the case when 
$P$ is the intersection of $r$ compactification divisors
having normal crossings at $P$. This situation can be represented by taking 
$S$ to be locally a Cartesian product $\Delta$ of $r$ punctured disks:
\be
\Delta=(D^*)^r~=\{z=(z_1~...~z_r) \in \C^r| 0<|z_i|<1,~\forall~i=1..r\}~~,
\ee
where $D^*=\{z\in \C| 0<|z|<1\}$. This has a natural compactification 
${\overline \Delta}=(D)^r$, with $D$ a copy of the unit disc. 
The compactification divisors
intersecting at $P=(0,0~...~0)$ are then locally given by the equations 
$z_i=0$. The fundamental group $\pi_1(\Delta)$ is 
generated by the $r$ circles 
$\gamma_i:\{~z_i=\frac{1}{2}e^{2\pi i t} ~(~t\in [0,1]~),~z_j=0 
\mbox{~for~} j\neq i\}$.
The {\em monodromy theorem} \cite{mon_theorem} 
assures us that  the 
automorphisms $T_i$ of $H^3(X)$ defined by transporting the 
cohomology classes around $\gamma_i$ are {\em quasi-unipotent},
i.e. each of them satisfies:
\be
\label{qunip}
(T_i^{a_i}-I)^{r_i}=0~~
\ee
for some pair of positive integers $(a_i,r_i)$.

To avoid lengthening this discussion, let us focus on the case when 
all $T_i$ are unipotent, i.e. one can take $a_i=1$ in (\ref{qunip}).
Then one can define operators 
$N_i={\rm log} T_i:=\sum_{k=1..r_i-1}{\frac{(-1)^{k-1}}{k}(T_i-I)^k}$
(the series defining the logarithm terminates due to the fact that 
$T_i$ is unipotent) as well as an operator-valued function:
\be
{\cal N}(z):=e^{-\frac{1}{2\pi i}\sum_{i=1..r}{({\rm log} z_i)~N_i}}~~
\ee
(this series again reduces to a finite sum, since the operators 
$N_i$ are obviously nilpotent).

We can now state another basic fact of relevance for us, namely the {\em 
nilpotent orbit theorem}. This states that one can use the monodromy operators
$T_i$ in order to generate an extension of ${\cal H}$ to a holomorphic 
vector bundle ${\overline {\cal H}}$ over ${\overline \Delta}$. More precisely,
given a flat local frame
\footnote{Flat with respect to the Gauss-Manin connection.} 
$(e_j(z))_{j=1..2r+2}$ of 
${\cal H}$, the sections $e_j(z)$ will generally have nontrivial monodromies 
around the curves $\gamma_i$ (which can be identified with the holonomy 
transformations of $\nabla $ along these curves). It follows that 
$e_i$ do not directly extend to ${\overline \Delta}$. 
However, the nilpotent orbit theorem states that the sections:
\be
\sigma_j(z):={\cal N}(z)e_j(z)~~
\ee
are single-valued (i.e. have trivial monodromy) around each of the curves 
$\gamma_i$. It follows that one can extend the bundle ${\cal H}$ to 
${\overline \Delta}$ trivially, i.e. by declaring that 
${\overline {\cal H}}$ is trivializable on a vicinity of each of $\gamma_i$ 
and extending $\sigma_i$ to a frame over that vicinity by continuity of their
local representations in that trivialization. Then a simple computation in the 
frame $\sigma_i$ shows that $\nabla $ extends to a connection with 
regular singular points on ${\overline {\cal H}}$~\footnote{Concretely,  
the connection matrix in a local frame of ${\overline {\cal H}}$
above the origin has poles along the curves $\gamma_i$. Abstractly, the 
extended connection gives a map ${\overline \nabla}:{\overline {\cal H}}
\rightarrow {\Omega}^1({\rm log} Z)\otimes {\overline {\cal H}}$, where 
${\Omega}^1({\rm log} Z)$ is the space of `1-forms in 
$d {\rm log} z_i=\frac{dz_i}{z_i}$' and 
$Z:={\overline \Delta}-\Delta$.}.

We can now describe the associated {\em monodromy weight filtration}. 
This is simply a filtration:
\be
0\subset W_0\subset W_1\subset ~...~ \subset W_{6}=H^3(X) 
\ee
of $H^3(X)$ with the property: 
\be
\label{filtration1}
N_iW_k\subset W_{k-2}~~,\mbox{~for~all~}k
\ee
and such that for all positive $a_1~...~a_r$, the powers of the 
operator $N=\sum_{i=1..r}{a_iN_i}$ induce isomorphisms:
\be
\label{filtration2}
N^p:W_{3+p}/W_{2+p}\rightarrow W_{3-p}/W_{2-p}~~ 
\ee
for all $p=0~...~3$, where $W_{-1}$ is defined to be the null vector space.
These properties can be used in order to determine $W$. 

Finally, let us recall some useful concepts introduced in 
\cite{morrison_quintic,morrison_fuchs,morrison_cpctfs}. The boundary point
$P$ is called {\em maximally unipotent} (or a 
{\em large complex structure limit point}) if:

(1)All monodromy transformations $T_j$ are unipotent

(2)${\rm dim} W_0={\rm dim} W_1=1$ and ${\rm dim} W_2=1+r$ 

(3) If $g_0 ~...~ g_r$ is a basis of $W_2$ such that $g_0\in W_0$, then the 
matrix $M:=(m_{ij})_{i,j=1..r}$ (defined by $N_ig_j=m_{ij}g_0$) is 
invertible. 

\noindent Then (\ref{filtration2}) gives 
$W_0=W_1\subset W_2=W_3\subset W_4=W_5\subset W_6=H^3(X)$ 
so we can work with the {\em reduced filtration}:
\be
\label{reduced_filtration}
0\subset {\cal W}_0\subset {\cal W}_1\subset {\cal W}_2 \subset {\cal W}_3 =
H^3(X)~~,
\ee
where ${\cal W}_k:=W_{2k}$ for all $k=0~...~3$.

For a one-parameter model ($r=1$), one has a single monodromy operator 
$T$ around $P$ and this definition is equivalent to the 
requirement that $T$ is maximally unipotent, i.e. $(T-I)^4=0$ and 
$(T-I)^3\neq 0$~\footnote{We assume a {\em compact} one-parameter model; the 
correct generalization for a non-compact case such as the $\C^3/\Z_3$ orbifold 
considered in Section 5 is to require $(T-I)^3=0$ and $(T-I)^2\neq 0$.}. 
Then (\ref{reduced_filtration}) can be  viewed as the 
natural filtration defined by the nilpotent operator $N={\rm log} (T-I)$. 
In other words, the fact that $T-I$ is nilpotent of maximal order means 
that the Jordan form of $N$ is:
\be
\label{jord}
N_J=\left[\begin{array}{cccc}
0&1&0&0\\
0&0&1&0\\
0&0&0&1\\
0&0&0&0
\end{array}\right]~~,
\ee
and ${\cal W}_k=<e_0~...~e_k>$, where $e_0..e_3$ is any basis of $H^4(X)$ in 
which the matrix of $N$ has this form. 
In terms of the associated 
Picard-Fuchs system, such a monodromy matrix corresponds to a basis 
of solutions $u_0~...~u_3$ with the property that~\footnote{If $N$ has the form
(\ref{jord}), then $T=e^{N}$ has the form
{\scriptsize $\left[\begin{array}{cccc}1 & 1&1/2& 1/6\\0&1&1&1/2\\0&0&1&1
\\0&0&0&1\end{array}\right]$}.}:
\be
u_j(z)=\sum_{0\leq i \leq j}{a_i(z){\rm log}(z)^j}~~,
\ee
where $a_i$ are single-valued around $z=0$. In other words, the spaces 
${\cal W}_j$ consist of the `${\rm log}^j$-monodromy periods', where the 
terminology indicates the {\em leading} logarithmic behaviour.

\section{Some useful identities}

For the convenience of the reader, we list a few identities which are 
important for simplifying some of the expressions encountered in this paper. 
A first group of identities can be obtained by starting with the basic 
equality:
\be
\label{basic}
\prod_{k=0}^{n-1}{\Gamma(x+k/n)}=(2\pi)^{\frac{n-1}{2}}~n^{1/2-nx}\Gamma(nx)~~,
\ee
which immediately leads to:
\bea
\sum_{k=0}^{n-1}{\psi(x+k/n)}-n\psi(nx)=-n{\rm log}(n)~~\nn\\
\sum_{k=0}^{n-1}{\psi^{(j)}(x+k/n)}-n^{j+1}\psi^{(j)}(nx)=0~~.
\eea
These relations can be used for $x=1/n$ to give:
\bea
\label{identities1}
\sum_{k=1}^{n-1}{\psi(k/n)}-(n-1)\psi(1)=-n{\rm log}(n)~~\nn\\
\sum_{k=1}^{n-1}{\psi^{(j)}(x+k/n)}=(n^{j+1}-1)\psi^{(j)}(1)~~.
\eea
\noindent We also recall that $\psi(1)=-\gamma$, where $\gamma$ is Euler's 
constant and that:
\be
\psi^{(j)}(1)=(-1)^{j-1}j!\zeta(j+1)~~,~~
\zeta(2k)~~=(-1)^{k-1}\frac{(2\pi)^{2k}}{2(2k)!}B_{2k}
\ee
with $B_j$ the Bernoulli numbers. In particular, we have
$\zeta(2)=\frac{\pi^2}{6}$~.

Another set of identities, useful for simplifying some of 
the expressions encountered in Section 6, is: 
{\footnotesize
\bea
\label{identities2}
\begin{array}{cccccccc}
\psi(1/4)&=&-\gamma-3~{\rm log}2-\frac{\pi}{2}~~&,&~~ 
\psi(3/4)&=&-\gamma-3~{\rm log}2+\frac{\pi}{2}\\ 
\psi(3/4) &=& -56~\zeta(3)+2\pi^3~~&,&~~
\psi(1/4) &=& -56~\zeta(3)-2\pi^3\\
\psi'(1/4)&+&\psi'(3/4) = 2\pi^2~~&,&~~
\psi''(3/4) &=& -56~\zeta(3)+2\pi^3\\
\psi''(1/4) &=& -56~\zeta(3)-2\pi^3~~&,&~~
\psi(1/2)&=&-\gamma-2~{\rm log}2\\
\psi'(1/2)&=&\frac{1}{2}\pi^2~~&,&~~
\psi''(1/2)&=&-14~\zeta(3)
\end{array}~~.
\eea}
\noindent These follow by applying (\ref{identities1}) for $n=4$ and $n=2$, 
together with the relation:
\be
\psi(x)-\psi(1-x)=-\pi {\rm ctg}(\pi x)~~
\ee
and the identities following from it by differentiation.

Finally, we list the limiting values of the quantities $\alpha_i(\nu,1)$ and 
$\beta_i(\nu,r,1)$ of Section 7 for $\nu\rightarrow 0$:
{\footnotesize 
\be
\alpha_1(0,1)=0~~,~~ \alpha_2(0,1)=1/3{\pi }^{2}
\ee}
and:
{\footnotesize
\be
\begin{array}{cccccccc}
\beta_1(0,1,1)&=&1/2i\pi +1/2\pi ~~&,&~~\beta_1(0,2,1)&=&i\pi \\
\beta_1(0,3,1)&=&1/2i\pi -1/2\pi~~&,&~~\beta_2(0,1,1)&=&-1/2i{\pi }^{2}-
2/3{\pi }^{2}\\
\beta_2(0,2,1)&=&1/3{\pi }^{2}~~&,&~~\beta_2(0,3,1)&=&-2/3{\pi }^{2}+
1/2i{\pi }^{2}\\
\beta_3(0,1,1)&=& 3/2i{\pi }^{3}-12\zeta (3)+1/2{\pi }^{3}~~&,&~~
\beta_3(0,2,1)&=&-12\zeta (3)\\
\beta_3(0,3,1)&=&-1/2{\pi }^{3}-12\zeta (3)+3/2i{\pi }^{3}&~.&~&~&~
\end{array}
\ee}

\end{document}